\documentclass[aps,prb,twocolumn,superscriptaddress,groupedaddress,showpacs]{revtex4-1}
\usepackage{graphicx,amsmath,color,placeins}
\def\k{\kappa}
\def\l{\lambda}
\def\d{\delta}
\def\w{\omega}
\def\W{\Omega}
\def\bk{{\bf k}}
\def\e{\epsilon}
\def\ve{\varepsilon}
\def\<{\langle}
\def\>{\rangle}
\def\D{\partial}
\def\e{\epsilon}
\let\hide\iffalse
\let\unhide\fi

\usepackage[caption=false]{subfig}

\begin{document}

\title{One-shot calculation of temperature-dependent optical spectra \\ and
       phonon-induced band-gap renormalization}

\author{Marios Zacharias}
\author{Feliciano Giustino}
\affiliation{Department of Materials, University of Oxford, Parks Road, Oxford OX1 3PH, United Kingdom}

\date{\today}

\begin{abstract}
Recently, Zacharias {\it et al.} [Phys. Rev. Lett. 115, 177401 (2015)] developed a new
{\it ab initio} theory of temperature-dependent optical absorption spectra and band gaps
in semiconductors and insulators. In that work the zero-point renormalization and the
temperature dependence were obtained by sampling the nuclear wavefunctions using a stochastic 
approach. In the present work, we show that the stochastic sampling of Zacharias {\it et al.} 
can be replaced by fully deterministic supercell calculations based on a single optimal 
configuration of the atomic positions. We demonstrate that a single calculation is able 
to capture the temperature-dependent band gap renormalization including quantum nuclear effects
in direct-gap and indirect-gap semiconductors, as well as phonon-assisted optical absorption 
in indirect-gap semiconductors. In order to demonstrate this methodology we calculate from 
first principles the temperature-dependent optical absorption spectra and the renormalization of direct
and indirect band gaps in silicon, diamond, and gallium arsenide, and we obtain good agreement
with experiment and with previous calculations. In this work we also establish the formal connection between
the Williams-Lax theory of optical transitions and the related theories of indirect absorption
by Hall, Bardeen, and Blatt, and of temperature-dependent band structures by Allen and Heine.
The present methodology enables systematic {\it ab initio} calculations of optical absorption 
spectra at finite temperature, including both direct and indirect transitions. This feature will 
be useful for high-throughput calculations of optical properties at finite temperature, and for 
calculating temperature-dependent optical properties using high-level theories such as GW 
and Bethe-Salpeter approaches.
\end{abstract}

\maketitle

\section{Introduction}\label{sec.1}
The electron-phonon interaction plays a central role in the optical properties of solids.
For example, electron-phonon couplings lead to the temperature dependence and the quantum zero-point 
renormalization of the critical point energies, to temperature-dependent broadening of 
light absorption and emission lineshapes, and to indirect optical transitions.

Recently, it has become possible to study these effects using {\it ab initio} calculations.
The phonon-induced renormalization of band gaps and band structures was investigated from
first principles in Refs.~\onlinecite{Allen_Cardona_1981, Marini_2008, 
FG_diamond,Gonze_2011,Cannuccia2011,Ponce_2014,Ponce_2014_2,Antonius2014,Monserrat2014, Ponce2015,Bartomeu_2015,Bartomeu_2016,Allen_arxiv}
starting from the theory of Allen and Heine (AH).\cite{Allen} The optical absorption in indirect-gap
semiconductors was studied from first principles in Ref.~\onlinecite{Noffsinger}
using the classic theory of Hall, Bardeen, and Blatt (HBB),\cite{hbb} and in 
Ref.~\onlinecite{Zacharias_2015} using the theory of Williams\cite{Williams} and Lax\cite{Lax} (WL).
A review of the standard formalism and the of most recent literature can be found in Ref.~\onlinecite{FG_review}.

In this manuscript we focus on the WL theory, and on how to perform accurate
and efficient {\it ab initio} calculations of temperature-dependent band gaps and
optical spectra in semiconductors and insulators using the WL formalism.
In its original formulation the WL theory was employed to study the vibrational
broadening of the photoluminescence spectra of defects in solids.\cite{Williams,Lax}
In a recent work we showed that the same theory can successfully be employed
for predicting temperature dependent optical spectra and band gaps in semiconductors,
including phonon-assisted indirect absorption.\cite{Zacharias_2015}  
The reason for this success is that a perturbative treatment of the WL theory 
naturally leads to the adiabatic approximations of the AH and the HBB theories.
In fact, as it was shown in Ref.~\onlinecite{Zacharias_2015}, the AH theory of temperature-dependent band structures 
can alternatively be derived from the WL theory by neglecting the optical matrix elements. Similarly, 
the adiabatic limit of the HBB theory of indirect optical absorption can be derived from 
the WL theory by retaining only one-phonon processes. The relations between the WL, the
HBB, and the AH theory will be analyzed in detail in Sec.~\ref{sec.general-theory}.

In the WL theory,\cite{Williams,Lax,Patrick2013,CEP_FG,Zacharias_2015} the
effect of quantum nuclear motion on the optical properties is described by first calculating
the optical spectrum evaluated at clamped nuclei, and then taking the expectation value
of this quantity over a given nuclear wavefunction. The temperature is introduced by
performing a canonical average over all possible nuclear quantum states.
Formally, this approach corresponds to a `semiclassical' Franck-Condon approximation,
whereby the initial quantum states are described by Born-Oppenheimer products of electronic
and nuclear wavefunctions, and the final quantum states are replaced by a classical
continuum. This approach is related to but does not coincide with the standard
adiabatic approximation. An extensive discussion of the formalism and its limit of
applicability can be found in Refs.~\onlinecite{Lax,CEP_FG}.

The key advantages of the WL method are (i) the calculations are simple and can be performed
as a post-processing step on top of any electronic structure code. (ii) The formalism
is {\it agnostic} of the level of theory used to describe optical excitations at clamped nuclei;
therefore the same procedure can be used with any level of theory
(e.g.~independent-particle approximation, random-phase approximation, GW/Bethe-Salpeter),
so long as the optical process can be described by means of Fermi's Golden Rule.
(iii) The method seamlessly combines the AH theory of temperature-dependent band structure
and the HBB theory, of phonon-assisted indirect optical absorption.

The two main disadvantages of the WL method are (i) the calculations require the use
of supercells in order to accommodate phonon wavevectors within the first Brillouin zone.
(ii) The evaluation of expectation values over the nuclear wavefunctions requires 
calculations for many nuclear configurations. In Refs.~\onlinecite{Patrick2013,Zacharias_2015}
the latter issue was addressed by using a stochastic approach based on importance-sampling
Monte Carlo integration.
In this manuscript we further improve the configurational averaging by replacing 
the stochastic approach of Ref.~\onlinecite{Zacharias_2015} with a fully {\it deterministic}
method. In particular, we demonstrate that it is possible to choose {\it a single} configuration 
of the nuclei yielding at once the band structure renormalization and indirect optical absorption 
at a given temperature.
In order to demonstrate this method we report applications to silicon, diamond and gallium arsenide.
Our calculated spectra and temperature dependent band gaps compare well with previous calculations
and with experiment. For completeness we also provide a detailed analysis of the relation between the WL theory,
the AH theory, and the HBB theory. 

The organization of the manuscript is as follows: in Sec.~\ref{sec.method-results} we briefly outline the
WL expression for the temperature-dependent dielectric function, and summarize the 
`one-shot' procedure for evaluating this expression using a single atomic configuration.
In this section we also show our main results for the optical absorption spectra of
Si, C, and GaAs in order to emphasize the simplicity and effectiveness of the formalism.
In Sec.~\ref{sec.deterministic} we develop the formalism which is used to select 
the optimal atomic configuration in the one-shot calculations of Sec.~\ref{sec.method-results}.
In particular, we prove that our optimal configuration yields exact results in the limit
of infinite supercell size.
In Sec.~\ref{sec.manyconf} we extend the concepts of Sec.~\ref{sec.deterministic} by showing
that it is possible to deterministically select further atomic configurations in order
to control and systematically reduce the error resulting from the configurational averaging.
In Sec.~\ref{sec.general-theory} we discuss the link between the WL theory of temperature-dependent
optical spectra, the AH theory of temperature-dependent band structures, and the HBB theory
of indirect optical absorption. 
In Sec.~\ref{sec.allresults} we present our calculations of temperature-dependent band gaps
for silicon, diamond, and gallium arsenide.
Section~\ref{sec.computational} reports all computational details of the calculations
presented in this manuscript. 
In Sec.~\ref{sec.conclusion} we summarize our key findings and indicate avenues for future work.
Lengthy formal derivations and further technical details are left to Appendices~A-D.

\section{One-shot method and main results}\label{sec.method-results}

In this section we outline the procedure for calculating temperature-dependent optical spectra
using one-shot frozen-phonon calculations. For clarity we also anticipate 
our main results on silicon, diamond, and gallium arsenide, leaving all computational 
details to Sec.~\ref{sec.computational}.

In the WL theory the imaginary part of the dielectric function of a solid at the temperature $T$
is given by:\cite{Zacharias_2015}
  \begin{equation}\label{eq.WL}
  \e_2(\w;T)  = Z^{-1} {\sum}_n \exp(-E_n /k_{\rm B}T) \< \e_2(\w;x) \>_n.
  \end{equation}
In this expression, $E_n$ denotes the energy of a nuclear quantum state evaluated in the
Born-Oppenheimer approximation, $k_{\rm B}$ is the Boltzmann constant, 
and $Z = \sum_n \exp(-E_n/k_{\rm B}T)$ is the canonical partition function.
The function $\e_2(\w;x)$ is the imaginary part of the macroscopic, electronic dielectric
function, evaluated at {\it clamped} nuclei. For notational simplicity we indicate the set 
of all atomic coordinates by $x$. In the following we denote by $N$ the total number
of atomic coordinates. In Eq.~(\ref{eq.WL}) each expectation value
$\< \cdots \>_n$ is taken with respect to the quantum nuclear state with energy $E_n$,
and involves a multi-dimensional integration over all atomic coordinates.
A detailed derivation of Eq.~(\ref{eq.WL}) can be found in Sec.~9.2 of Ref.~\onlinecite{CEP_FG}.

In order to focus on quantum nuclear effects and temperature shifts, we here describe
the dielectric function at clamped nuclei using the simplest possible approximations,
namely the independent-particle approximation and the electric dipole approximation:
  \begin{equation}\label{eq.eps}
  \epsilon_2(\w;x) = \frac{2 \pi }{ m_{\rm e} N_{\rm e} } \frac{\w_{\rm p}^2}{\,\w^2}
  \sum_{cv} | p_{cv}^x|^2 \delta(\ve_c^x-\ve_v^x-\hbar\w).
  \end{equation}
In this expression $m_{\rm e}$ is the electron mass, $N_{\rm e}$ is the number of electrons
in the crystal unit cell, $\w_{\rm p}$ is the plasma frequency, and $\w$ the photon frequency.
The factor 2 is for the spin degeneracy. The sum extends to the occupied Kohn-Sham
states $|v^x\>$ of energy $\ve_v^x$, as well as the unoccupied states $|c^x\>$ of energy $\ve_c^x$.
The superscripts are to keep in mind that these states are evaluated for nuclei clamped
in the configuration labelled by $x$. The matrix elements of the momentum operator along
the polarization direction of the photon is indicated as $p^x_{cv}$.
In the present case we use non-local pseudopotentials and a scissor operator, therefore
the momentum matrix elements are modified following Ref.~\onlinecite{Starace_1971},
as described in Sec.~\ref{sec.computational}. 
In all the calculations presented in this manuscript the dielectric functions are
obtained by first evaluating Eqs.~(\ref{eq.WL}) and (\ref{eq.eps}) for each Cartesian direction,
and then performing the isotropic average over the photon polarizations.

In principle Eq.~(\ref{eq.WL}) could be evaluated using the nuclear wavefunctions
obtained from the solution of the nuclear Schr\"odinger equation with electrons in their
ground state. This choice would lead to the automatic inclusion of {\it anharmonic} effects.
However, for conciseness, in the present work we restrict the discussion to the {\it harmonic} 
approximation.

In the harmonic approximation, every many-body nuclear quantum state can be expressed as 
a product of Hermite functions, and the atomic displacements can be written
as linear combinations of normal coordinates.~\cite{Sakurai} By exploiting the property of Hermite
polynomials and Mehler's formula,~\cite{Watson1933} the summation in Eq.~(\ref{eq.WL}) 
is exactly rewritten as follows:\cite{CEP_FG}
  \begin{equation}\label{eq.wl}
  \e_2(\w;T)  = {\prod}_\nu \int\! dx_\nu \frac{\exp(-x_\nu^2/2\sigma_{\nu,T}^2)}{\sqrt{2\pi}\sigma_{\nu,T}}
  \e_2(\w;x).
  \end{equation}
Here the product runs over all the normal coordinates $x_\nu$. In this and all following expressions 
it is understood that the three translational modes with zero vibrational frequency are skipped in the sums.
We indicate the vibrational frequency of the $\nu$-th normal mode by $\W_\nu$. The corresponding
zero-point vibrational amplitude is given by $l_\nu = (\hbar/2 M_{\rm p} \W_{\nu})^{1/2}$, 
where $M_{\rm p}$ is a reference mass that we take equal to the proton mass. Using these
conventions, the Gaussian widths in Eq.~(\ref{eq.wl}) are given by:
  \begin{equation}\label{eq.sigma}
   \sigma^2_{\nu,T}  = (2n_{\nu,T}+1) \,l_\nu^2, 
  \end{equation}
where $n_{\nu,T} = [\exp(\hbar\w_\nu/k_{\rm B}T)\!-\!1]^{-1}$ is the Bose-Einstein 
occupation of the $\nu$-th mode. In the reminder of this manuscript we will concentrate
on the expression for the WL dielectric function given by Eq.~(\ref{eq.wl}).

The configurational average appearing in Eq.~(\ref{eq.wl}) was evaluated in Ref.~\onlinecite{Zacharias_2015}
using importance-sampling Monte Carlo integration.\cite{Patrick2013} More specifically, the Monte Carlo
estimator of the integral\cite{Caflisch}
was evaluated by averaging over a set of atomic configurations in a Born-von K\'arm\'an supercell.
Each configuration in the set was generated according to the importance function 
$\exp(-x_\nu^2/2\sigma_{\nu,T}^2)/\sqrt{2\pi}\sigma_{\nu,T}$.
In Ref.~\onlinecite{Zacharias_2015} it was remarked that, in the case of the optical spectrum 
of silicon, $<10$ random samples were sufficient in order to converge the integral in Eq.~(\ref{eq.wl}). 
Furthermore, calculations performed using a {\it single} sample were found to be of comparable accuracy to
fully-converged calculations.
Motivated by these observations, we decided to investigate in detail why the stochastic
evaluation of Eq.~(\ref{eq.wl}) requires only very few samples.

In Sec.~\ref{sec.deterministic} we provide a formal proof of the fact that, in the limit of large supercell, 
only one atomic configuration is enough for evaluating Eq.~(\ref{eq.wl}).
In the reminder of this section we only give the optimal configuration and outline the calculation procedure, 
so as to place the emphasis on our main results.

\begin{figure}[t!]
  \begin{center}
  \subfloat{\includegraphics[width=0.355\textwidth]{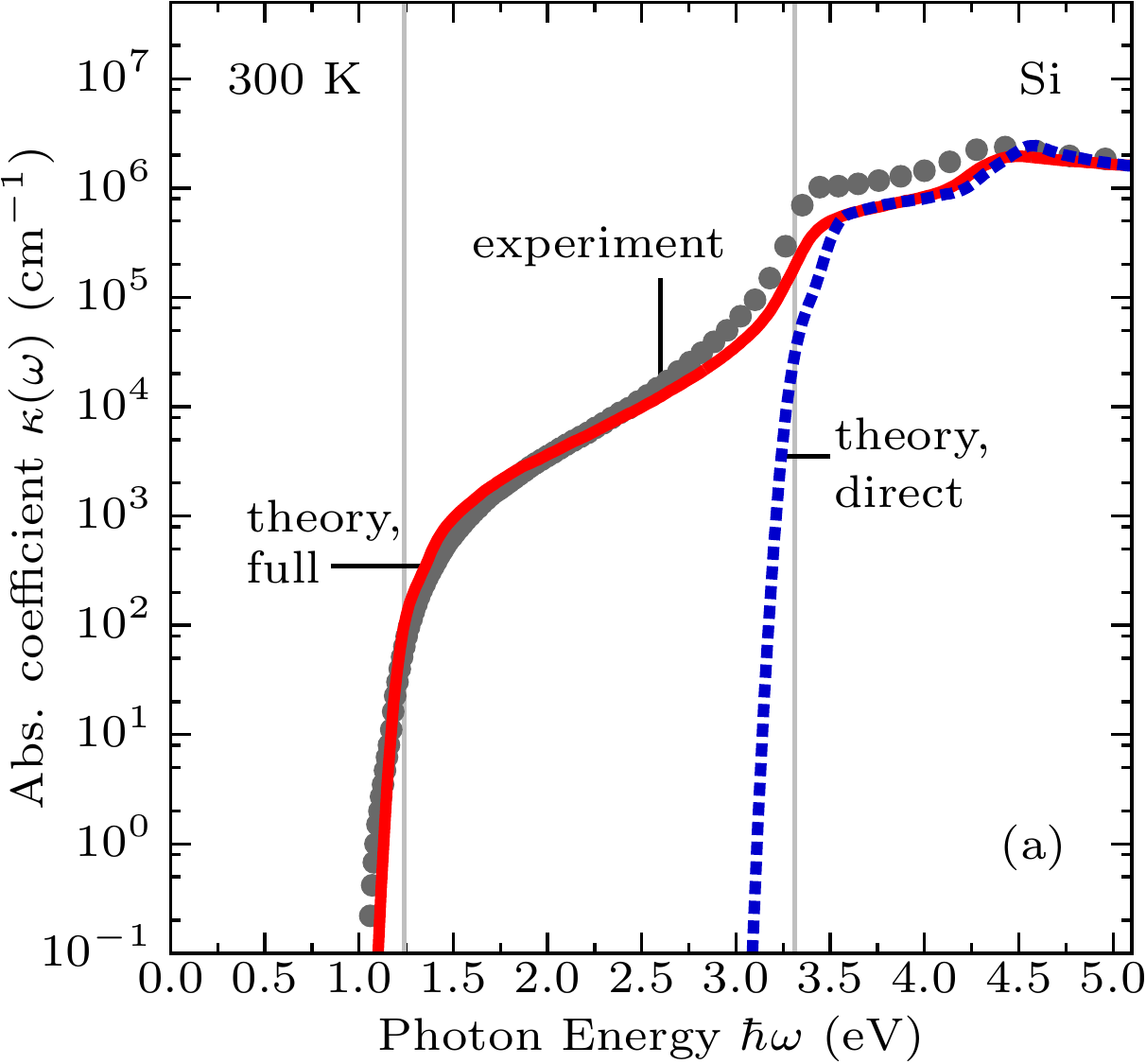}} 
  \vspace*{-0.25cm}
  \newline
  \subfloat{\includegraphics[width=0.355\textwidth]{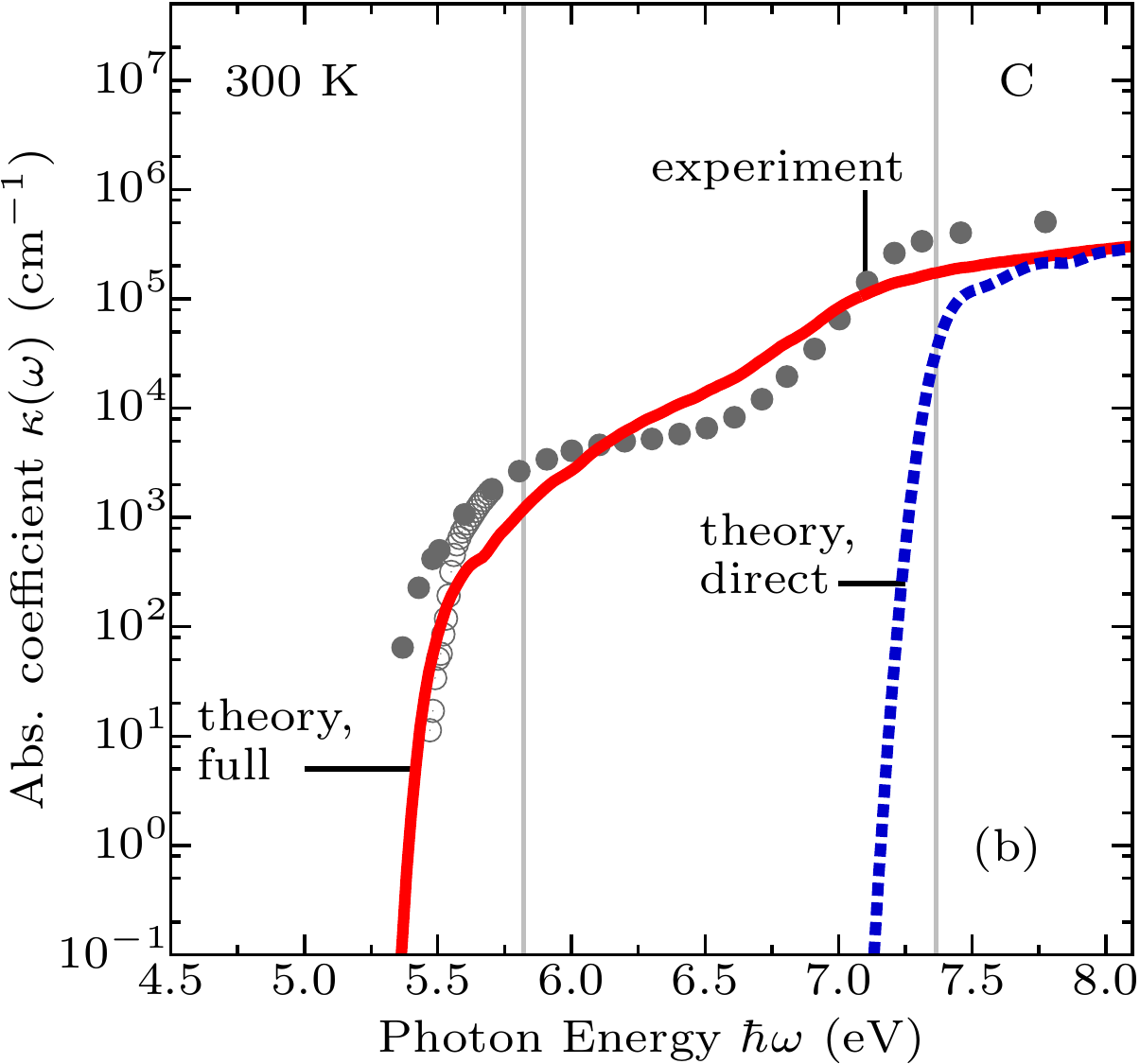}} 
  \vspace*{-0.0cm}
  \newline
  \hspace*{-0.8cm}
  \subfloat{\includegraphics[width=0.352\textwidth]{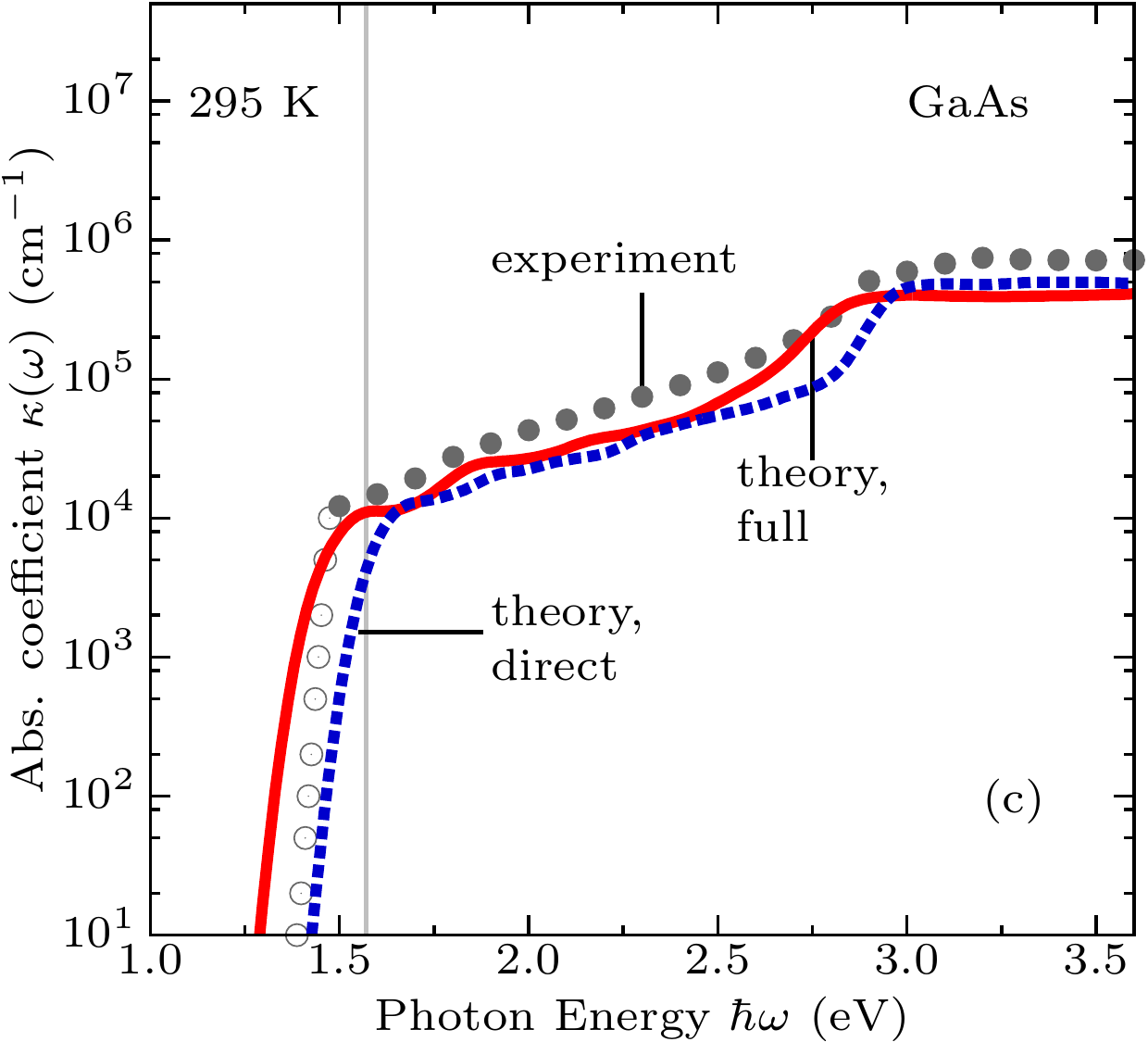}}
  \end{center}
  \vspace*{-0.48cm}
  \caption{\label{fig1}
  Absorption coefficient of (a) Si, (b) C, and (c) GaAs at room temperature.
  Calculations with the atoms clamped at their equilibrium positions are shown as blue dashed lines.
  Calculations using the WL method in the atomic configuration specified by Eq.~(\protect\ref{eq.conf_opt})
  are shown as red solid lines. The experimental data for Si are from Ref.~\onlinecite{M_Green3} (grey discs), those  for C are from Refs.~\onlinecite{Phillip_1964} (grey discs) and~\onlinecite{Clark312} (grey circles).
  Experimental data for GaAs  are from Refs.~\onlinecite{Aspnes_exp} (grey discs) and~\onlinecite{Sturge_294K} 
  (grey circles).
  The thin vertical lines indicate the direct and indirect band gaps calculated for nuclei in
  their equilibrium positions. The calculations were performed using $8\times 8\times 8$ supercells,
  using a Gaussian broadening of 30~meV for Si and C, and of 50~meV for GaAs.}
  \end{figure}

In order to calculate the optical absorption spectrum (including band gap renormalization)
at finite temperature using a {\it one-shot} frozen-phonon calculation, we proceed as follows:
\begin{enumerate}
\item 
We consider a $m\!\times\!m\!\times\!m$ supercell of the primitive unit cell.
We determine the interatomic force constants~\cite{FG_review} by means of 
density-functional perturbation theory calculations in the {\it primitive} unit cell, 
using a $m\!\times\!m\!\times\!m$ Brillouin-zone grid.\cite{Baroni2001,QE}
\item By diagonalizing the dynamical matrix obtained from the matrix of force constants,
we determine the vibrational eigenmodes $e_{\kappa \alpha, \nu}$ and eigenfrequencies
$\Omega_{\nu}$ ($\kappa$ and $\alpha$ indicate the atom and the Cartesian direction, respectively).
\item 
For a given temperature $T$, we generate {\it one} distorted atomic configuration
by displacing the atoms from equilibrium by an amount $\Delta \tau_{\kappa \alpha}$, with:
  \begin{equation}\label{eq.conf_opt}
  \Delta \tau_{\kappa \alpha} =  (M_{\rm p}/M_\kappa)^{\frac{1}{2}} \,{\sum}_\nu\, 
  (-1)^{\nu-1} e_{\kappa \alpha, \nu}\, \sigma_{\nu,T}.\!\!\!\!\!
\end{equation}
In this expression $M_\kappa$ is the mass of the $\kappa$-th nucleus, and the sum runs
over all normal modes.
The vibrational modes are assumed to be sorted in ascending order with respect to their frequencies.
 In order to enforce the same choice of gauge for each vibrational mode, 
the sign of each eigenvector is chosen so as to have the first non-zero element positive.
The prescription given by Eq.~(\ref{eq.conf_opt}) will be motivated in Sec.~\ref{sec.deterministic}.
\item 
We calculate the dielectric function using the atomic configuration specified by Eq.~(\ref{eq.conf_opt}).
The result will be the temperature-dependent dielectric function at the temperature $T$.
\item 
We check for convergence by repeating all previous steps using increasingly larger supercells.
\end{enumerate}
In Fig.~\ref{fig1} we present the room-temperature optical absorption coefficients of Si, C, and GaAs
calculated using the procedure just outlined (red solid lines), and we compare our results with experiment
(grey discs). For completeness we also show the absorption coefficients evaluated with the
atoms clamped at their equilibrium positions (blue solid line). The calculations were performed 
on $8\!\times\!8\!\times\!8$ supercells, 
using density-functional theory (DFT) in the local density approximation (LDA) and a scissor
correction; all computational 
details are provided in Sec.~\ref{sec.computational}. 
Our present results for Si in Fig.~\ref{fig1}(a), obtained using a single atomic configuration,
are in perfect agreement with those reported in Ref.~\onlinecite{Zacharias_2015} using
importance-sampling Monte Carlo integration.
In Fig.~\ref{fig1}(a) we see that the spectrum calculated with the nuclei clamped
in their equilibrium positions exhibit absorption only above the direct gap, as expected.
At variance with these calculations, our one-shot calculation based on the WL theory
correctly captures indirect transitions. In particular, this calculation is in good agreement with experiment
throughout a wide range of photon energies.\cite{M_Green3,Phillip_1964} This agreement
surprisingly extends over seven orders of magnitude of the absorption coefficient.
However, we should point out that our adiabatic theory does not capture the fine structure 
close to the absorption onset: there the non-adiabatic HBB theory gives two different slopes
for phonon absorption and emission processes.\cite{Macfarlane,Noffsinger}
Near the onset for direct transitions, our calculation underestimates the experimental
data. This behavior is expected since we are not including electron-hole interactions,
which are known to increase the oscillator strength of the $E_1$ peak near 3.4~eV.~\cite{Benedict}

In Fig.~\ref{fig1}(b) we show our WL calculation for diamond. In this case our method
correctly captures the absorption in the indirect range, however we observe more pronounced
deviations between theory and experiment than in the case of Si. We assign the residual discrepancy to the inability
of DFT/LDA to accurately describe the joint density of states of diamond. In fact,
in contrast to the case of silicon, in diamond a simple scissor correction is not enough
to mimic quasiparticle corrections.\cite{Henry_FG} For example, the GW corrections to the
$X_{1c}$, $L_{1c}$, and $L_{3c}$ states of Si are all in the narrow range between 0.66 and 0.75~eV; instead
the GW corrections to the same states of diamond span a broader range, between 1.47 and 2.04~eV.\cite{Henry_FG}
This is expected to lead to a significant redistribution of spectral weight precisely in
the range of photon energies considered in Fig.~\ref{fig1}(b).
Also in this case the strength of the $E_1$ peak near 7.3~eV is underestimated due to
our neglecting electron-hole interactions.\cite{Benedict} In Fig.~\ref{fig1}(b) 
we are reporting two sets of experimental data.~\cite{Phillip_1964,Clark312} 
These data exhibit different intensities near the absorption edge (at energies $<5.7$~eV). 
According to Ref~\onlinecite{Phillip_1964}, the intensity of the
absorption coefficient for energies below 6.5~eV is not fully reliable. 
Our calculated spectrum is in closer agreement with the data from Ref~\onlinecite{Clark312},
which exhibit a sharper absorption edge. This comparison suggests that our present 
method might prove useful for the validation of challenging experiments, 
especially near the weak absorption edge. 
As in the case of Fig.~\ref{fig1}(a), the fine structure features close to the absorption 
onset are absent in our calculation.

In Fig.~\ref{fig1}(c) we compare the absorption coefficient calculated for GaAs with experiment.~\cite{Aspnes_exp,Sturge_294K}
This example clearly demonstrates that our WL calculation correctly describes the absorption
spectrum of a direct-gap semiconductor. In this case the shape of the absorption coefficient
is not altered, as expected, but the spectrum is redshifted as a result of the zero-point 
renormalization of the band structure.
Also in the case of GaAs the calculated absorption coefficient underestimates the measured
values. This is partially a consequence of our neglecting of excitonic effects,\cite{Rohlfing_1998}
but most importantly it is a consequence of the inability of DFT/LDA to accurately describe the effective
masses of GaAs. In fact, according to the standard theory of absorption in direct-gap semiconductors,\cite{Parravicini}
the absorption coefficient scales as $(m^*)^{3/4}$, where $m^*$ is the average isotropic effective mass.
Since DFT calculations for GaAs yield masses which are up to a factor of 2 smaller than in experiment,
we expect a corresponding underestimation of the absorption coefficient by up to a factor of $2^{3/4}\sim 2$. 
This estimate is in line with our results in Fig.~\ref{fig1}(c). We note that this issue
is not found in the case of Si [Fig.~\ref{fig1}(a)], since the DFT/LDA effective masses of
silicon are in surprising agreement with experiment.\cite{FG_book}
We also note that for GaAs our calculation correctly predicts phonon-assisted absorption below 
the direct gap. This phenomenon is related to the Urbach tail.\cite{Urbach_1953}

\begin{figure}[t!]
  \subfloat{\includegraphics[width=0.37\textwidth]{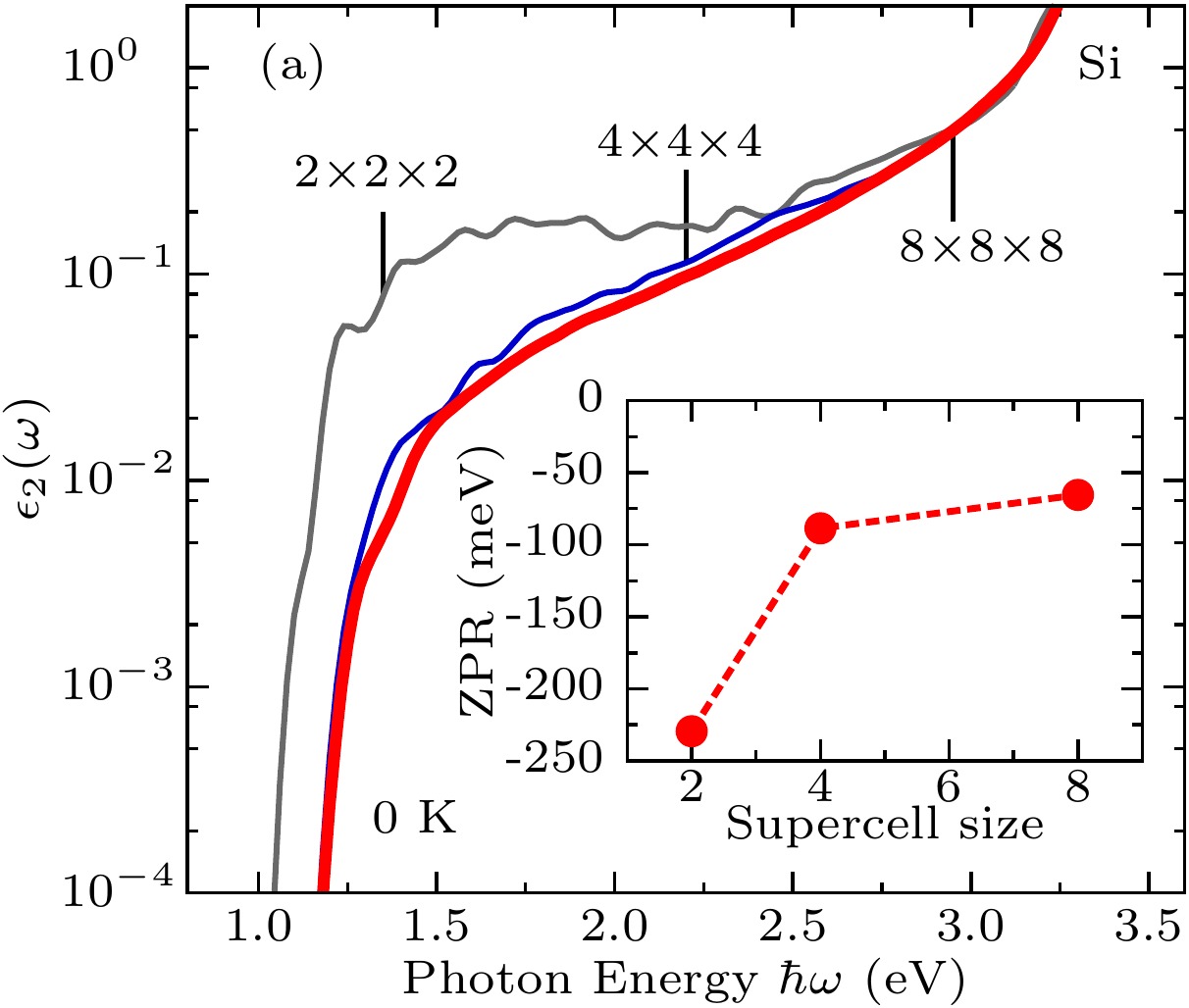}}
  \vspace*{-0.25cm}
  \newline
  \subfloat{\includegraphics[width=0.37\textwidth]{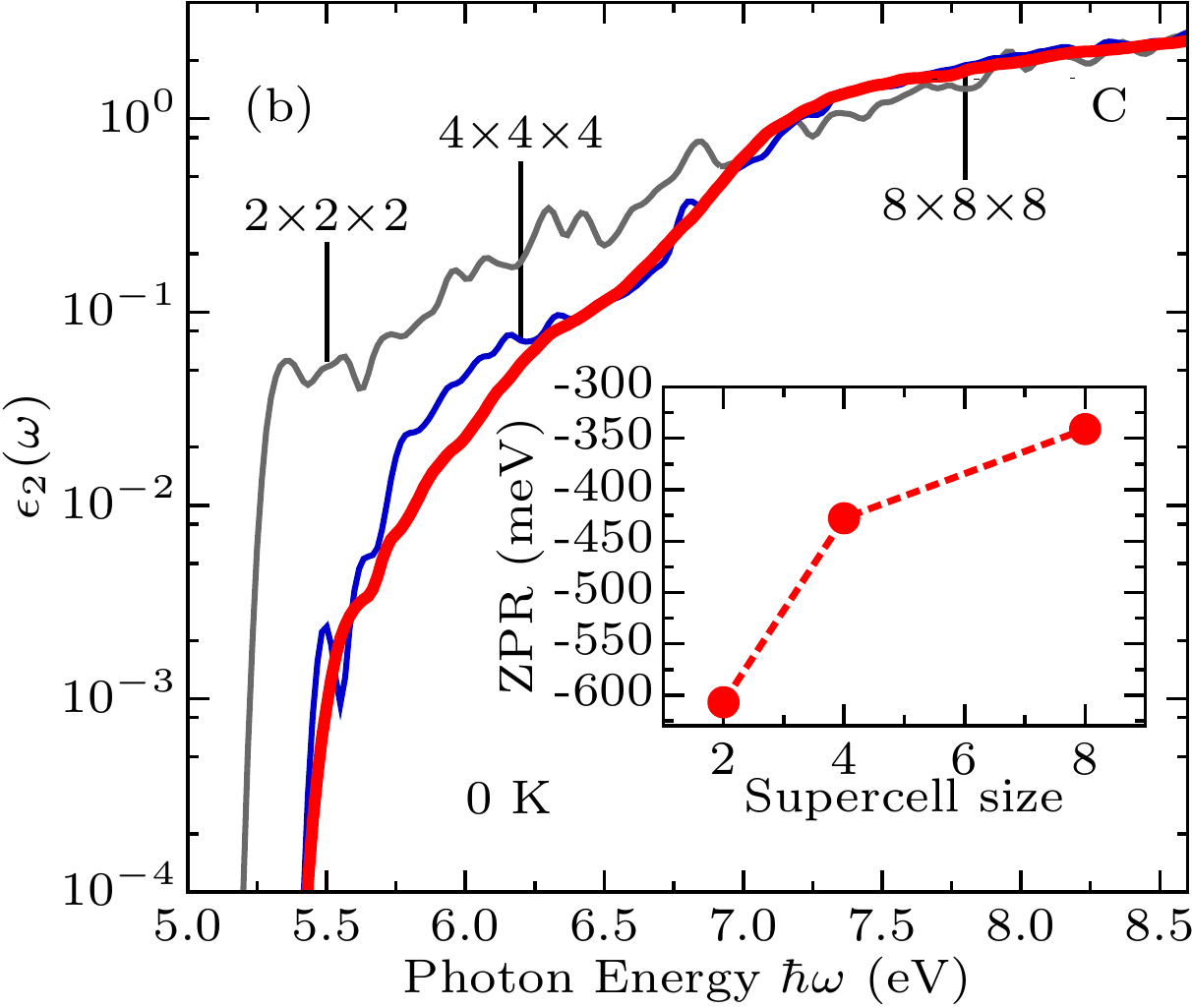}}
  \vspace*{-0.0cm}
  \newline
  \hspace*{-0.8cm}
  \subfloat{\includegraphics[width=0.37\textwidth]{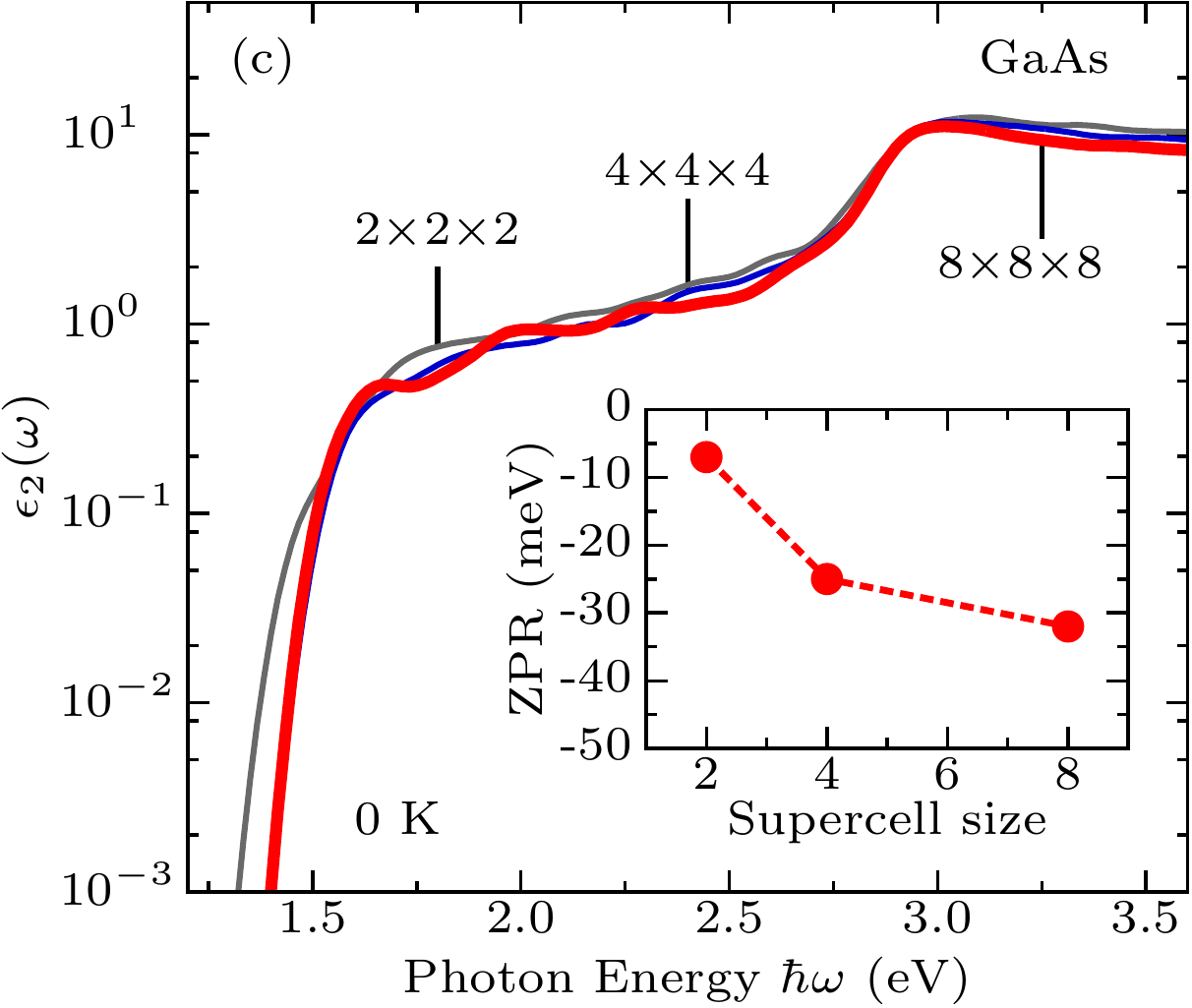}}
  \vspace*{-0.3cm}
  \caption{\label{fig2}
  Convergence of the WL dielectric function as a function of supercell size: (a) Si, (b) C, and
  (c) GaAs. The grey curves are for calculations using $2\times2\times2$ supercells; blue solid lines
  indicate calculations using $4\times4\times4$ supercells; thick red solid lines are for $8\times8\times8$
  supercells. Each curve was obtained using a single calculation, in the configuration specified by
  Eq.~(\protect\ref{eq.conf_opt}). For Si and C the electronic Brillouin zone was sampled using
  1920, 240, and 30 {\bf k}-points for the $2\times2\times2$, $4\times4\times4$, and $8\times8\times8$
  supercells, respectively. In the case of GaAs a finer sampling was required, and we used
  6400, 800 and 100 {\bf k}-points, respectively. The calculations were performed
  using a Gaussian broadening of 30~meV for Si and C, and of 50~meV for GaAs.
  The insets show the band gap extracted from the Tauc plots. These values automatically include
  zero-point renormalization.
  }
  \end{figure}

After discussing the comparison between our calculations and experiment, we briefly comment
on the computational effort and the numerical convergence.
Figure~\ref{fig2} shows the imaginary part of the WL dielectric function of Si, C, and GaAs 
evaluated at zero temperature using the procedure outlined in the previous page.
In order to achieve convergence we performed calculations for supercells of increasing size,
from $2\times 2 \times 2$ to $8\times 8 \times 8$. It is clear that large supercells are
required in order to obtain converged results. Increasing the supercell size has the twofold
effect of refining the sampling of the electron-phonon coupling in the 
equivalent Brillouin zone, and of approaching the limit where Eq.~(\ref{eq.conf_opt}) 
becomes exact.
In Fig.~\ref{fig2} we also report the band gaps of Si, C, and GaAs as extracted from 
$\epsilon_2(\w)$ using the standard Tauc plots.\cite{Tauc} 
These results are in agreement with previous work and will be
discussed in Sec.~\ref{sec.allresults}. From this figure we see that our methodology correctly describes the
zero-point renormalization of the band gap of both direct- and indirect-gap semiconductors.

From Figs.~\ref{fig1} and \ref{fig2} it should be apparent that, apart from the inherent
deficiencies of the DFT/LDA approximation, with our new method it is possible to compute
optical spectra and band gaps in both direct and indirect semiconductors including 
electron-phonon interactions, at the cost of {\it a single} supercell calculation
with clamped nuclei.

In the following sections 
we develop the theory underlying our computational approach, and we provide 
an extensive set of benchmarks.

\section{Formal justification of Eq.~(\ref{eq.conf_opt})}\label{sec.deterministic}

In this section we provide the rationale for the choice of the optimal configuration
given by Eq.~(\ref{eq.conf_opt}). We start from a heuristic argument, and then
we provide a formal justification.

\subsection{Heuristic approach}
According to Eq.~(\ref{eq.wl}), the WL dielectric function can be interpreted as
the average of $\epsilon_2(\omega;x)$ over the $N$ standard normal random variables 
$x_\nu/\sigma_{\nu,T}$. Let us consider the sum of the squares of these variables, 
$q^2=\sum_\nu (x_\nu/\sigma_{\nu,T})^2$. The random variable $q^2$ follows by construction 
the $\chi^2$ distribution.\cite{Koch} Owing to the central limit theorem,
the $\chi^2$ distribution tends to a normal distribution for $N\rightarrow \infty$.
More specifically, in the limit of large $N$ the variable $(q^2-N)/(2N)^{1/2}$ follows
a standard normal distribution. 
As a result, as $N$ increases, the variable $q^2$ becomes strongly
peaked at~$N$, with a standard deviation $N^{1/2}$.\cite{Abramowitz}
As a sanity check, we verified these limits numerically
by generating one million random atomic configurations for a $4\times 4 \times 4$ supercell
of diamond.
Based on these considerations, we infer that for large $N$ the integration
in Eq.~(\ref{eq.wl}) is dominated by atomic configurations such that
$\sum_\nu (x_\nu/\sigma_{\nu,T})^2 = N$.
This same conclusion can alternatively be reached be rewriting the integral in Eq.~(\ref{eq.wl})
as the product of an integral over the `radial' variable $q$, and an integral
over a generalized `angular' variable which runs over the $(N\!-\!1)$-dimensional
sphere of radius $q$.

In the absence of information about the electron-phonon coupling constants of
each vibrational mode, we must assume that all modes are equally important in the
evaluation of the integral in Eq.~(\ref{eq.wl}).
Therefore the most representative sets of
coordinates for evaluating the integral in Eq.~(\ref{eq.wl}) are those satisfying
the condition $x_\nu/\sigma_{\nu,T} = \pm 1$ with $\nu = 1, \dots N$.
This is precisely what we observed in the importance-sampling Monte Carlo
calculations reported in Ref.~\onlinecite{Zacharias_2015}. 
A similar conclusion was reached in Ref.~\onlinecite{Bartomeu_2016}, where the concept of
`thermal lines' was introduced. Here we do not follow up on the idea of thermal
lines since, as we prove below, there exists {\it one} atomic configuration which yields
the exact temperature-dependent dielectric function in the limit of large~$N$.
If we were to choose a single configuration to evaluate the integral in Eq.~(\ref{eq.wl}), in absence
of information about the electron-phonon couplings the least-biased choice
would correspond to taking random signs for each normal coordinate.
This reasoning formed the {\it heuristic} basis for the choice made in Eq.~(\ref{eq.conf_opt}).

\subsection{Formal proof}

We now proceed to demonstrate that Eq.~(\ref{eq.conf_opt}) is not just
a sound approximation, but it is indeed {\it the} optimal configuration for 
evaluating Eq.~(\ref{eq.wl}) using a one-shot calculation.
To this aim we perform a Taylor expansion of $\epsilon_2(\omega;x)$ in the variables $x_\nu$,
and then evaluate each integral $\int\! d x_\nu$ in Eq.~(\ref{eq.wl}) analytically. The result is:
\hide
  \begin{eqnarray}\label{eq.expansion-exact}
  && \e_2(\w,T) = \e_2(\w) + \sum_\nu \frac{\D \e_2}{\D n_\nu} (n_\nu+1/2) \nonumber \\
  &&\qquad+\frac{1}{2}\sum_{\nu\mu} \frac{\D^2 \e_2}{\D n_\nu\D n_\mu}(n_\nu+1/2)(n_\mu+1/2) +
  \mathcal{O}(\sigma^6).\,\,\,\,\,
  \end{eqnarray}
  In this expression $\e_2(\w)$ denotes the dielectric function evaluated for nuclei
  clamped in their equilibrium positions, and we have defined:
  \begin{equation}
  \frac{\D}{\D n_\nu} = l_\nu^2 \left. \frac{\D }{\D x_\nu^2}\right|_{x_\nu=0},
  \end{equation}
  where $l_\nu$ is given by Eq.~(\ref{eq.zp}). The term $\mathcal{O}(\sigma^6)$
  is a short-hand notation to indicate all terms of the kind $\sigma_{\nu,T}^6$ and higher powers.
  Intuitively Eq.~(\ref{eq.expansion-exact}) expresses the temperature-dependent
  dielectric function as a Taylor expansion in the phonon occupations. 
\unhide
  \begin{equation}\label{eq.expansion-exact}
   \e_2(\w,T) = \e_2(\w) + \frac{1}{2}\sum_\nu \frac{\D^2 \e_2(\w;x)}{\D x_\nu^2} \sigma_{\nu,T}^2  
  + \mathcal{O}(\sigma^4).
  \end{equation}
In this expression $\e_2(\w)$ denotes the dielectric function evaluated for nuclei
clamped in their equilibrium positions, and the term $\mathcal{O}(\sigma^4)$
is a short-hand notation to indicate all terms of the kind $\sigma_{\nu,T}^4$ and higher powers.

We now consider the dielectric function calculated with the nuclei clamped in the
positions specified by Eq.~(\ref{eq.conf_opt}). We denote this function by $\e_2^{1{\rm C}}(\w;T)$,
with `1C' standing for `one configuration'. Another Taylor expansion 
in the normal mode coordinates yields:
    \begin{eqnarray}\label{eq.expansion-1conf}
   &&\e_2^{1{\rm C}}(\w;T) = \e_2(\w) +  
      \sum_\nu (-1)^{\nu-1} \frac{\D \e_2(\w;x)}{\D x_\nu}\sigma_{\nu,T} \nonumber \\
    &&\qquad+
      \frac{1}{2}\sum_{\nu\mu} (-1)^{\nu+\mu-2} \frac{\D^2 \e_2(\w;x)}{\D x_\nu x_\mu} 
       \sigma_{\nu,T} \sigma_{\mu,T} \nonumber \\
    &&\qquad+
       \frac{1}{6}\sum_{\nu\mu\l}(-1)^{\nu+\mu+\l-3} \frac{\D^3 \e_2(\w;x)}{\D x_\nu x_\mu x_\l}
         \sigma_{\nu,T} \sigma_{\mu,T} \sigma_{\l,T}
    \nonumber \\ && \qquad+ \mathcal{O}(\sigma^4).
  \end{eqnarray}
By comparing Eqs.~(\ref{eq.expansion-exact}) and (\ref{eq.expansion-1conf}) we see that
$\e_2^{1{\rm C}}(\w;T)$ and $\e_2(\w;T)$ do coincide up to $\mathcal{O}(\sigma^4)$ if
the following conditions hold: (i) the summations on the first and third lines of 
Eq.~(\ref{eq.expansion-1conf}) vanish; (ii) all terms $\nu\ne \mu$ in the second line 
of the same equation vanish.

In general the conditions (i) and (ii) do not hold, therefore calculations of
$\e_2^{1{\rm C}}(\w;T)$ and $\e_2(\w;T)$ will yield very different results.
However, these conditions are fulfilled in the limit $N\rightarrow \infty$, as we show in the following.
To this aim let us consider the summation on the first line of Eq.~(\ref{eq.expansion-1conf}).
We focus on two successive terms in the sum, $\nu$ and $\nu+1$.
Assuming a vibrational density of states (vDOS) which is nonvanishing up to the highest vibrational
frequency, when $N\rightarrow \infty$, then $\Omega_{\nu+1}-\Omega_\nu \rightarrow 0$, therefore
these modes are effectively degenerate, and hence must exhibit the same electron-phonon
coupling coefficients. Under these conditions we can write:
  \begin{eqnarray}\label{eq.proof.1}
   &&  (-1)^{\nu-1} \frac{\D \e_2(\w;x)}{\D x_\nu}\sigma_{\nu,T} +
    (-1)^{\nu+1-1} \frac{\D \e_2(\w;x)}{\D x_{\nu+1}}\sigma_{\nu+1,T} 
    \nonumber \\
   && \qquad \simeq \frac{\D \e_2(\w;x)}{\D x_\nu}\sigma_{\nu,T} (-1)^{\nu-1} \left[ 1+(-1)\right] 
    = 0.
  \end{eqnarray}
This reasoning can be repeated for every pair of vibrational modes $(\nu,\nu+1)$
appearing in the first line of Eq.~(\ref{eq.expansion-1conf}). If $N$ is even, then
this proves that the sum on the first line vanishes. If $N$ is odd, then there is one
mode left out, but the contribution of this one mode is negligibly small for $N\rightarrow\infty$.
If there are gaps in the vDOS, then the above reasoning remains valid by
considering separately the frequency ranges where the vDOS is nonzero.
This completes the proof that the sum in the first line of Eq.~(\ref{eq.expansion-1conf}) vanishes
in the limit of large $N$.

The summation in the third line of Eq.~(\ref{eq.expansion-1conf}) can be analyzed 
along the same lines, after noticing that one can rearrange the sum as follows:
    \begin{equation}\label{eq.proof.2}
    \sum_\nu \left[ \frac{1}{6}\sum_{\mu\l}(-1)^{\mu+\l-3} \frac{\D^3 \e_2(\w;x)}{\D x_\nu x_\mu x_\l}
         \sigma_{\nu,T} \sigma_{\mu,T} \sigma_{\l,T} \right] (-1)^\nu.
  \end{equation}
Also in this case we can consider any pair of successive eigenmodes $\nu$ and $\nu+1$
and repeat the reasoning made above for the first line of Eq.~(\ref{eq.expansion-1conf}).
The result is that for $N\rightarrow \infty$ the entire sum must vanish.

If we now consider the second line of Eq.~(\ref{eq.expansion-1conf}), 
the sum of the terms with $\nu\ne \mu$ must vanish for large~$N$.
In fact, the second derivatives
$\D^2 \e_2(\w;x)/\D x_\mu\D x_\nu$ and $\D^2 \e_2(\w;x)/\D x_\mu\D x_{\nu+1}$
enter the sum with opposite signs, therefore their contribution vanishes in the
limit $N$~$\rightarrow$~$\infty$. On the other hand, when $\nu=\mu$, two successive terms in the sum
contribute with the same sign, yielding $2 \frac{1}{2}\D^2 \e_2(\w;x)/\D x_\nu^2$.
These contributions lead precisely to the second term on the r.h.s. of Eq.~(\ref{eq.expansion-exact}).

Taken together, Eqs.~(\ref{eq.expansion-exact})-(\ref{eq.proof.2}) and the above discussion 
demonstrate that our single-configuration dielectric function, $\e_2^{1{\rm C}}(\w;T)$, and the exact
WL dielectric function, $\e_2(\w;T)$, do coincide to $\mathcal{O}(\sigma^4)$ 
for $N\rightarrow \infty$. It is not difficult to see that this result can
be generalized to all orders in $\sigma_{\nu,T}$, therefore the following general
statement holds true: 
{\it in the limit of large supercell 
the dielectric function evaluated with the nuclei
clamped in the configuration specified by Eq.~(\ref{eq.conf_opt}) approaches 
the WL dielectric function, Eq.~(\ref{eq.wl})}. In symbols:
  \begin{equation}\label{eq.equiv}
  \lim_{\vphantom{\int_0^1}N\rightarrow \infty} \,\, \e_2^{1{\rm C}}(\w;T) \,=\, \e_2(\w;T).
  \end{equation}
This result forms the basis for the methodology presented in this manuscript.
The importance of the equivalence expressed by Eq.~(\ref{eq.equiv}) resides in that
it allows us to calculate dielectric functions at finite temperature using a single
atomic configuration. This represents a significant advance over alternative
techniques such as for example path-integrals molecular dynamics, importance-sampling 
Monte Carlo, or the direct evaluation of each term in Eq.~(\ref{eq.expansion-exact})
using frozen-phonon calculations for each vibrational mode.

We emphasize that, while we have proven the limit in Eq.~(\ref{eq.equiv}), we have no
information about the convergence rate of $\e_2^{1{\rm C}}(\w;T)$ towards the exact
result $\e_2(\w;T)$. In principle this rate could be estimated {\it a priori} by 
inspecting the convergence of the Eliashberg function~\cite{Allen_Cardona_1981} with the sampling of the Brillouin zone
in a calculation within the primitive unit cell. In practice we found it easier
to directly calculate $\e_2^{1{\rm C}}(\w;T)$ for supercells of increasing size.
Convergence tests for Si, C, and GaAs were reported in Fig.~\ref{fig2}. It is seen
that, for these tetrahedral semiconductors, converged results are obtained for 
$8\times8\times8$ supercells. We emphasize that in Fig.~\ref{fig2}, 
$\e_2^{1{\rm C}}(\w;T)$ is given in logarithmic scale: in a linear scale the differences
between a $4\times4\times4$ and an $8\times8\times8$ calculation would be barely discernible.

Calculations using the largest supercells in Fig.~\ref{fig2} are obviously time-consuming.
However, one should keep in mind that each line in this figure corresponds to a {\it single}
calculation of the dielectric function at clamped nuclei, therefore this method
enables the incorporation of temperature at low computational cost, and 
removes the need of configurational sampling.

In the insets of Fig.~\ref{fig2} we also presented the zero-point renormalization of the fundamental
gaps of Si, C, and GaAs. These gaps were directly obtained from the calculated $\e_2^{1{\rm C}}(\w;T)$
using the standard Tauc plots.\cite{Tauc}
Details will be discussed in Sec.~\ref{sec.allresults};
here we only point out that also the band gaps are calculated by using a single
atomic configuration in each case. To the best of our knowledge this is the first time
that calculations of temperature-dependent band gaps using a single atomic configuration
have been reported.

\section{Improvements using configurational averaging}\label{sec.manyconf}

There might be systems for which calculations using large supercells are prohibitively
time-consuming, and the limit in Eq.~(\ref{eq.equiv}) is practically beyond reach.
For these cases it is advantageous to extend the arguments presented in Sec.~\ref{sec.deterministic}
to calculations using more than one atomic configuration. In this section we show
how it is indeed possible to construct a hierarchy of atomic configurations
in order to systematically improve the numerical evaluation of Eq.~(\ref{eq.wl}) at fixed
supercell size.

We restart by considering atomic configurations specified by the normal coordinates 
$x_\nu=s_\nu \sigma_{\nu,T}$, where the signs $s_\nu = +1$ or $-1$ are yet to be specified.
A Taylor expansion as in Eq.~(\ref{eq.expansion-1conf}) yields:
    \begin{eqnarray}\label{eq.expansion-manyconf}
   &&\e_2^S(\w;T) = \e_2(\w) +
      \sum_\nu s_\nu \frac{\D \e_2(\w;x)}{\D x_\nu}\sigma_{\nu,T} \nonumber \\
    &&\qquad+
      \frac{1}{2}\sum_{\nu\mu} s_\nu s_\mu \frac{\D^2 \e_2(\w;x)}{\D x_\nu x_\mu}
       \sigma_{\nu,T} \sigma_{\mu,T} \nonumber \\
    &&\qquad+
       \frac{1}{6}\sum_{\nu\mu\l} s_\nu s_\mu s_\l \frac{\D^3 \e_2(\w;x)}{\D x_\nu x_\mu x_\l}
         \sigma_{\nu,T} \sigma_{\mu,T} \sigma_{\l,T}
    \nonumber \\ && \qquad+ \mathcal{O}(\sigma^4).
  \end{eqnarray}
Here the superscript $S$ denotes the entire set of $N$ signs, $S = (s_1,\cdots,s_N)$.
With this notation, the choice expressed by Eq.~(\ref{eq.conf_opt}) corresponds to setting
$S = ( + \, - \, + \, - \, + \, - \, \cdots )$. 

In the language of stochastic
sampling, the configurations specified by $S$ and $-S$ are called
an `antithetic pair'.\cite{Caflisch}
It is immediate to see that the dielectric function calculated using both $S$
and $-S$ does not contain any odd powers of $\sigma_{\nu,T}$. In fact from 
Eq.~(\ref{eq.expansion-manyconf}) we have:
  \begin{eqnarray}\label{eq.anti}
  && \frac{1}{2}\left[\e_2^{S}(\w;T) + \e_2^{-S}(\w;T)\right] = \nonumber \\
  && \,\,\,\,\, \e_2(\w) + \frac{1}{2}\sum_{\nu\mu} s_\nu s_\mu \frac{\D^2 \e_2(\w;x)}{\D x_\nu x_\mu}
       \sigma_{\nu,T} \sigma_{\mu,T} + \mathcal{O}(\sigma^4).\,\,\,\,\,\,
  \end{eqnarray}
This result is already very close to the exact expansion in Eq.~(\ref{eq.expansion-exact}),
independent of the size of the supercell.
\begin{figure*}[t!]
  \subfloat{\includegraphics[width=0.35\textwidth]{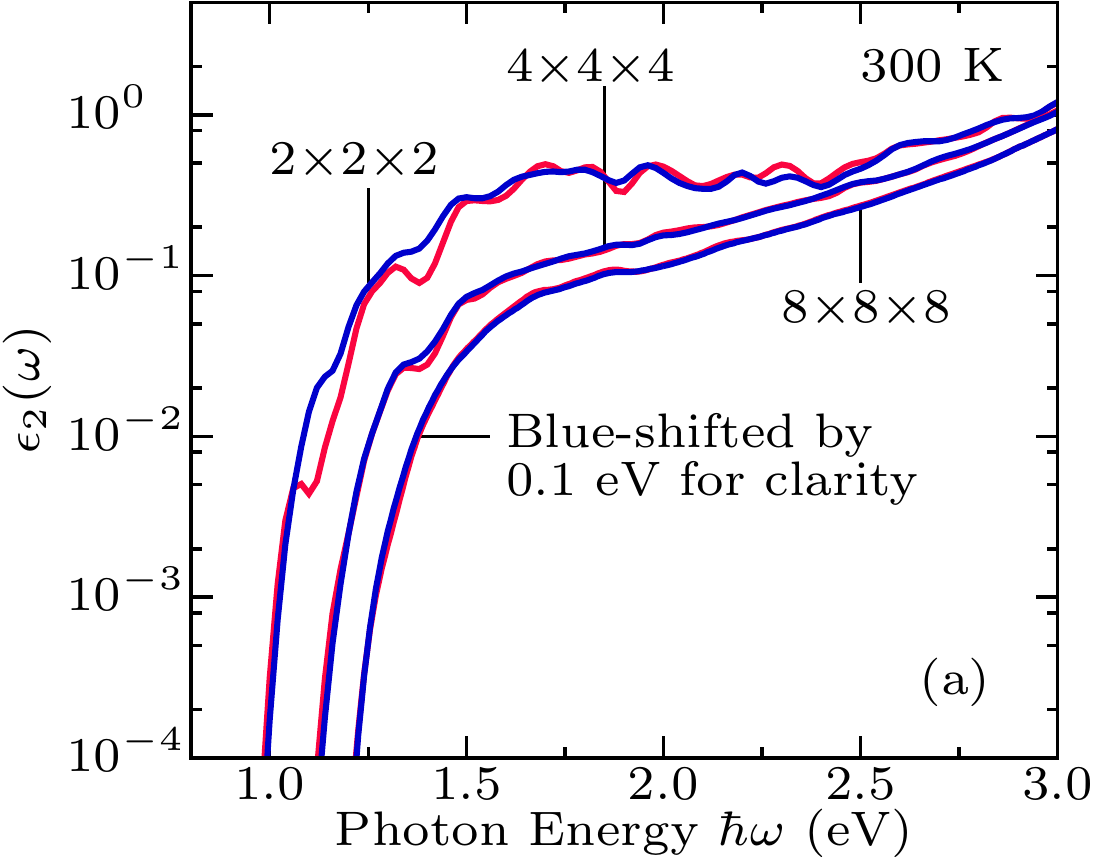}}
  \subfloat{\includegraphics[width=0.35\textwidth]{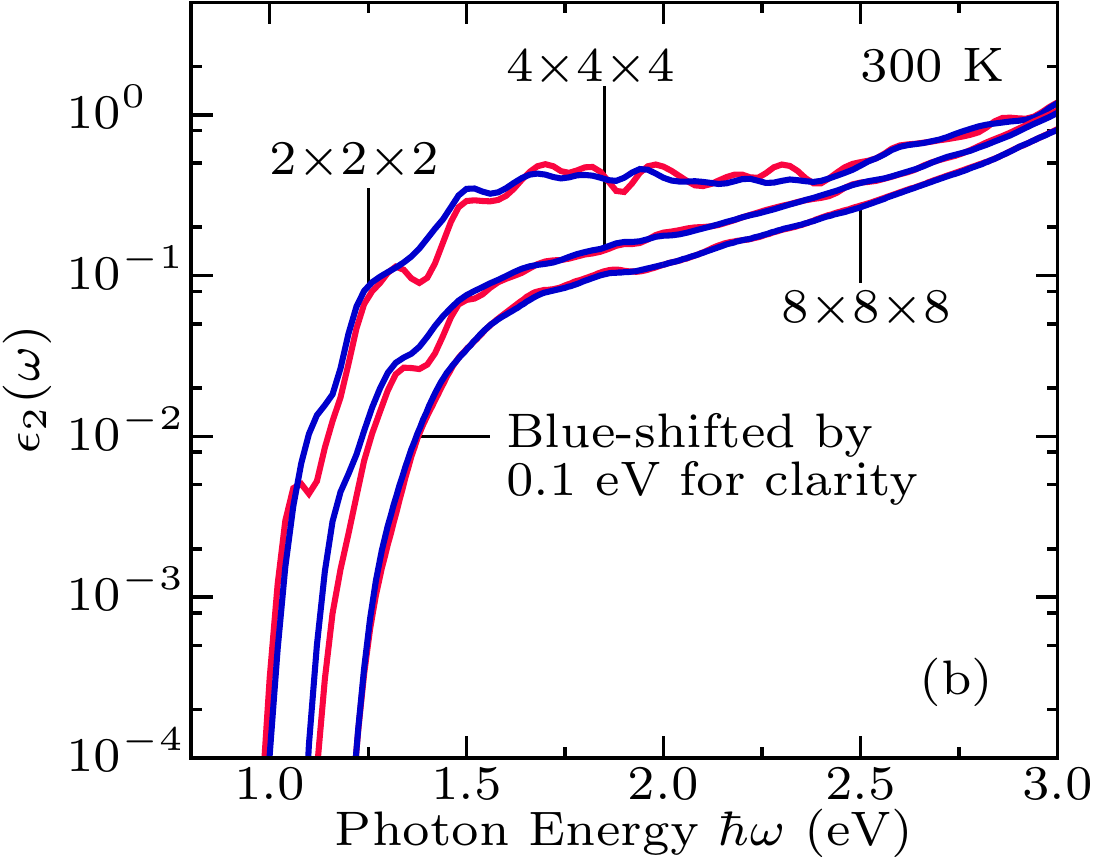}}
  \caption{\label{fig3}
  Effect of using multiple atomic configurations in the evaluation of the WL dielectric function
  of silicon. (a) Comparison between calculations performed using a single atomic configuration
  as specified by Eq.~(\protect\ref{eq.conf_opt}) (red solid lines), and calculations using
  this configuration and its antithetic pair from Eq.~(\protect\ref{eq.anti}) (blue solid line).
  We report calculations for increasing supercell size.
  (b) Comparison between calculations performed using a single configuration as in (a)
  (red solid lines), and those performed using 4 configurations, as specified by
  Eqs.~(\protect\ref{eq.conf_opt}), (\protect\ref{eq.anti}), and (\protect\ref{eq.swap}).
  The curves corresponding to the $8\times8\times8$ supercell have been {\it rigidly shifted}
  by 0.1~eV: without such a shift the curves would overlap with
  those obtained using a $4\times4\times4$ supercell.
}
\end{figure*}
In order to obtain Eq.~(\ref{eq.expansion-exact}) exactly, one would need to further 
eliminate all terms $\nu\ne \mu$ in the sum on the second line of Eq.~(\ref{eq.anti}).
This elimination can be achieved systematically by considering an additional configuration
$S'$ defined as follows:
  \begin{equation}\label{eq.swap}
  \begin{tabular}{lcccc|ccccl}
  $S\phantom{'}=($ & $+$ & $-$ & $+$ & $-$ & $+$ & $-$ & $+$ & $-$ & $)$, \\
  $S'=($ & $+$ & $-$ & $+$ & $-$ & $-$ & $+$ & $-$ & $+$ & $)$,
  \end{tabular}
  \end{equation}
where for the sake of clarity we considered the case $N=8$. In practice $S'$
is simply obtained by swapping all the signs of the second-half of the vector $S$.
It is immediate to verify that the dielectric function calculated by averaging the 4 atomic 
configurations specified by $S$, $-S$, $S'$, and $-S'$
contains only half of the terms $\mu\ne\nu$ that appear in Eq.~(\ref{eq.anti}). 
Since each individual configuration satisfies the asymptotic relation in Eq.~(\ref{eq.equiv}),
it is clear that the above calculation using 4 atomic configurations will approach the
exact WL dielectric function faster as the supercell size increases.

This strategy can also be iterated by partitioning each subset in Eq.~(\ref{eq.swap}) in two
halves, and applying a sign swap on two out of four of the resulting subsets. At each level
of iteration the number of atomic configurations doubles, and the number of remaining terms
$\mu\ne\nu$ in Eq.~(\ref{eq.anti}) halves. For example, it is easy to verify that, by using 4, 16, and 64
configurations generated in this way, it is possible to eliminate 50\%, 87.5\%, and 96.875\%
of the terms $\mu\ne\nu$ in Eq.~(\ref{eq.anti}), respectively. 

In Fig.~\ref{fig3}(a) we show the effect of using Eqs.~(\ref{eq.conf_opt}) 
and (\ref{eq.anti}) for the calculation
of the dielectric function of silicon at 300~K via 2 distinct atomic configurations.
In Fig.~\ref{fig3}(b) we repeat the calculations, this time using 4 distinct configurations,
according to Eqs.~(\ref{eq.conf_opt}), (\ref{eq.anti}), and (\ref{eq.swap}).
Here we see that increasing the number of atomic configurations suppresses spurious
fluctuations in the spectra; the effect is most pronounced for the smallest
supercell, which corresponds to $2\times 2\times 2$ Si unit cells. 
We note that, in Fig.~\ref{fig3}, the curves corresponding to the $8\times8\times8$ 
supercell have been {\it rigidly shifted} for clarity: without such a shift the curves 
are almost indistinguishable from the those calculated using a $4\times4\times4$ supercell.

Figure~\ref{fig3} clearly shows that, irrespective of the configurational averaging,
too small supercells may not be enough to accurately evaluate
the WL dielectric function. This is easily explained 
by considering that, in order to correctly describe an indirect
absorption onset, we need a supercell which can accommodate phonons 
connecting the band extrema. 

\section{Relation between the Williams-Lax theory,
the Allen-Heine theory, and the Hall-Bardeen-Blatt theory}
\label{sec.general-theory}

Having described the conceptual basis of our methodology, we now establish the 
link between the WL dielectric function given by Eq.~(\ref{eq.wl}) and the
standard theory of Hall, Bardeen, and Blatt of indirect optical absorption,\cite{hbb} 
as well as the theory of temperature-dependent band structures of Allen and Heine.\cite{Allen}
In this section we only present the main results, leaving the mathematical details 
to Appendices~\ref{app.WL} and \ref{app.WL-ind}.

The HBB theory describes indirect optical absorption by means of time-dependent perturbation theory,
and the final expression [see Eq.~(\ref{eq.adiab-hbb}) below] involves the momentum
matrix elements evaluated for the nuclei clamped in their equilibrium positions, $p_{cv}$, 
and the linear electron-phonon matrix elements:
   \begin{equation}
   g_{mn\nu} = l_\nu \,\< m | \frac{\D V}{\D x_\nu} | n\>.
   \end{equation} 
Here $|n\>$ denotes a Kohn-Sham state and $\D V/\D x_\nu$ is the variation of the Kohn-Sham
potential with respect to the normal mode coordinate $x_\nu$. In order to make these quantities
explicit in Eq.~(\ref{eq.eps}), we expand the momentum matrix elements $p_{cv}^x$
to first order in the atomic displacements, and the energies $\ve_n^x$ to second order in
the displacements. Using Raleigh-Schr\"odinger perturbation theory we find:
  \begin{equation}\label{eq.pcv}
  p_{cv}^x= p_{cv} + {\sum_{\nu n}}^\prime \left[\frac{p_{cn}\, g_{nv\nu}}{\ve_v-\ve_n}
          +\frac{g_{cn\nu}\, p_{nv}}{\ve_c-\ve_n}\right]
  \frac{x_{\nu}}{l_\nu} + \mathcal{O}(x^2),
  \end{equation}
where the primed summation indicates that we skip terms such that $n=v$ or $n=c$. 
Similarly the expansion of the energies yields, for example:
  \begin{eqnarray}\label{eq.E-exp}
  \ve_c^x = \ve_c & + & \sum_\nu g_{cc\nu} \frac{x_\nu}{l_\nu} +
  {\sum_{\mu\nu n}}^\prime \left[ \frac{ g_{cn\mu} g_{nc\nu} }{\ve_c -\ve_n} + h_{c\mu\nu}
  \right] \frac{x_\mu x_\nu}{l_\mu l_\nu} \nonumber \\ & +& \mathcal{O}(x^3), 
  \end{eqnarray}
  with $h_{c\mu\nu}$ being the `Debye-Waller' electron-phonon matrix element:\cite{Allen,Allen_Cardona_1981,CEP_FG,FG_review} 
  \begin{equation}
  h_{c\mu\nu} = \frac{1}{2}l_\mu l_\nu\< c | \frac{\partial^2 V}{\partial x_\mu\partial x_\nu} | c\>.
  \end{equation}
To make contact with the AH theory of temperature-dependent band structures
and with the HBB theory of indirect absorption, we analyze separately the cases 
of (i) direct gaps and (ii) indirect gaps.

\subsection{Direct gaps}\label{sec.dirgap}
When the gap is direct, the optical matrix elements $p_{cv}^x$ in Eq.~(\ref{eq.pcv}) are dominated by the
term~$p_{cv}$. In fact,
if we denote by $g$ the characteristic electron-phonon matrix element and by $E_{\rm g}$ the minimum gap,
and we note that $x_\nu/l_\nu \sim 1$, then
the terms in the square brackets are $\sim (g/E_{\rm g})p_{cv}$, and hence can be neglected
next to $p_{cv}$.
This approximation is implicitly used in {\it all} calculations of optical absorption spectra
which do not include phonon-assisted processes.
By using this approximation in Eqs.~(\ref{eq.eps}) and (\ref{eq.wl}) we obtain:
 \begin{eqnarray}\label{eq.eps-dir-tmp1}
  &&\!\!\!\e_2(\w;T)  
   = \frac{2 \pi }{ m_{\rm e} N_{\rm e} } \frac{\w_{\rm p}^2}{\,\w^2} \sum_{cv} | p_{cv}|^2
   \nonumber \\
   && \,\,\,\times {\prod}_\nu \int\! dx_\nu \frac{\exp(-x_\nu^2/2\sigma_{\nu,T}^2)}{\sqrt{2\pi}\sigma_{\nu,T}}
  \delta(\ve_{cv} + \Delta\ve_{cv}^x -\hbar\w),\,\,\,\,\,
  \end{eqnarray}
where we have defined $\ve_{cv}^x=\ve_c^x-\ve_v^x$, $\ve_{cv}=\ve_c-\ve_v$, and 
$\Delta\ve_{cv}^x = \ve_{cv}^x - \ve_{cv}$. Equation~(\ref{eq.eps-dir-tmp1}) can be rewritten
by exploiting the Taylor expansion of the Dirac delta distribution in powers of $\Delta\ve_{cv}^x$. 
The derivation is laborious and is reported in Appendix~\ref{app.WL}. The final result is:
  \begin{eqnarray}\label{eq.dir.3}
   && \!\!\!\e_2(\w;T) = \frac{2 \pi }{ m_{\rm e} N_{\rm e} } \frac{\w_{\rm p}^2}{\,\w^2}
     \sum_{cv} | p_{cv}|^2 \nonumber \\
    && \,\,\times \frac{1}{\sqrt{2\pi}\Gamma_{cv}}
    \exp\left[-\frac{\left(\ve_{c,T}^{\rm AH}-\ve_{v,T}^{\rm AH}-\hbar\w\right)^2}{2 \,\Gamma_{cv}^2}\right]
   \!+\! \mathcal{O}(\sigma^4).\,\,
  \end{eqnarray}
In this expression, $\ve_{m,T}^{\rm AH}$ denotes the temperature-dependent electron energy in the Allen-Heine 
theory:\cite{Allen}
   \begin{equation}\label{eq.E.AH}
   \ve_{m,T}^{\rm AH} = \ve_m +
     \sum_\nu \left[ {\sum_{n}}^\prime \frac{ |g_{mn\nu}|^2 }{\ve_m -\ve_n} +h_{m\nu\nu}\, \right]
          (2 n_\nu + 1).
   \end{equation}
In particular, the first term in the square brackets is the Fan self-energy correction,
while the second term is the Debye-Waller correction.\cite{Antoncik_1955,Walter_Cohen_1970,Allen,Allen_Cardona_1981,Cardona_1} 
The quantity $\Gamma_{cv}$ in Eq.~(\ref{eq.dir.3}) is the width of the optical transition, 
and is defined as follows:
    \begin{equation}\label{eq.gamma.main}
   \Gamma_{cv}^2 = 
     {\sum}_\nu \left|g_{cc\nu}-g_{vv\nu}\right|^2 (2 n_{\nu,T}+1).
    \end{equation}
Equation~(\ref{eq.dir.3}) shows that, in the case of direct absorption processes,
the WL theory yields a dielectric function which exhibits normalized peaks 
at the temperature-dependent excitation energies $\ve_{cv,T}^{\rm AH}$. 
We note that the width of the optical transitions $\Gamma_{cv}$ 
is similar to but does {\it not} coincide with the electron-phonon 
linewidth obtained in time-dependent perturbation theory, compare for example with 
Eq.~(169) of Ref.~\onlinecite{FG_review}. This subtle difference is a direct consequence 
of the semiclassical approximation upon which the Williams-Lax theory is based.

To the best of our knowledge the connection derived here between the Williams-Lax theory
and the Allen-Heine theory, as described by Eqs.~(\ref{eq.dir.3}) and (\ref{eq.E.AH}),
is a novel finding.
In particular, the present analysis demonstrates that, to lowest order in perturbation
theory, the WL theory of optical spectra yields the adiabatic version of the AH theory 
of temperature-dependent band structures.

\subsection{Indirect gaps}\label{sec.indirgap}
In the case of indirect semiconductors, owing to the momentum selection rule,
optical transitions near the fundamental gap are forbidden in the absence of phonons.\cite{Cardona_Book}
By consequence, the optical matrix elements at equilibrium must vanish, and we can 
set $p_{cv}=0$ in Eq.~(\ref{eq.pcv}).
This observation represents the starting point of the classic HBB theory of indirect optical processes.\cite{hbb}
In this case Eq.~(\ref{eq.wl}) becomes:
  \begin{eqnarray}\label{eq.eps-dir-tmp2}
  &&\e_2(\w;T)
   = \frac{2 \pi }{ m_{\rm e} N_{\rm e} } \frac{\w_{\rm p}^2}{\,\w^2} \sum_{cv} 
     {\prod}_\nu \int\! dx_\nu \frac{\exp(-x_\nu^2/2\sigma_{\nu,T}^2)}{\sqrt{2\pi}\sigma_{\nu,T}} \nonumber \\
  && \times \left|  {\sum_{\mu n}}^\prime \left[\frac{p_{cn}\, g_{nv\mu}}{\ve_v-\ve_n}
          +\frac{g_{cn\mu}\, p_{nv}}{\ve_c-\ve_n}\right] \!\frac{x_{\mu}}{l_\mu} \right|^2
  \delta(\ve_{cv} + \Delta\ve_{cv}^x -\hbar\w) \nonumber \\ &&+ \mathcal{O}(\sigma^4).
  \end{eqnarray}
We stress that in this expression we are only retaining those terms in Eq.~(\ref{eq.pcv})
which are linear in the normal coordinates. This choice corresponds to considering only {\it one-phonon}
processes, as it is done in the HBB theory. While multi-phonon processes are automatically included in the
WL theory, in the following we do not analyze them explicitly.

\begin{figure*}[ht!]
	\subfloat{\includegraphics[width=0.38\textwidth]{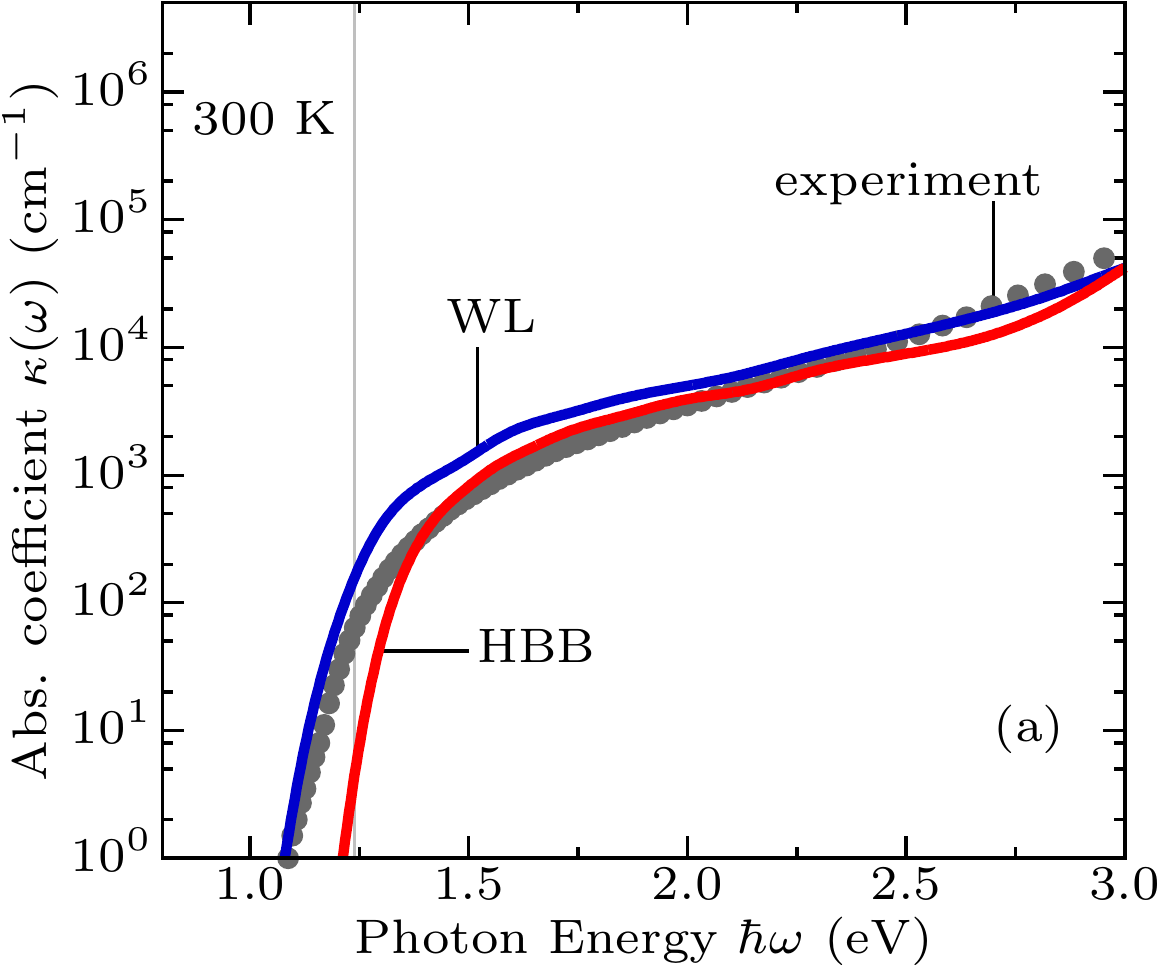}}
	\hspace{0.05cm}
	\subfloat{\includegraphics[width=0.344\textwidth]{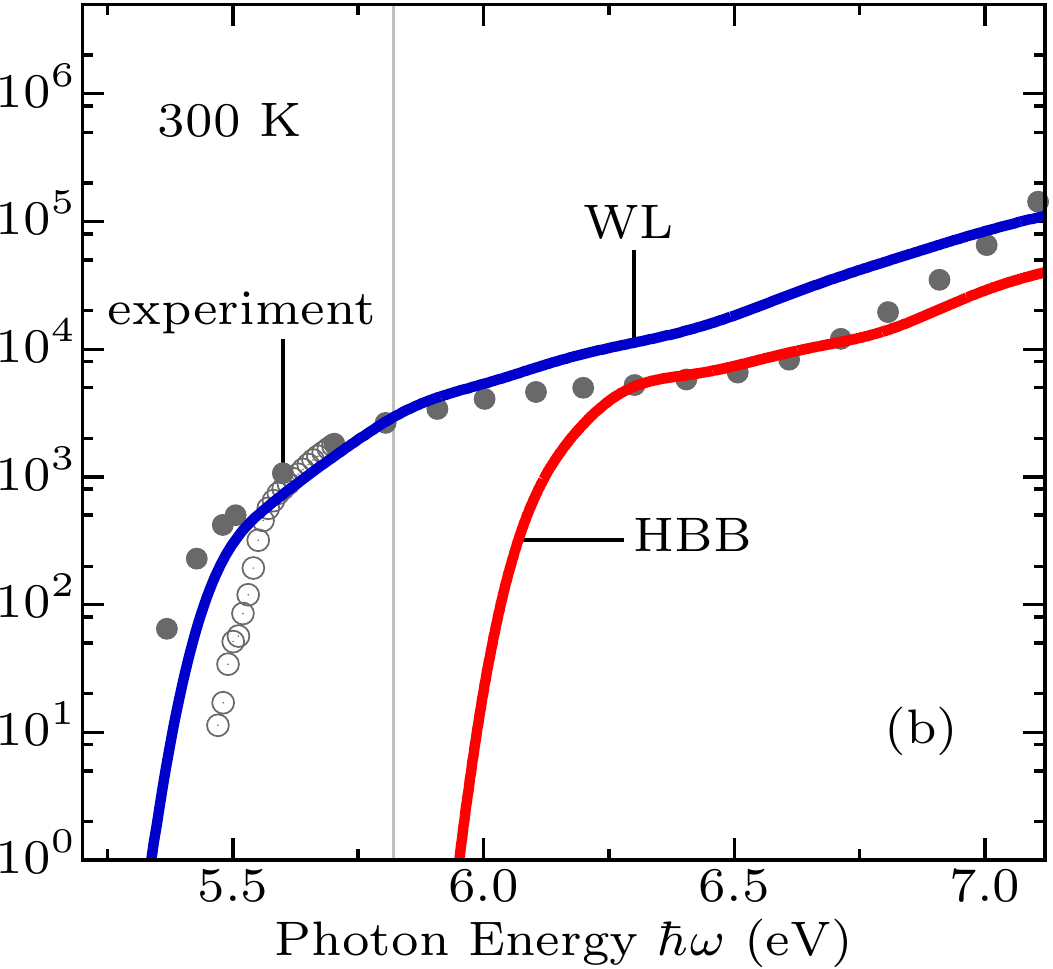}}
	\caption{\label{fig4}
		(a) Comparison between measured and calculated optical absorption coefficients of silicon at 300~K.
		The grey dots are experimental data from Ref.~\onlinecite{M_Green3}, the red line is a calculation
		using the Hall-Bardeen-Blatt theory, and the blue line is a calculation using the Williams-Lax theory.
		(b) Comparison between measured absorption coefficient of diamond at 300~K [data from
		Refs.~\onlinecite{Phillip_1964} (discs) and~\onlinecite{Clark312} (circles)] and our
		calculations using the HBB theory or the WL theory.
		The color code is the same as in (a).
		In both panels the calculations were performed on $4\times4\times4$ supercells, using 65~random
		points in the electronic Brillouin zone (i.e.~$>4000$ points in the Brillouin zone of the
		crystalline unit cell), and a Gaussian smearing of 50~meV.
		The thin vertical lines indicate the energy of the band gaps calculated at the equilibrium geometries.
	}
\end{figure*}

By performing a Taylor expansion of the Dirac delta appearing in Eq.~(\ref{eq.eps-dir-tmp2}) 
in powers of $\Delta\ve_{cv}^x$, as we did for the case of direct gaps, we arrive at the following 
expression:
  \begin{eqnarray}\label{eq.ind-final}
 &&\e_2(\w;T)
   = \frac{2 \pi }{ m_{\rm e} N_{\rm e} } \frac{\w_{\rm p}^2}{\,\w^2} \sum_{cv\nu}
    \left| {\sum_n}^\prime \left[\frac{p_{cn}\, g_{nv\nu}}{\ve_v-\ve_n}
          +\frac{g_{cn\nu}\, p_{nv}}{\ve_c-\ve_n}\right] 
    \right|^2 \nonumber \\
   &&\,\,\,\,\,\times 
   \frac{1}{\sqrt{2\pi}\Gamma_{cv}}
    \exp\left[-\frac{\left(\ve_{c,T}^{\rm AH}-\ve_{v,T}^{\rm AH}-\hbar\w\right)^2}{2 \,\Gamma_{cv}^2}\right]
    (2n_{\nu,T}+1)
      \nonumber \\ &&\,\,\,\,\,+ \mathcal{O}(\sigma^6).
 \end{eqnarray}
The derivation of this result is lengthy, and is reported for completeness in Appendix~\ref{app.WL-ind}.

Equation~(\ref{eq.ind-final}) demonstrates that the WL theory correctly describes indirect optical
absorption. In fact, if we neglect the broadening $\Gamma_{cv}$ and the temperature
dependence of the band structure, we obtain the theory of Hall, Bardeen,
and Blatt:\cite{hbb,Cardona_Book, Noffsinger}
    \begin{eqnarray}\label{eq.adiab-hbb}
 \e^{\rm HBB}_2(\w;T)
  & = &\frac{2 \pi }{ m_{\rm e} N_{\rm e} } \frac{\w_{\rm p}^2}{\,\w^2} \sum_{cv\nu}
    \left| {\sum_n}^\prime \left[\frac{p_{cn}\, g_{nv\nu}}{\ve_v-\ve_n}
          +\frac{g_{cn\nu}\, p_{nv}}{\ve_c-\ve_n}\right]
    \right|^2 \nonumber \\
   &\times &\d\left(\ve_{c}-\ve_{v}-\hbar\w\right) (2n_{\nu,T}+1).
 \end{eqnarray}
The only difference between this last expression and the original HBB theory is that here
the Dirac delta function does not contain the phonon energy. Physically this corresponds
to stating that the WL theory provides the {\it adiabatic} limit of the HBB theory of indirect
absorption.

To the best of our knowledge, this is the first derivation of the precise formal connection between 
the WL theory of indirect absorption, as expressed by Eq.~(\ref{eq.ind-final}), and the theories 
of Hall, Bardeen, and Blatt and of Allen and Heine, as given by Eqs.~(\ref{eq.adiab-hbb}) and (\ref{eq.E.AH}).

We emphasize that the adiabatic HBB theory does {\it not} incorporate the temperature dependence
of the band structure, as it can be seen from Eq.~(\ref{eq.adiab-hbb}). Instead, the WL theory
includes band structure renormalization by default, see Eq.~(\ref{eq.ind-final}). This point
is very important in view of performing predictive calculations at finite temperature.

From Eq.~(\ref{eq.ind-final}) we can also tell that, in order to generalize the HBB theory
to include temperature-dependent band structures, one needs to incorporate the temperature
dependence only in the energies corresponding to {\it real} transitions,
i.e.~in the Dirac deltas in Eq.~(\ref{eq.adiab-hbb}),
and not in the energies corresponding to {\it virtual} transitions, i.e.~the energy denominators in the
same equation.

In order to illustrate the points discussed in Secs.~\ref{sec.dirgap} and \ref{sec.indirgap},
we show in Fig.~\ref{fig4} a comparison between the absorption 
spectra of silicon and diamond calculated using either the HBB theory or the WL theory.
For the HBB calculations we employed the adiabatic approximation and we recast 
Eq.~(\ref{eq.adiab-hbb}) in the following equivalent form:
  \begin{equation}\label{eq.adiab-hbb-equiv}
 \e^{\rm HBB}_2(\w;T) =  \sum_\nu \frac{\D^2 \ve_{2}(\w;x)}{\D x_\nu^2} \sigma_{\nu,T}^2.
 \end{equation}
The equivalence between this expression and Eq.~(\ref{eq.adiab-hbb}) is readily proven
by using Eqs.~(\ref{eq.pcv}) and (\ref{eq.b3}). For the evaluation of the second
derivatives for each vibrational mode we used finite-difference formulas; this
operation requires $2 N$ frozen-phonon calculations.
In the examples shown in Fig.~\ref{fig4} we employed $4\times 4 \times 4$ supercells,
corresponding to 768 calculations on supercells containing 128 atoms.
We emphasize that the WL spectrum requires instead only 1 calculation.

\begin{figure*}[ht]
	\subfloat{\includegraphics[width=0.315\textwidth]{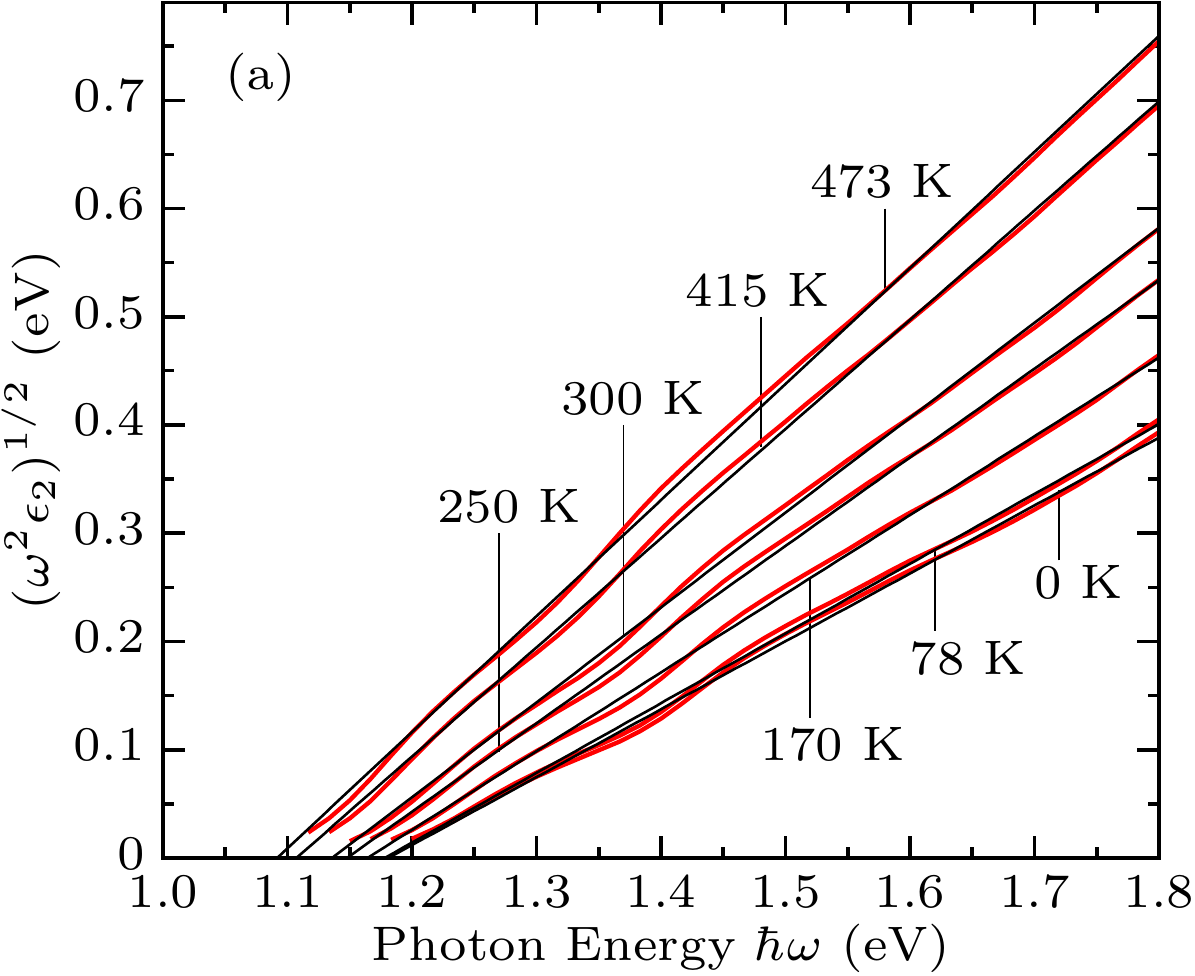}}
	\subfloat{\includegraphics[width=0.32\textwidth]{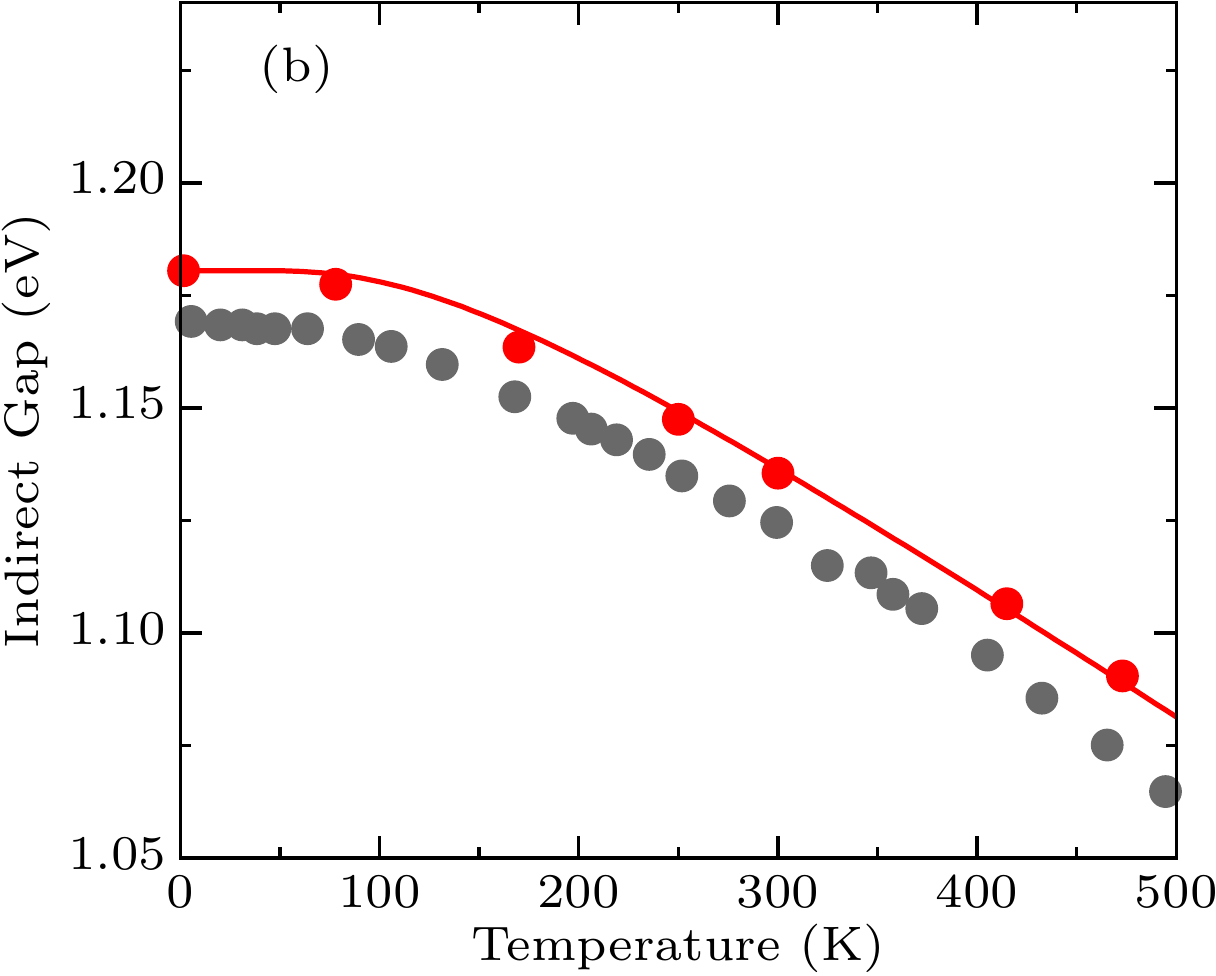}}
	\subfloat{\includegraphics[width=0.32\textwidth]{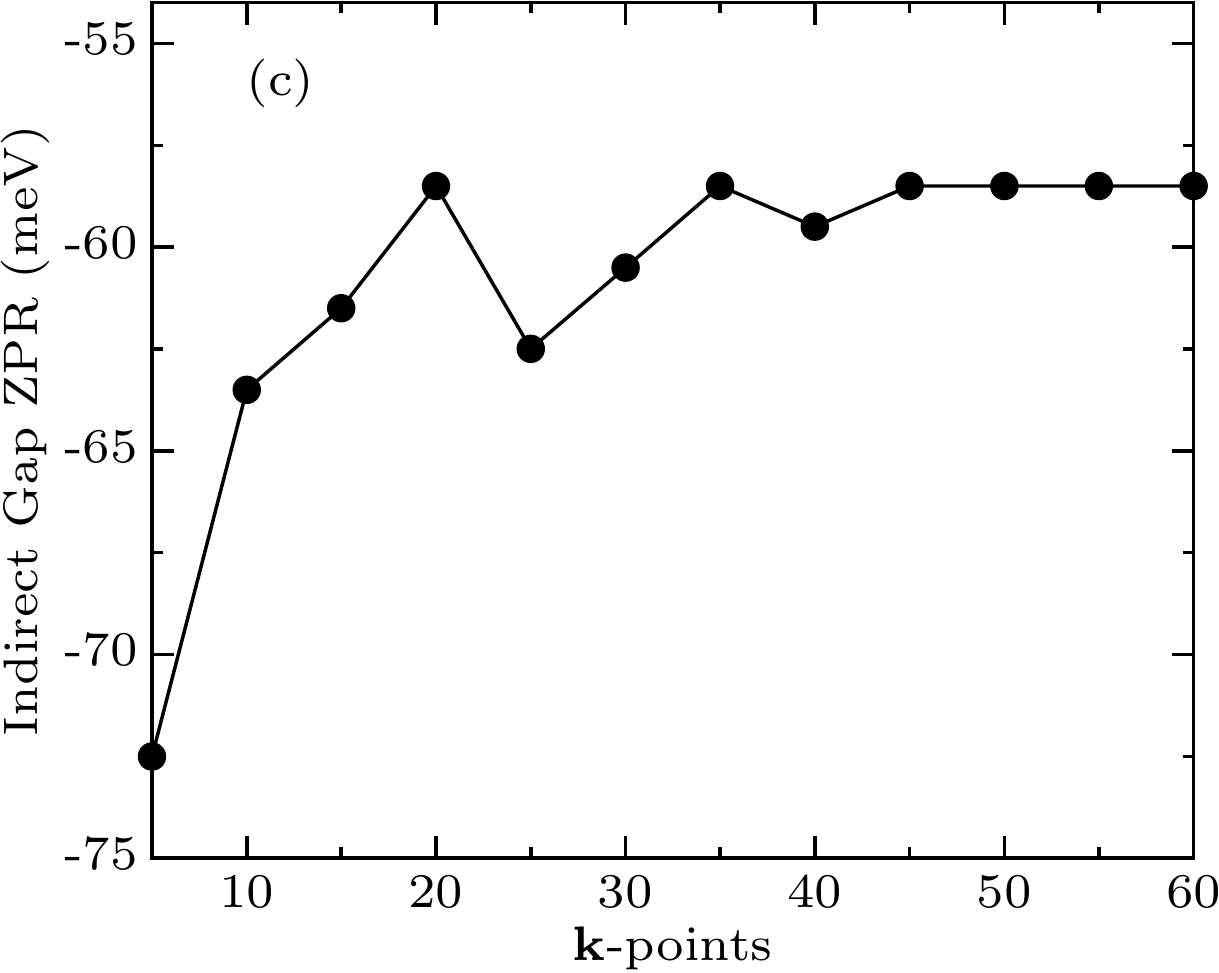}}
	\newline
	\vspace*{-0.0cm}
	\hspace*{-0.6cm}
	\subfloat{\includegraphics[width=0.322\textwidth]{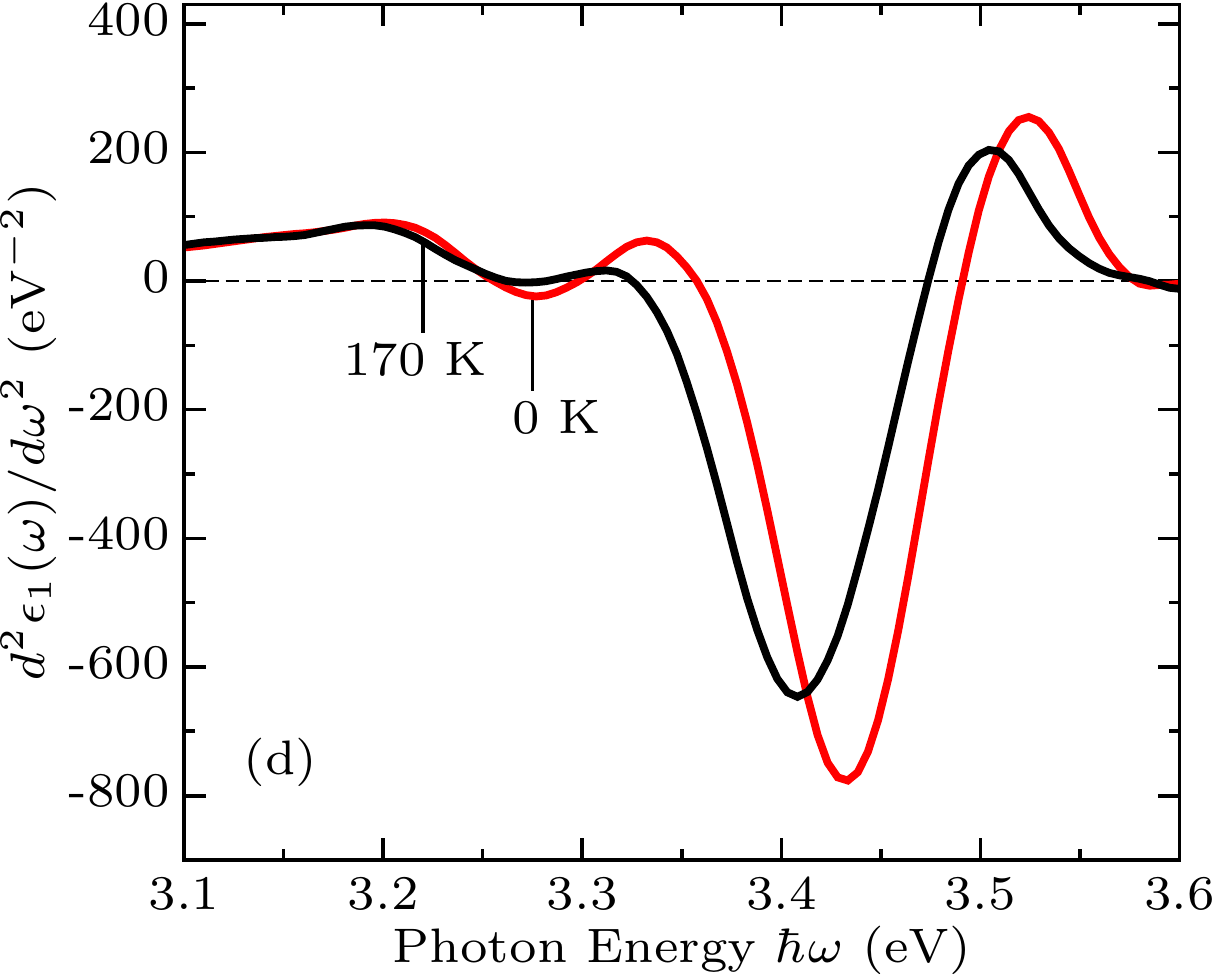}}
	\hspace*{-0.1cm}
	\subfloat{\includegraphics[width=0.314\textwidth]{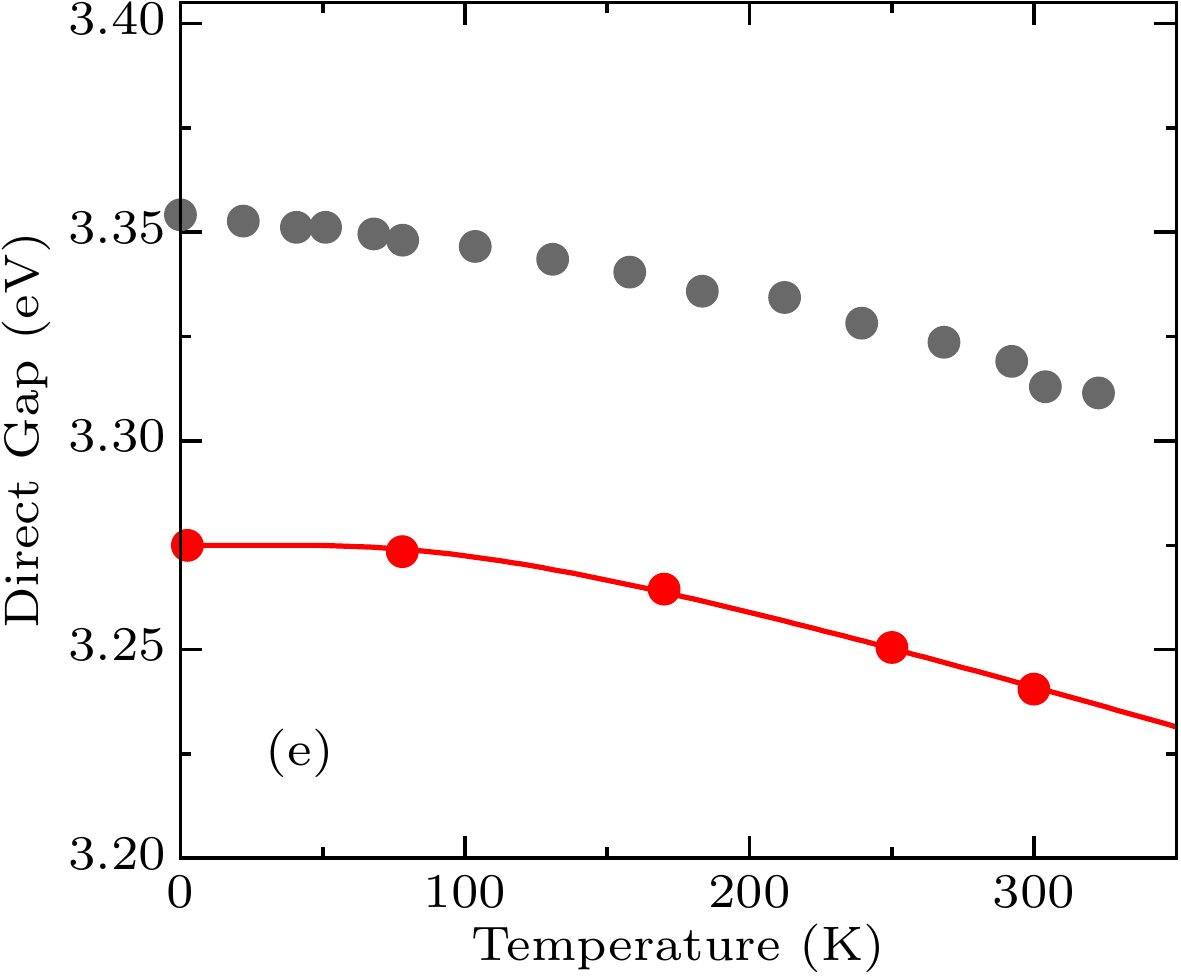}}
	\hspace*{0.1cm}
	\subfloat{\includegraphics[width=0.322\textwidth]{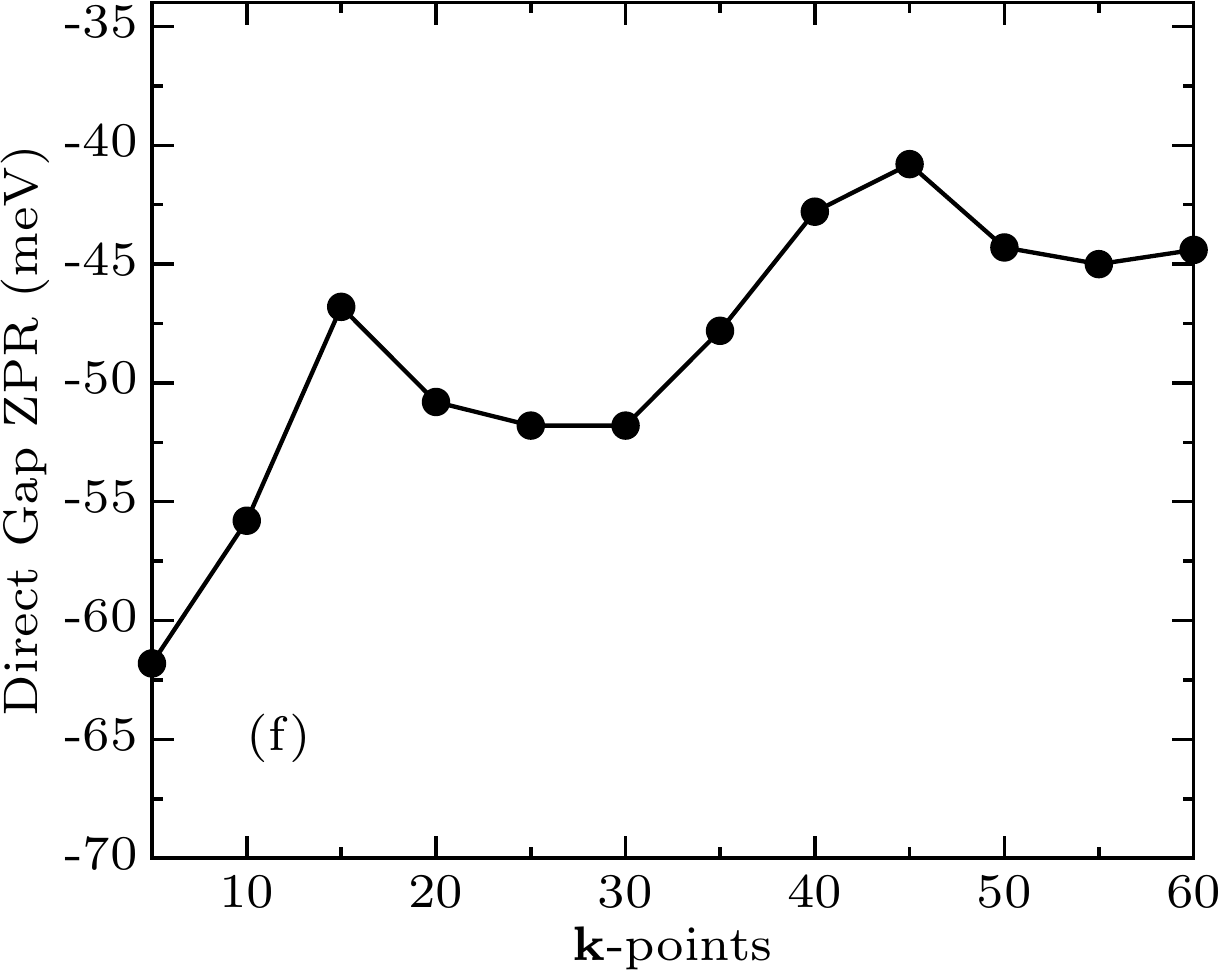}}
	\vspace*{-0.1cm}
	\caption{\label{fig5}
		(a) Tauc plot for determining the indirect band gap of silicon as a function of temperature.
		The red lines show $(\w^2\ve_2)^{1/2}$, and the thin black lines are the corresponding
		linear fits at each temperature. The indirect gap is obtained from the intercept with the
		horizontal axis.
		(b) Temperature-dependence of the indirect band gap of silicon: present theory (red discs)
		and experimental data from Ref.~\onlinecite{Alex} (grey discs). The solid line is a guide to the eye.
		The theoretical values were corrected to match the
		zero-point renormalization calculated as discussed in Appendix~~\ref{app.equil_JDOS}.
		(c) Convergence of the zero-point renormalization of the indirect gap of silicon with respect
		to the Brillouin-zone sampling of an 8$\times$8$\times$8 supercell.
		(d) Calculated second-derivatives of the real part of the dielectric function of silicon.
		(e) Temperature-dependence of the direct gap of silicon: present calculations (red discs)
		and experimental data from Ref.~\onlinecite{Lautenschlager_Si}. The solid line is a guide to the eye.
		(f) Convergence of the zero-point renormalization of the direct gap of silicon
		with respect to the Brillouin-zone sampling.
		All calculations in this figure were performed using an 8$\times$8$\times$8 supercell.
	}
\end{figure*}

Figure~\ref{fig4}(a) compares the measured absorption coefficient of silicon (grey dots) with 
those calculated using the HBB theory (red line) and the WL theory (blue line). All data
are for the temperature $T=300$~K. Here we see that the WL and HBB calculations yield similar
spectra. However, while the optical absorption onset in the HBB theory coincides with the
indirect band gap of silicon at equilibrium, the WL spectrum is {\it red-shifted} by an amount
$\sim$0.1~eV, corresponding to the zero-point renormalization and the temperature shift of the gap. 
As a result, the WL calculation is in better agreement with experiment. 
We emphasize again that the agreement between theory and experiment
remarkably extends over a range spanning 6 orders of magnitude, and the curves shown in 
Fig.~\ref{fig4} do not carry any empirical scaling factors.

The effect of band gap renormalization becomes more spectacular in the case of diamond,
as shown in Fig.~\ref{fig4}(b). In this case the HBB calculation misses the absorption
onset by as much as $\sim$0.5~eV, while the WL theory yields an onset in agreement
with experiment. The discrepancy between the prediction of the adiabatic HBB theory 
and experiment is understood as the result of the very large electron-phonon renormalization of the
band gap of diamond\cite{FG_diamond,Cannuccia2011,Antonius2014}. In addition,
the adiabatic version of the HBB theory also misses the small redshift associated with
phonon-emission processes.\cite{Noffsinger}

We stress that the WL theory does not come without faults. The main
shortcoming is that, unlike the HBB theory, it does not capture 
the fine structure corresponding to phonon absorption and emission processes near the absorption 
onset. This is a direct consequence of the semiclassical approximation underpinning the WL theory,
whereby the quantization of the final vibrational states is replaced by a classical continuum.\cite{Lax}

\section{Band gap renormalization} \label{sec.allresults}

\subsection{Comparison between calculations using the Williams-Lax theory
and Hall-Bardeen-Blatt theory}

In Sec.~\ref{sec.general-theory} we demonstrated that the imaginary part of the
temperature-dependent dielectric function in the WL theory exhibits an absorption
onset at the temperature-dependent band gap given by the AH theory, see Eqs.~(\ref{eq.dir.3})
and (\ref{eq.ind-final}).
This result suggests that it should be possible to extract
temperature-dependent band gaps directly from calculations of $\ve_2(\w;T)$ or $\kappa(\w;T)$, 
as it is done in experiments.

 \begin{figure*}[htb]
	 \subfloat{\includegraphics[width=0.317\textwidth]{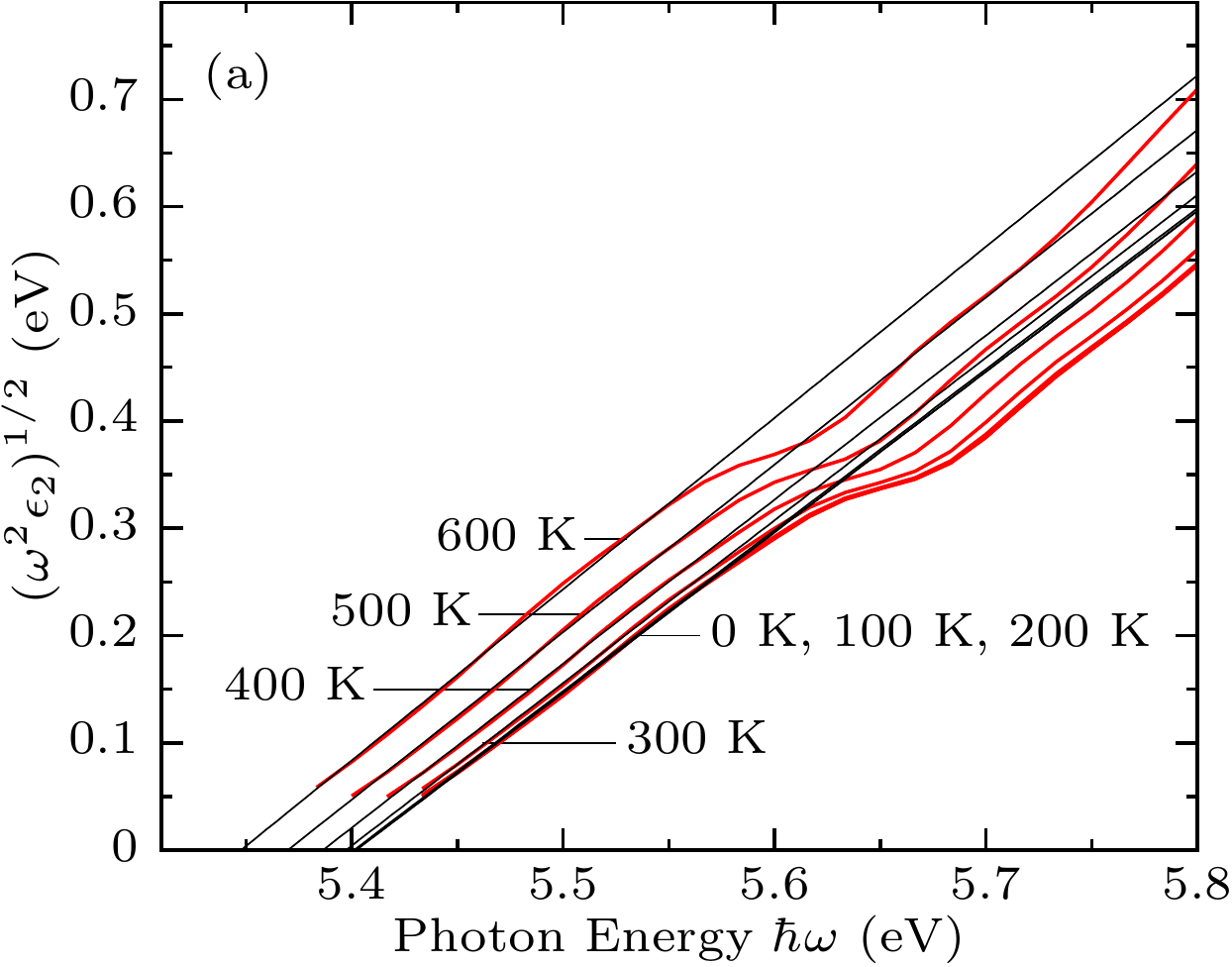}}
	 \subfloat{\includegraphics[width=0.32\textwidth]{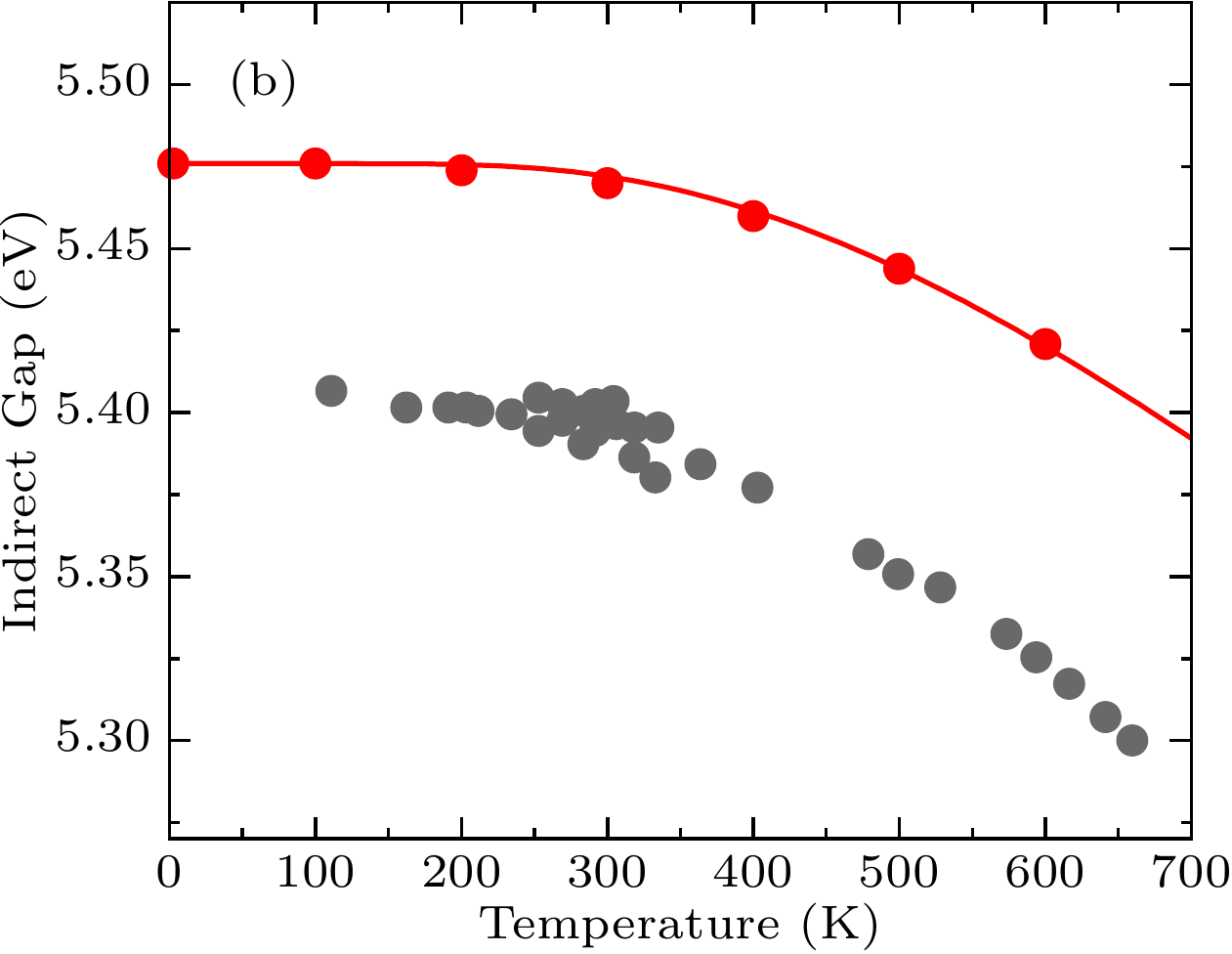}}
	 \subfloat{\includegraphics[width=0.32\textwidth]{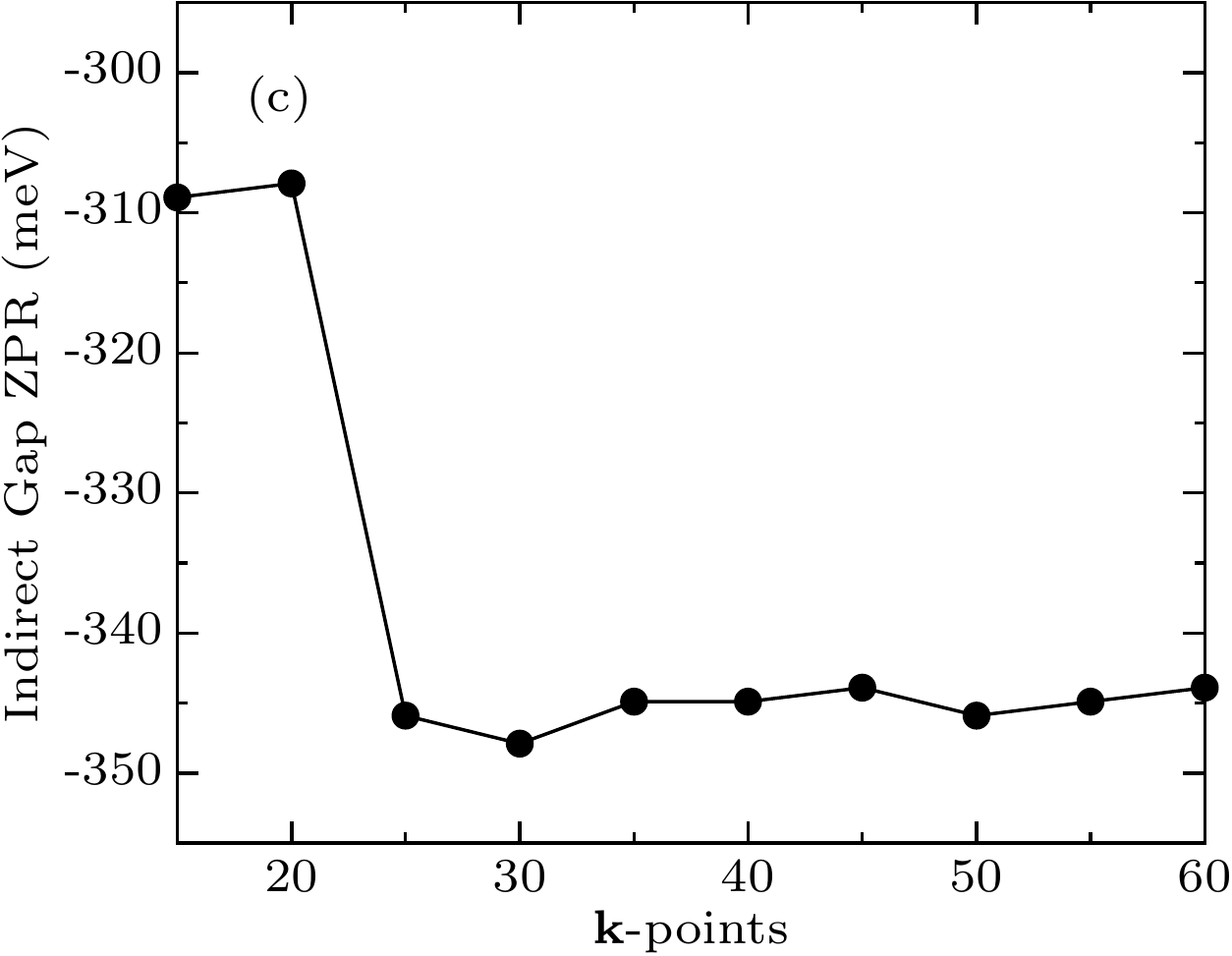}}
	 \vspace*{-0.2cm}
	 \newline
	 \vspace*{-0.2cm}
	  \hspace*{-0.3cm}
	  \subfloat{\includegraphics[width=0.308\textwidth]{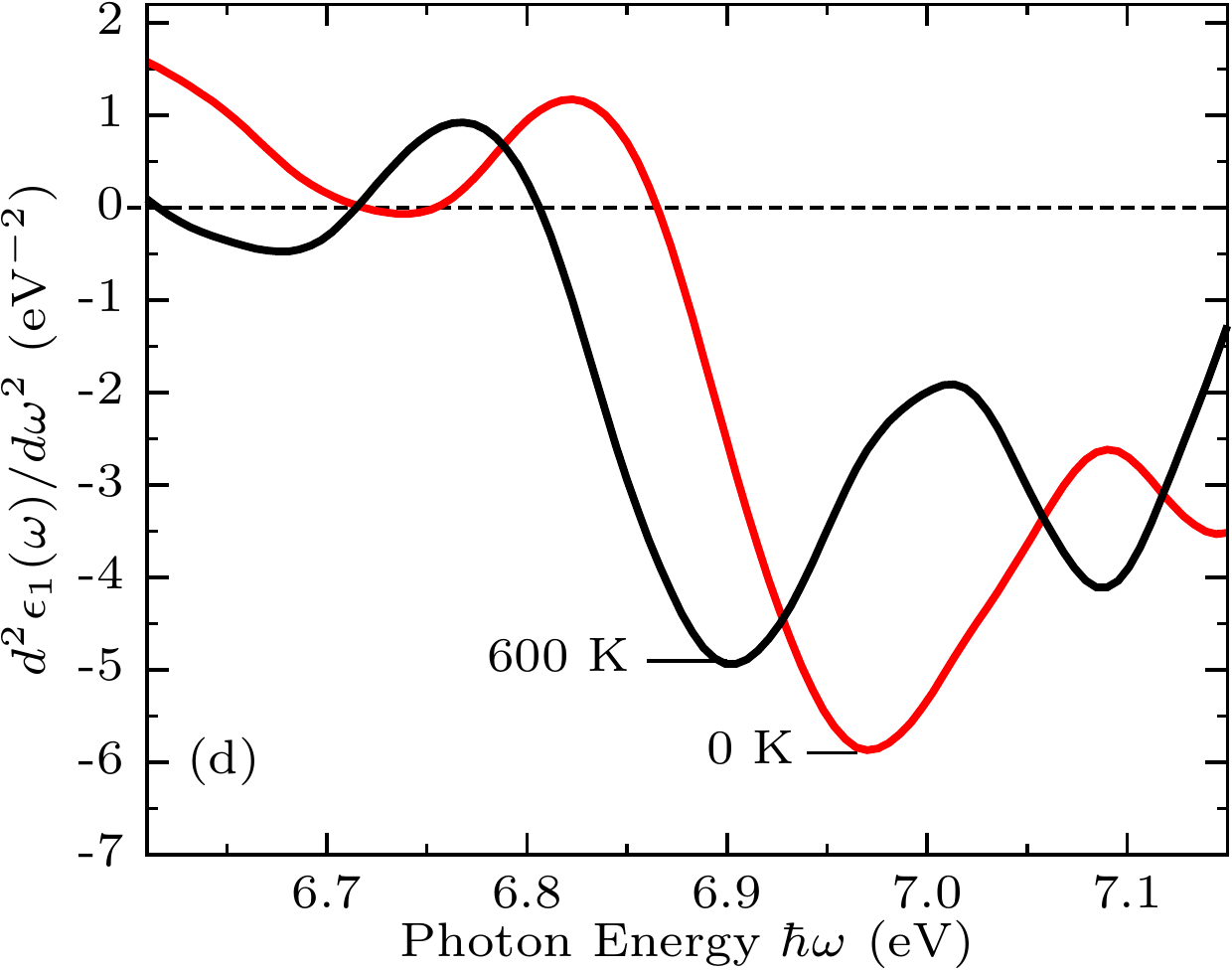}}
	    \hspace*{0.15cm}
	    \subfloat{\includegraphics[width=0.32\textwidth]{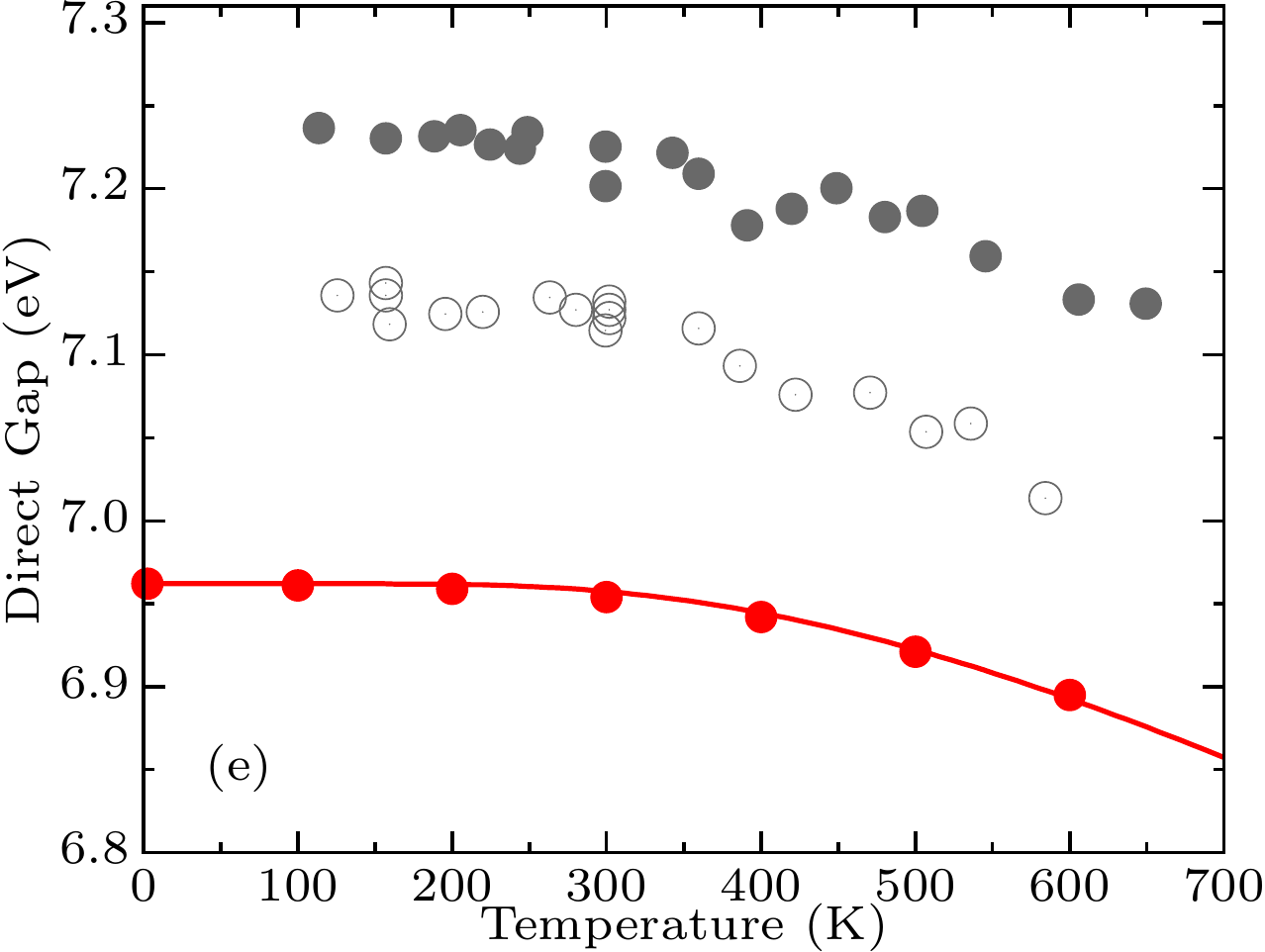}}
	    \subfloat{\includegraphics[width=0.31\textwidth]{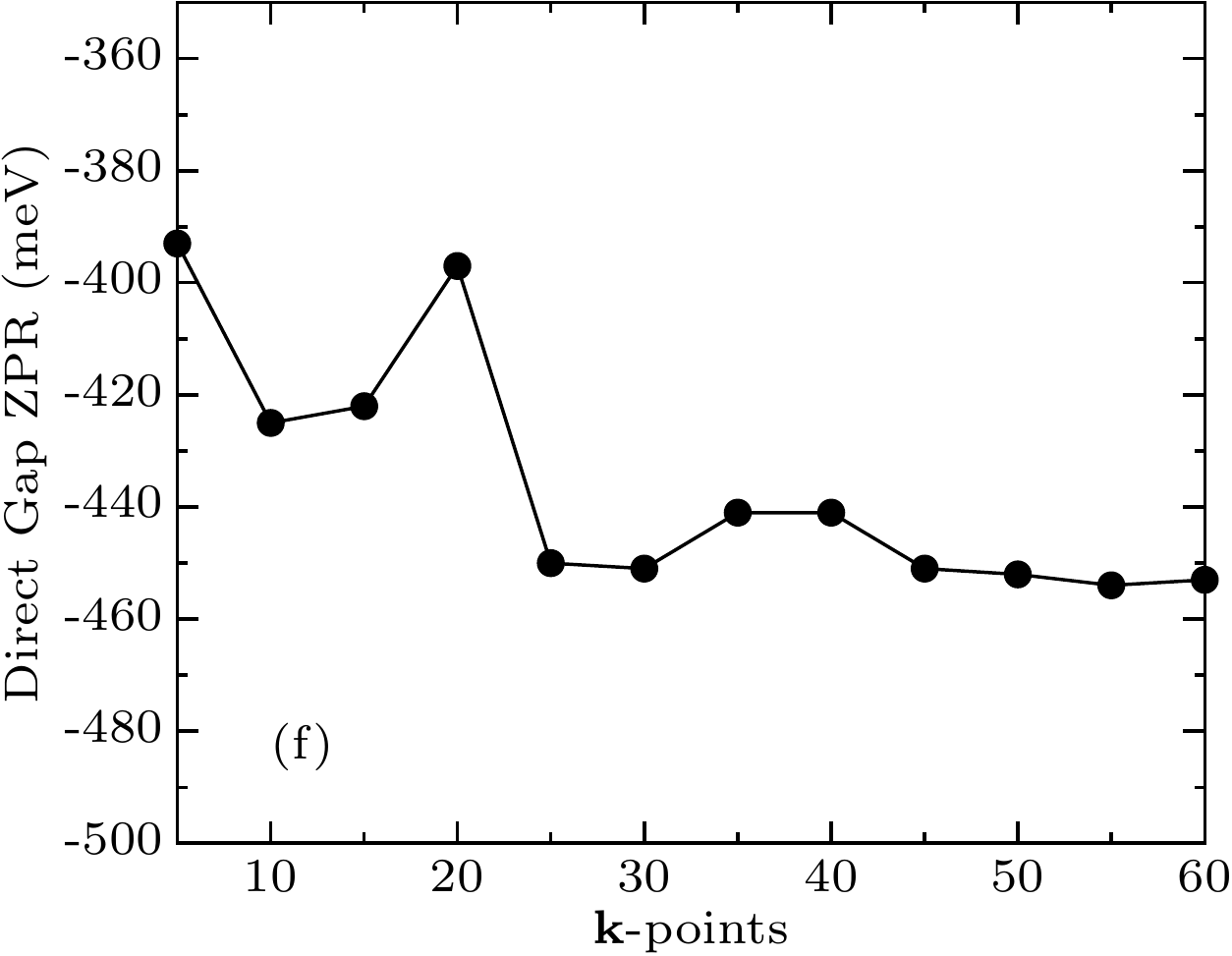}}
	    \caption{\label{fig6}
(a) Tauc plot to determine the indirect gap of diamond vs temperature. The red/black lines
indicate $(\w^2\ve_2)^{1/2}$ and the corresponding linear fits at each temperature. The indirect
gap is obtained from the intercept with the horizontal axis. In this case the linear fits
were performed in the range of photon energies $\hbar\w=0$-5.6~eV.
(b) Temperature-dependence of the indirect gap of diamond: this work (red discs)
and experimental data from Ref.~\onlinecite{Clark312} (grey discs). The solid line is a guide
to the eye. The theoretical values were corrected to match the 
zero-point renormalization calculated as discussed in Appendix~~\ref{app.equil_JDOS}.
(c) Convergence of the zero-point renormalization of the indirect gap of diamond with respect
to the Brillouin-zone sampling (for an 8$\times$8$\times$8 supercell).
(d) Calculated $\D^2 \ve_1(\w;T)/ \D\w^2$ for diamond at two different temperatures.
The $E_0'$ transition is identified using the deep minimum, following Ref.~\onlinecite{Logothetidis_1992}.
(e) Temperature-dependent direct band gap of diamond: current calculations (red discs)
and experimental data from Ref.~\onlinecite{Logothetidis_1992}. We report the experimental
data corresponding to both diamond samples II$a$ (grey circles) and II$b$ (grey discs) used in Ref.~\onlinecite{Logothetidis_1992}.
(f) Convergence of the zero-point renormalization of the direct band gap of diamond
with respect to Brillouin-zone sampling. All calculations were performed using an
8$\times$8$\times$8 supercell.
}
    \end{figure*}
Using Eqs.~(\ref{eq.dir.3}) and (\ref{eq.ind-final}) and the standard parabolic approximation
for the band edges of three-dimensional solids,\cite{Parravicini} the following relations can
be obtained after a few simple manipulations:
  \begin{eqnarray}
\mbox{direct:}\,\,\,\,\,\,\,\,&& \left[\w^2 \ve_2(\w;T) \right]^2 = {\rm const}\times
\left(\hbar\w - E_{{\rm g},T}\right), \label{eq.tauc.1}\\
\mbox{indirect:}\,\,\,&& \left[\w^2 \ve_2(\w;T) \right]^\frac{1}{2} \!= {\rm const}\times
\left(\hbar\w - E_{{\rm g},T}\right).\label{eq.tauc.2}
\end{eqnarray}
Here $E_{{\rm g},T}$ is the temperature-dependent band gap,
and these relations are valid near the absorption onset.
Equations~(\ref{eq.tauc.1}) and (\ref{eq.tauc.2}) simply reflect the joint-density
of states of semiconductors, and form the basis for the standard Tauc
plots which are commonly used in experiments in order to determine band gaps.\cite{Tauc}
These relations are employed as follows: after having determined $\ve_2(\w;T)$, one plots
$(\w^2 \ve_2 )^2$ for direct-gap materials, or
$(\w^2 \ve_2 )^{1/2}$ for indirect gaps. The plot should be linear
near the absorption onset, and the intercept with the horizontal axis
gives the band gap $E_{{\rm g},T}$.

In the case of indirect-gap semiconductors it is not possible to use Eq.~(\ref{eq.tauc.1})
in order to determine the {\it direct} band gap. Nevertheless, in these cases one can still
determine the gap by analyzing second-derivative spectra of the real part of the
dielectric function, $\D^2 \ve_1(\w;T)/ \D\w^2$.
These spectra exhibit characterisic dips, which can be used to identify the direct gap.
This is precisely the procedure employed in experiments in order to measure direct
band gaps.\cite{Lautenschlager_Si,Logothetidis_1992,Lautenschlager_GaAs}
In the following we use this lineshape analysis to calculate the band gaps of Si, C, and GaAs.

\subsection{Indirect semiconductors: silicon}\label{sec.indirect-Si}

Figure~\ref{fig5}(a) shows the Tauc plots obtained for silicon using the WL theory.
As expected from Eq.~(\ref{eq.tauc.1}), we obtain straight lines over an energy range of almost
1~eV from the absorption onset. From the intercept of linear fits taken in the range 0-1.8~eV
we obtain the indirect band gaps at several temperatures; the results are shown in Fig.~\ref{fig5}(b)
as red discs, and compared to experiment (grey discs).
The agreement with experiment is good, with the exception of a constant offset
which relates to our choice of scissor correction for the band structure of silicon
(cf.~Sec.~\ref{sec.computational}). 

In order to minimize numerical noise, we determine the zero-point renormalization of the band 
gap, $\Delta E_{\rm g}$, as the offset between the square root of the joint density of states
calculated at equilibrium and that at $T=0$~K. This refinement is discussed in 
Appendix~\ref{app.equil_JDOS}. 
In this case we find $\Delta E_{\rm g}=57$~meV.
Small changes of this value are expected for larger supercells and denser Brillouin-zone sampling.
For completeness we show in Fig.~\ref{fig5}(c) the convergence of the zero-point
renormalization of the indirect gap with the number of $\bk$-points in the supercell.

\begin{figure}[ht!]
  \subfloat{\includegraphics[width=0.45\textwidth]{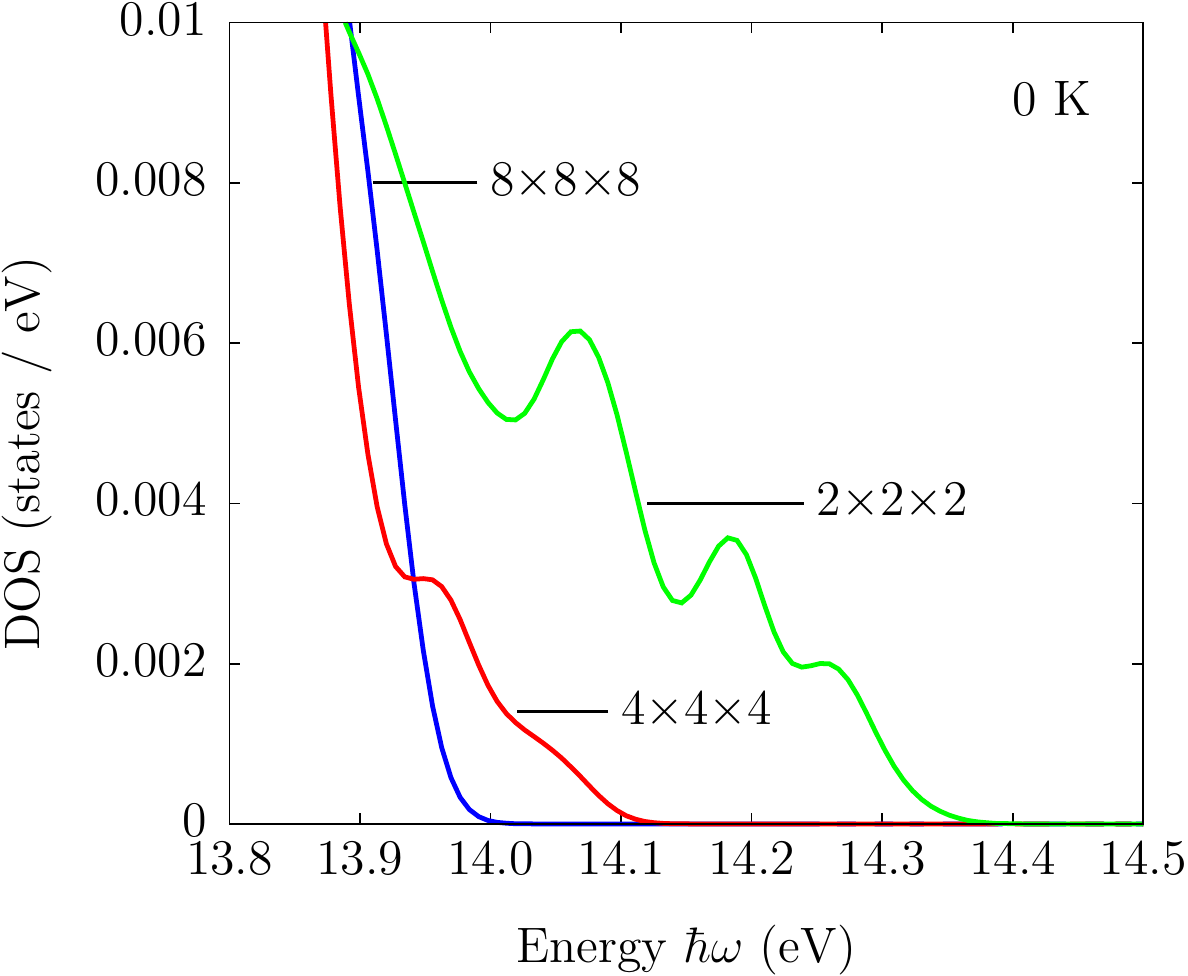}}
 \caption{\label{fig7}
Density of states near the valence band edge of diamond at 0~K. Calculations were performed
on  $2\times2\times2$ (green), $4\times4\times4$ (red) and $8\times8\times8$ (blue) supercells
using the one-shot WL method. We used 2560, 320 and 40 {\bf k}-points, respectively, and
a Gaussian broadening of 15~meV. The broken degeneracy of the valence band top, which can
be seen for the $2\times2\times2$ supercell, tends to vanish with increasing supercell size.
}
\end{figure}

Our calculated zero-point renormalization of the indirect gap of silicon is in good
agreement with previous calculations and with experiment. In fact,
Ref.~\onlinecite{Monserrat2014} reported a renormalization of 60~meV using finite-differences
supercell calculations; Ref.~\onlinecite{Bartomeu_2016} reported 58~meV using Monte Carlo
calculations in a supercell.
Ref.~\onlinecite{Ponce2015} obtained a zero-point
renormalization of 56~meV using the perturbative Allen-Heine approach.
Measured values of the renormalization range between 62 and 64~meV.\cite{Cardona_2001,
Cardona20053, Cardona_1}

\begin{figure*}[t!]
	\subfloat{\includegraphics[width=0.32\textwidth]{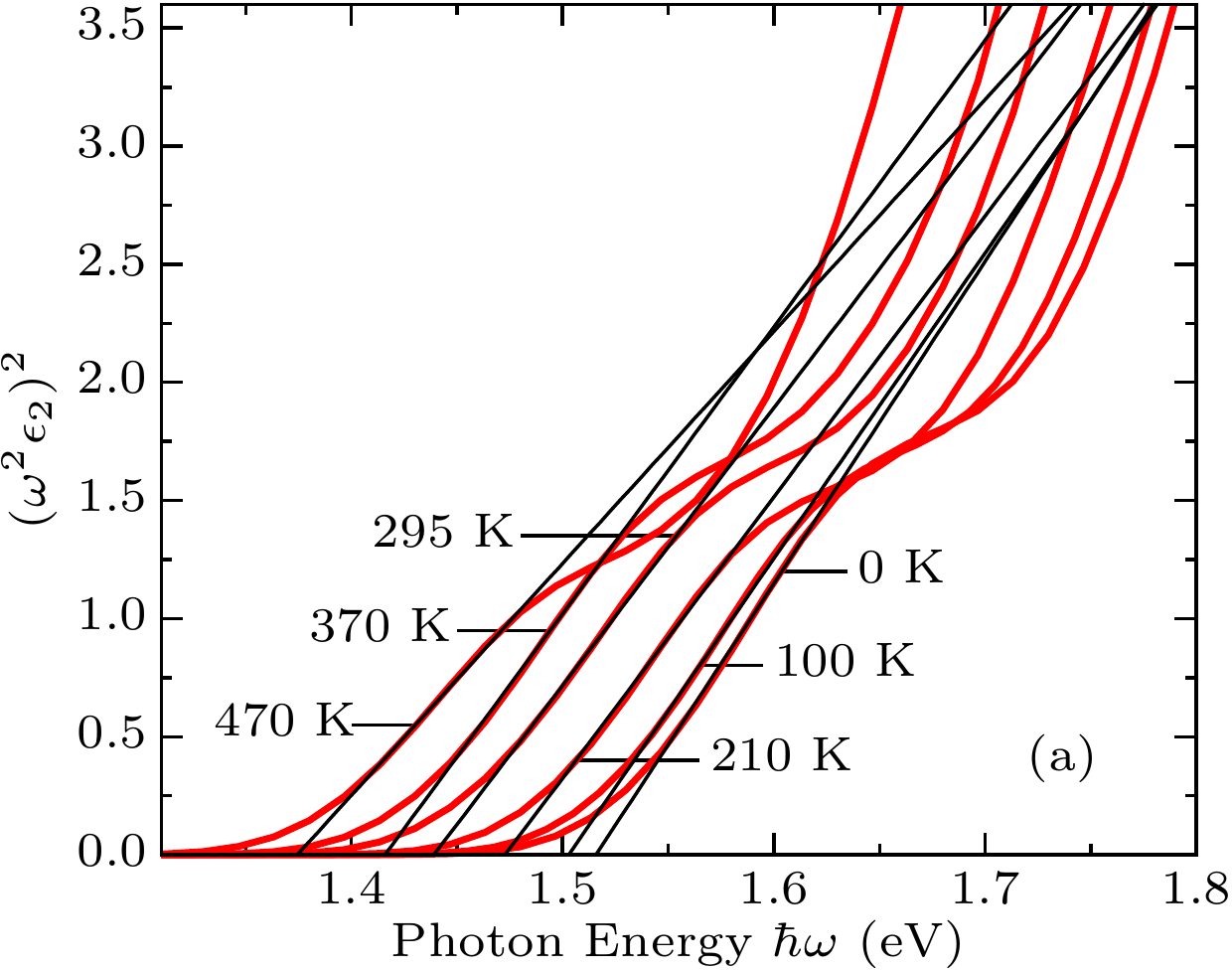}}
	\subfloat{\includegraphics[width=0.32\textwidth]{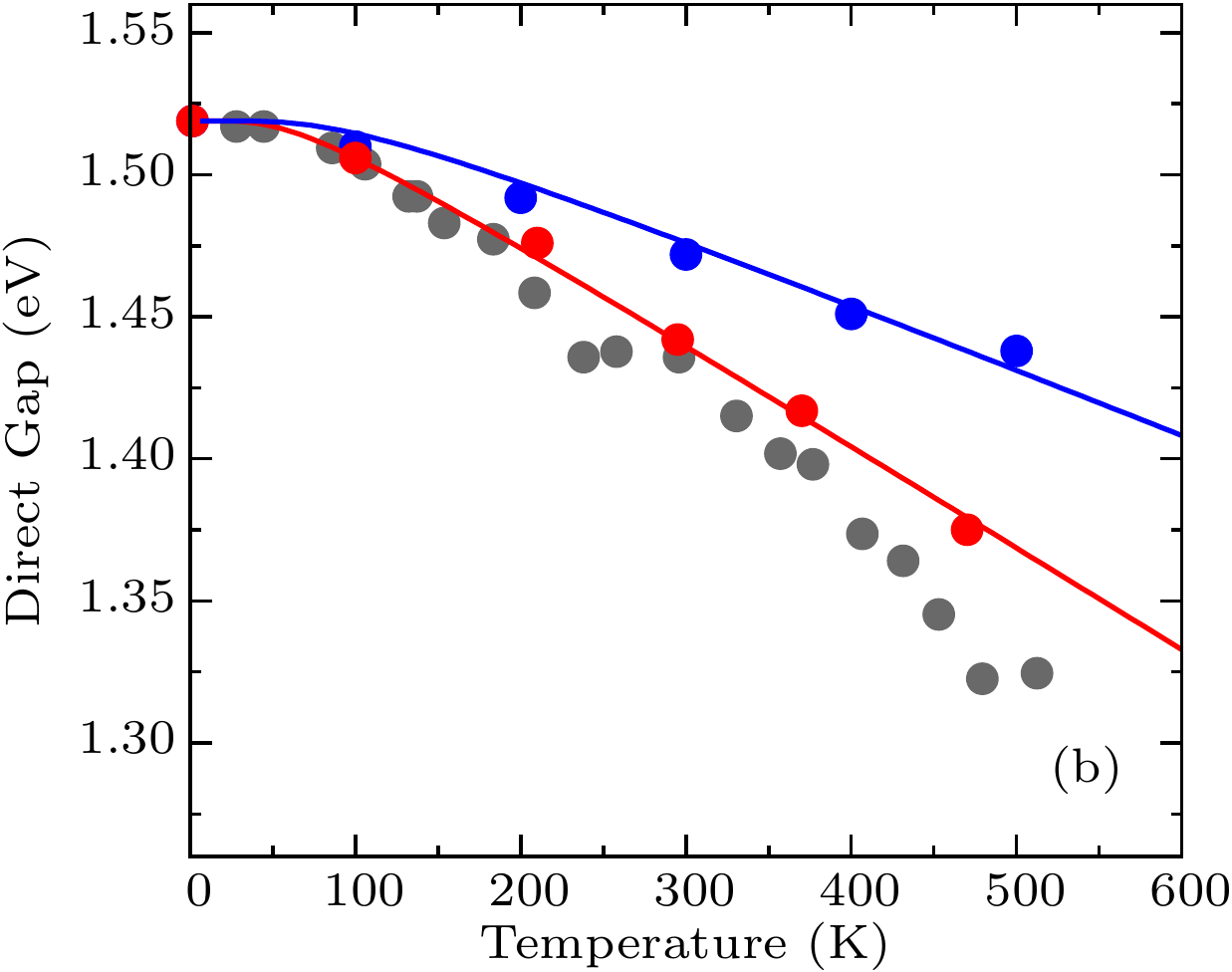}}
	\subfloat{\includegraphics[width=0.32\textwidth]{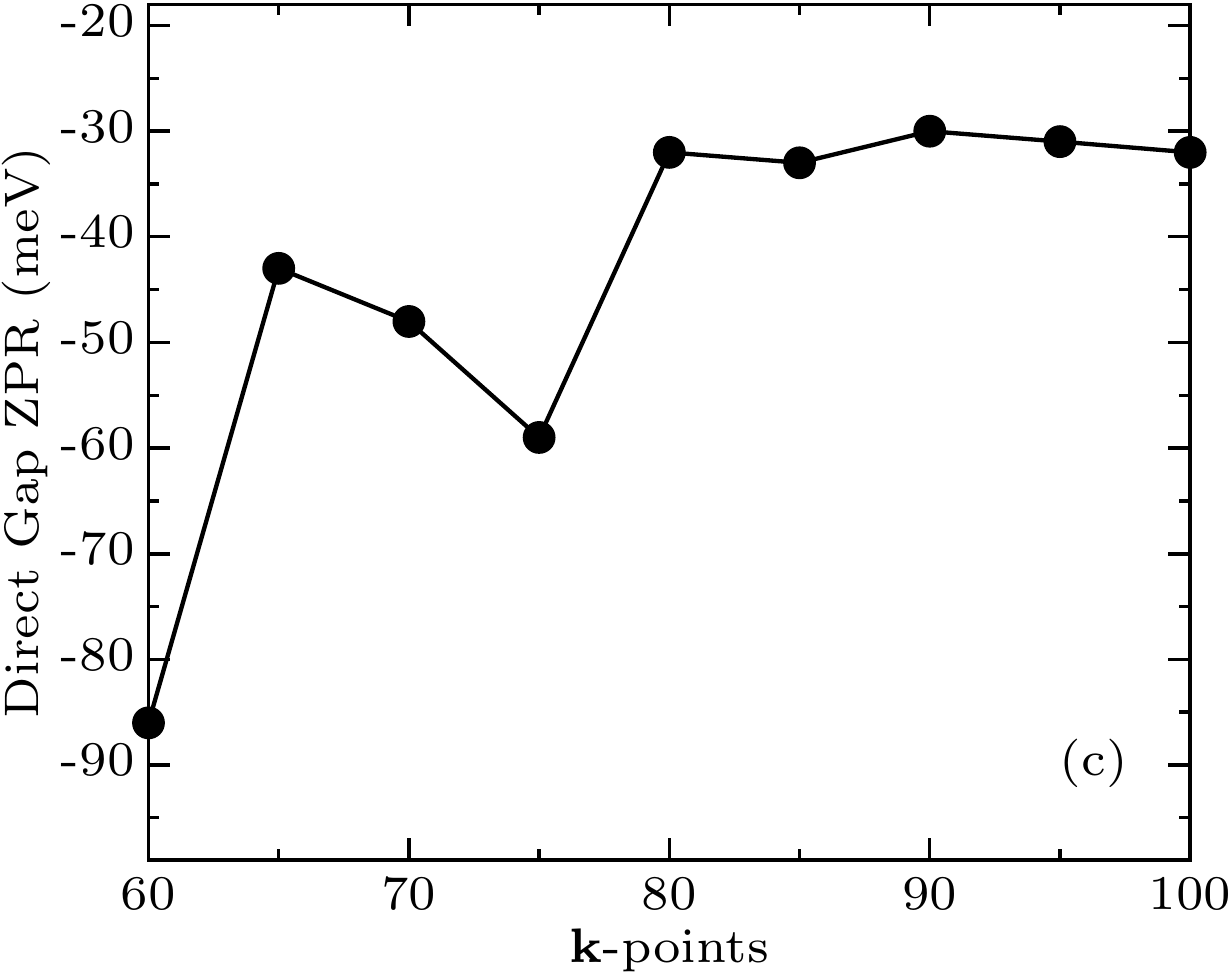}}
	\vspace*{-0.2cm}
	\caption{\label{fig8}
		(a) Direct optical absorption onset in GaAs for various temperatures, as calculated using
		the WL theory in a 8$\times$8$\times$8 supercell. The calculations were performed in the
		quasi-harmonic approximation. The red curves correspond to $(\w^2\ve_2)^2$ and the thin
		black lines are the corresponding linear fits (within the range of photon energies 1.42-1.62~eV).
		(b) Temperature-dependence of the band gap of GaAs:
		calculations in the quasi-harmonic approximation (red discs), calculations without considering
		the lattice thermal expansion (blue discs), and experimental data from
		Ref.~\onlinecite{Lautenschlager_GaAs} (grey discs). The lines are guides to the eye.
		(c) Convergence of the calculated zero-point renormalization with respect to the
	number of $\bk$-points, in a 8$\times$8$\times$8 supercell.}
\end{figure*}

Figure~\ref{fig5}(d) shows the second-derivative spectra, $\D^2 \ve_1(\w;T)/ \D\w^2$,
calculated for silicon at two temperatures. Following Ref.~\onlinecite{Lautenschlager_Si},
we determine the energy of the $E_0'$ transition using the first dip in the spectra.
In Fig.~\ref{fig5}(e) we compare the direct band gaps thus extracted with the experimental
values. Apart from the vertical offset between our data and experiment, which relates
to the choice of the scissor correction (cf.~Sec.~\ref{sec.computational}), the agreement
with experiment is good.
In order to determine the zero-point renormalization of the direct gap,
we took the difference between the dips of the second-derivative spectra
calculated at equilibrium and using the WL theory at $T=0$~K.
We obtained a zero-point renormalization of $44$~meV,
which compares well with the experimental range $25\pm17$~meV.\cite{Lautenschlager_Si}
Values in the same range were reported in previous calculations, namely 28~meV from
Ref.~\onlinecite{Monserrat_2016_GW} and 42~meV from Ref.~\onlinecite{Ponce2015}.
We note that, in Ref.~\onlinecite{Monserrat_2016_GW}, a GW calculation on a 4$\times$4$\times$4
supercell yielded a renormalization of 53 meV.
For completeness we show in Fig.~\ref{fig5}(f) the convergence of the zero-point
renormalization with respect to Brillouin-zone sampling.

\subsection{Indirect semiconductors: diamond}\label{sec.indirect-C}

Figure~\ref{fig6}(a) shows the Tauc plots calculated for diamond using the WL approach.
As in the case of silicon, we determined the indirect gap as a function of temperature
by means of the intercept of the linear fits with the horizontal axis; the results are
shown in Fig.~\ref{fig6}(b) together with the experimental data of Ref.~\onlinecite{Clark312}.
From this comparison we see that the agreement with experiment is reasonable,
however the calculations underestimate the measured renormalization. This effect
has been ascribed to the fact that DFT/LDA underestimates the electron-phonon matrix
elements as a consequence of the DFT band gap problem.\cite{Antonius2014}
Our calculated zero-point renormalization of the indirect gap is $345$~meV.
This result was obtained following the procedure in Appendix~\ref{app.equil_JDOS}.
Our value for the zero-point renormalization is compatible with previously reported values based on DFT/LDA,
namely 330~meV,\cite{Ponce2015} 334~meV,\cite{Monserrat2014}  343~meV,\cite{Bartomeu_2015} and 344~meV.\cite{Bartomeu_2016}. 
Our calculations  are in good agreement with the experimental values of 340~meV and 370~meV
reported in Refs.~\onlinecite{Cardona_2001,Cardona20053}. However, the use
of a more recent extrapolation of the experimental data yields a renormalization
of 410~meV,~\cite{Monserrat2014_b} which is 65~meV larger than our result.
Since it is known that GW quasiparticle corrections do increase the zero-point
renormalization as compared to DFT/LDA,\cite{Antonius2014,Monserrat_2016_GW}
we expect that by repeating our WL calculations in a GW framework our results will 
be in better agreement with the experimental data. 
For completeness in Fig.~\ref{fig6}(c) we show the convergence of the zero-point
renormalization with Brillouin-zone sampling (for an 8$\times$8$\times$8 supercell).

Figure~\ref{fig6}(d) shows the second-derivative spectra of the real part of the
dielectric function of diamond, as calculated from the WL theory. The direct
gap of diamond was obtained in Ref.~\onlinecite{Logothetidis_1992} using the deep
minimum in the experimental curves; here we follow the same approach, and
we report our results in Fig.~\ref{fig6}(e). For comparison we also show the
experimental data from Ref.~\onlinecite{Logothetidis_1992}.
Apart from the vertical offset which reflects our choice of scissor correction,
the calculations are in reasonable agreement with experiment. In particular
we determined a zero-point renormalization of 450~meV, 
to be compared with the experimental ranges of $180\pm150$~meV (sample II$a$ of Ref.~\onlinecite{Logothetidis_1992}) 
and $450\pm370$~meV (sample II$b$ of Ref.~\onlinecite{Logothetidis_1992}).
Our value of the zero-point renormalization is compatible with previous calculations at the DFT/LDA level,
for example Ref.~\onlinecite{Bartomeu_2015}
reported 400~meV (when using an 8$\times$8$\times$8 Supercell), Ref.~\onlinecite{Ponce_2014_2} reported 409~meV, 
Ref.~\onlinecite{Monserrat_2016_GW} reported 410~meV (value extracted from Fig.3)
and Ref.~\onlinecite{Ponce2015} reported 416~meV.

We also point out that a previous work by one of us on the electron-phonon renormalization
in diamond reported a correction of 615~meV for the direct gap using DFT/LDA.\cite{FG_diamond}
The origin of the overestimation obtained in Ref.~\onlinecite{FG_diamond} might be related
to numerical inaccuracies when calculating electron-phonon matrix elements for unoccupied Kohn-Sham
states at very high-energy, although this point is yet to be confirmed.
Recent work demonstrated that GW quasiparticle corrections lead to an increase
of the zero-point renormalization of the direct gap.\cite{Monserrat_2016_GW,Antonius2014} 
Calculations of optical spectra using the WL theory and the GW method are 
certainly desirable, but lie beyond the scope of the present work.

For completeness, in Appendix~\ref{app.WL_AH} we also investigate multi-phonon
effects. In particular, we prove that WL calculations correctly yield the generalization
of the adiabatic AH theory to the case of two-phonon processes, and we demonstrate that
multi-phonon effects provide a negligible contribution to the band gap renormalization.

We also note incidentally that, in the case of small supercells, the band gap 
renormalization includes an additional spurious contribution when the band extrema are degenerate.
This effect arises from a linear-order electron-phonon coupling which lifts the band degeneracy,
and has already been observed in path-integral molecular dynamics calculations on
diamond using a supercell with 64 atoms.~\cite{Hernandez_2006}
This effect can easily be seen in the density of states in Fig.~\ref{fig7} as three separate 
peaks near the valence band edge. In the limit of large supercells this effect vanishes,
since it arises from zone-center phonons, whose weight becomes negligibly small as $N\rightarrow\infty$.

\subsection{Direct semiconductors: gallium arsenide} \label{sec.direct}

Figure~\ref{fig8} shows our WL calculations of the temperature-dependent absorption
onset and band gap of GaAs. Since GaAS is polar, we included the non-analytical part
of the dynamical matrix in the calculations of the vibrational frequencies and eigenmodes.\cite{Wang_2012}
In this case we also took into account the thermal expansion of the lattice, which is not
negligible for this semiconductor.\cite{Walter_Cohen_1970} To this aim we performed calculations 
within the {\it quasi-harmonic} approximation,\cite{Baroni_2001} that is we repeated the
calculations of the normal modes for each temperature, by varying the lattice constant 
according to the measured thermal lattice expansion coefficient.\cite{CLEC'H,Pierron}

\begin{table*}[th]
 \caption{
 Vibrational zero-point renormalization of the band gaps of silicon, diamond, and gallium
 arsenide. Our present results based on the WL theory and Tauc plots are compared with
 previous DFT/LDA calculations and with experiment. The grids within the
 brackets indicate the sampling of the Brillouin-zone or the supercell size employed in previous 
 calculations.
 }
 \centering
 {\footnotesize
 \begin{ruledtabular}
 \begin{tabular*}{0.5\textwidth}{l  *{3}c  *{3}c } 
 &   \multicolumn{3}{c}{Indirect gap} & \multicolumn{3}{c}{Direct gap}   \\  
	  &  Present & Previous &  Experiment& Present  & Previous   & Experiment       \\ [0.05 cm]  \hline \\
	  [-0.62 cm]\\
	  Si  &  57  & 56\footnote{Ref.~\onlinecite{Ponce2015} (75$\times$75$\times$75)}, 58\footnote{Ref.~\onlinecite{Bartomeu_2016} (6$\times$6$\times$6)}, 60\footnote{Ref.~\onlinecite{Monserrat2014} (5$\times$5$\times$5)}    & 62\footnote{Ref.~\onlinecite{Cardona_2001}}, 64\footnote{Ref.~\onlinecite{Cardona20053}}
	   & 44 & 28\footnote{Ref.~\onlinecite{Monserrat_2016_GW} (4$\times$4$\times$4)}, 42\footnotemark[1]   & $25\pm17$\footnote{Ref.~\onlinecite{Lautenschlager_Si}}   \\ [0.05 cm] 
	  C   &  345  & 330\footnotemark[1], 334\footnote{Ref.~\onlinecite{Monserrat2014} (6$\times$6$\times$6)}, 343\footnote{Ref.~\onlinecite{Bartomeu_2015} (48$\times$48$\times$48)}, 344\footnotemark[2]   & 370\footnotemark[5], 410\footnote{Ref.~\onlinecite{Monserrat2014_b}}&
	   450 & 409\footnote{Ref.~\onlinecite{Ponce_2014_2}}, 410\footnotemark[6], 416\footnotemark[1], 430\footnotemark[9]  &$180\pm150$\footnote{Ref.~\onlinecite{Logothetidis_1992}}, $450\pm370$\footnotemark[12]  \\ [0.05 cm]
	  GaAs &      &    &  & 32  & 23\footnote{Ref.~\onlinecite{Antonius2014} (4$\times$4$\times$4)} & $57\pm29$\footnote{Ref.~\onlinecite{Lautenschlager_GaAs}} \\ [0.05 cm]
 \end{tabular*}
 \end{ruledtabular}}
 \label{table.1}
\end{table*}

In Fig.~\ref{fig8}(a) we show the direct absorption onset. In this case we extract
the gap from straight-line fits of $(\w^2\ve_2)^2$, after Eq.~(\ref{eq.tauc.1}).
Since the range where the
function $(\w^2\ve_2)^2$ is linear is rather narrow (due to the presence
of low-lying conduction band valleys), 
we refine the calculation of the zero-point renormalization using the more
accurate procedure discussed in Appendix~\ref{app.equil_JDOS}.

In Fig.~\ref{fig8}(b) we show the calculated band gaps as a function of temperature
and we compare our results to experiment. For completeness we report calculations
performed without considering the thermal expansion of the lattice.
Here we see that our calculations are in reasonable agreement with experiment,
and that lattice thermal expansion is definitely not negligible.
Using an 8$\times$8$\times$8 supercell we obtained a zero-point renormalization of 
32~meV. 
Our value is in line with previous calculations,
yielding 23~meV,\cite{Antonius2014} as well as with the experimental
range of $57\pm29$~meV. We expect that GW quasiparticle corrections
will further increase the zero-point correction by $\sim$10~meV.\cite{Antonius2014}
For completeness in Fig.~\ref{fig8}(c) we show the convergence of the zero-point
renormalization with the sampling of the Brillouin zone (in an 8$\times$8$\times$8 supercell).

Recently it was pointed out that, in the case of polar semiconductors, calculations based
on the Allen-Heine theory exhibit a spurious divergence when the
quasiparticle lifetimes are set to zero.\cite{Ponce2015,Allen_arxiv} The origin of this artifact
relates to the Fr\"olich electron-phonon coupling,\cite{Verdi_2015,Sjakste_2015} and
has been discussed in Ref.~\onlinecite{FG_review}. In the present calculations it is
practically impossible to test whether we would have a singularity in the limit of 
very large spercells. However, we speculate that our calculations should not diverge,
since the Born-von K\'arm\'an boundary conditions effectively short-circuit the long-range electric 
field associated with longitudinal optical phonons. This aspect will require  
separate investigation.

In Table~\ref{table.1} we summarize all our calculations of zero-point renormalization
for Si, C, and GaAs, and compare our present results with previous theory and experiment.

\section{Computational setup} \label{sec.computational}

All calculations were performed within the local density approximation to density functional 
theory.\cite{CA1980,PZ1981} We used norm-conserving pseudopotentials,\cite{Fuchs199967} 
as implemented in the {\tt Quantum ESPRESSO} distribution.\cite{QE} The Kohn-Sham wavefunctions 
were expanded in planewave basis sets with kinetic energy cutoffs of 40~Ry, 50~Ry, and 120~Ry 
for Si, GaAs and C, respectively. 
The interatomic force constants in the Born-von K\'arm\'an supercell were obtained as the
Fourier transforms of the dynamical matrices calculated in the primitive unit cells via
density functional perturbation theory.\cite{Baroni2001} 

All calculations of band gap renormalization were performed by using two atomic configurations:
our optimal configuration, given by Eq.~(\ref{eq.conf_opt}), and its antithetic pair, as obtained
by exchanging the signs of all normal coordinates. This choice guarantees high accuracy in the
lineshapes near the absorption onset.
The absorption coefficients shown in Fig.~\ref{fig1} were calculated using $\k(\w;T) = \w\, \e_2(\w;T)/c\, n(\w,T)$,
where $c$ is the speed of light and $n(\w,T)$ the refractive index calculated as
$n(\w,T) = \left[ \sqrt{\e_1^2(\w;T)+\e_2^2(\w;T)}+\e_1(\w;T) \right]^{1/2}/\sqrt{2}$.
The optical matrix elements including the commutators with the non-local components of the pseudopotential
\cite{Starace_1971} were calculated using {\tt Yambo}.\cite{Marini20091392}
In order to compensate for the DFT band gap problem, we rigidly shifted the conduction bands
so as to mimic GW quasiparticle corrections. The scissor corrections were taken from previous
GW calculations performed using the same computational setup for Si and C,\cite{Henry_FG}
and from Ref.~\onlinecite{Remediakis_1999} for GaAs. In particular,
we used $\Delta=0.74$~eV, 1.64~eV, and 0.53~eV for Si, C, and GaAs, respectively.
The non-locality of the scissor operator was correctly taken into account in the oscillator 
strengths \cite{Starace_1971} via the renormalization factors $(\ve_c-\ve_v)/(\ve_c-\ve_v
+\Delta)$; this ensures that the $f$-sum rule is fulfilled. 

The dielectric functions were calculated by replacing the Dirac deltas in Eq.(\ref{eq.eps}) 
with Gaussians of width 30~meV for Si and C, and 50~meV, for GaAs. 
All calculations presented in this work were performed using 8$\times$8$\times$8 supercells of the 
primitive unit cell, unless specified otherwise.
The sampling of the Brillouin zone of each supercell was performed using random $\bk$-points,
with weights determined by Voronoi triangulation.\cite{Voronoi}~In order to
sample the 8$\times$8$\times$8 supercells of Si, C and GaAs we used 40, 40, and 100 random 
$\bk$-points, respectively.

The expansion of the crystalline lattice with temperature was only considered in the case
of GaAs. This choice is justified by the fact that the calculated band gaps of silicon and diamond
change by less than 2~meV when using the lattice parameters at $T=0$~K and $T=300$~K.
The temperature-dependence of the lattice parameters of Si and C was taken from 
Ref.~\onlinecite{Okada1984,Sato_2002}.

\section{Conclusions and outlook} \label{sec.conclusion}

In this manuscript we developed a new {\it ab initio} computational method for calculating
the temperature-dependent optical absorption spectra and band gaps of semiconductors,
including quantum zero-point effects. The present work significantly expands the scope of
our previous investigation in Ref.~\onlinecite{Zacharias_2015}, by completely removing the
need for stochastic sampling of the nuclear wavefunctions. In particular we demonstrated,
both using a formal proof and by means of explicit first-principles calculations, that in order
to compute dielectric functions and band gaps at finite temperature it is sufficient
to perform a {\it single} supercell calculation with the atoms in a well-defined configuration,
as given by Eq.~(\ref{eq.conf_opt}).

Using this new technique we reported the first calculations of the complete optical
absorption spectra of diamond and gallium arsenide including electron-phonon interactions,
and we confirmed previous results obtained for silicon in Refs.~\onlinecite{Noffsinger,Zacharias_2015}.
Our calculations are in good agreement with experiment at the
level of lineshape, location of the absorption onset, and magnitude of the absorption
coefficient. From these calculations we extracted the temperature-dependence and
the zero-point renormalization of the direct band gaps of Si, C, and GaAs, and of the indirect
gaps of Si and C. Our calculations are in good agreement with previous theoretical studies.

Our present work relies on the Williams-Lax theory of optical transitions including
nuclear quantum effects.\cite{Williams,Lax} For completeness we investigated in detail the formal relation between
the WL theory, the Allen-Heine theory of temperature-dependent band structures, and the
Hall-Bardeen-Blatt theory of phonon-assisted optical absorption. We demonstrated that
both the AH theory and the HBB theory can be derived as low-order approximations of the
WL theory.

We emphasize that our present approach enables calculations of {\it complete} optical
spectra at finite temperature, including seamlessly direct and indirect optical absorption.
This feature is useful in order to calculate spectra which are directly comparable to
experiment in absolute terms, i.e.~without arbitrarily rescaling the absorption coefficient 
or shifting the absorption onset to fit experiment. Our methodology will also be useful for 
{\it predictive} calculations of optical spectra, for example in the context of high-throughput
computational screening of materials.

It is natural to think that the present approach could be upgraded with calculations of electronic
structure and optical properties based on the GW/Bethe-Salpeter
approach. Indeed, as it should be clear from Eq.~(\ref{eq.wl}), our methodology holds unchanged
irrespective of the electronic structure technique employed to describe electrons
at clamped nuclei. Since the present approach requires only one calculation in a large
supercell, it is possible that complete GW/Bethe-Salpeter calculations of optical
spectra including phonon-assisted processes will soon become feasible.
In this regard we note that, recently, Ref.~\onlinecite{Bartomeu_2015} proposed the so-called
`non-diagonal' supercells in order to perform accurate supercell calculations at
a dramatically reduced computational cost. In the future, it will be interesting to 
investigate how to take advantage of nondiagonal supercells in order to make
the present methodology even more efficient.

Finally, it should be possible to generalize our present work to other
important optical and transport properties. In fact, the WL theory is completely
general and can be used with any property which can be described
by the Fermi Golden Rule. For example, we expect that generalizations to properties
such as photoluminescence or Auger recombination\cite{Kioupakis_2015}
should be within reach.

\acknowledgments
The research leading to these results has received funding from the 
the UK Engineering and Physical Sciences Research Council 
(DTA scholarship of M.Z. and grants No. EP/J009857/1 and EP/M020517/1),
the Leverhulme Trust (Grant RL-2012-001), and the Graphene Flagship (EU FP7 grant no. 604391).
The authors acknowledge the use of the University of Oxford Advanced 
Research Computing (ARC) facility (http://dx.doi.org/
10.5281/zenodo.22558), the ARCHER UK National 
Supercomputing Service under the `AMSEC' Leadership project 
and the PRACE for awarding us access to the dutch 
national supercomputer 'Cartesius’.

\appendix
\numberwithin{figure}{section}
\section{Williams-Lax expression for direct absorption}\label{app.WL}

In this Appendix we outline the steps leading from Eq.~(\ref{eq.eps-dir-tmp1}) to Eq.~(\ref{eq.dir.3}).
We start by introducing the compact notation:
  \begin{equation}\label{eq.dex}
  \Delta \ve_{cv}^x =  \sum_\nu A_{cv\nu} x_\nu + \sum_{\mu\nu} B_{cv\mu\nu} x_\mu x_\nu + \mathcal{O}(x^3),
  \end{equation}
where the coefficients $A_{cv\nu}$ and $B_{cv\mu\nu}$ are obtained from Eq.~(\ref{eq.E-exp}):
  \begin{eqnarray}
   A_{cv\nu} &=& \frac{1}{l_\nu}\left(g_{cc\nu}-g_{vv\nu}\right), \label{eq.coeff.A}\\
   B_{cv\mu\nu} &=& \frac{1}{l_\mu l_\nu}
     {\sum_n}^\prime \left[ \left( \frac{ g_{cn\mu} g_{nc\nu} }{\ve_c -\ve_n} +h_{c\mu\nu}\, \right)
         \right. \nonumber \\
      && \left. \hspace{1.15cm} -\left(  \frac{ g_{vn\mu} g_{nv\nu} }{\ve_v -\ve_n} +h_{v\mu\nu}\right)  \right].
        \label{eq.coeff.B}
  \end{eqnarray}
Equation~(\ref{eq.eps-dir-tmp1}) can be simplified by using the Taylor expansion of the Dirac delta
with respect to the energy argument. In general we have:\cite{William_book}
  \begin{equation}\label{eq.delta-exp}
  \d(\e+\eta) = \sum_{n=0}^\infty\frac{(-1)^n}{n!}\left.\frac{\D^n \d}{\,\,\D\e^n}\right|_{\e} \eta^n,
  \end{equation}
therefore we can set $\ve=\ve_{cv}-\hbar\w$ and $\eta=\Delta\ve_{cv}^x$, 
and replace inside Eq.~(\ref{eq.eps-dir-tmp1}). We find:
  \begin{eqnarray}
  \e_2(\w;T) & =& \frac{2 \pi }{ m_{\rm e} N_{\rm e} } \frac{\w_{\rm p}^2}{\,\w^2} \sum_{cv} | p_{cv}|^2
  \sum_{n=0}^\infty\frac{(-1)^n}{n!}\left.\frac{\D^n \d}{\D(\hbar\w)^n}\right|_{\hbar\w-\ve_{cv}}
  \nonumber \\
  &\times& {\prod}_\nu \int\! dx_\nu \frac{\exp(-x_\nu^2/2\sigma_{\nu,T}^2)}{\sqrt{2\pi}\sigma_{\nu,T}}
  (\Delta\ve_{cv}^x)^n.  
  \end{eqnarray} 
Now we replace $\Delta\ve_{cv}^x$ from Eq.~(\ref{eq.dex}) and carry out the integrals in the coordinates
$x_\nu$.
The resulting expression does not contain the normal coordinates any more, and the various derivatives
of the delta function can be regrouped using Eq.~(\ref{eq.delta-exp}) in reverse. The result is:
     \begin{eqnarray}\label{eq.dir.1}
   && \e_2(\w;T) = \frac{2 \pi }{ m_{\rm e} N_{\rm e} } 
    \frac{\w_{\rm p}^2}{\,\w^2} \sum_{cv} | p_{cv}|^2
        \nonumber \\
    && \times \left[
        \d\!\left(\ve_{cv,T}^{\rm AH}- \hbar\w \right) +
         \frac{1}{2}\sum_\nu A_{cv\nu}^2 \sigma_{\nu,T}^2
        \frac{\D^2 \d(\hbar\w-\ve_{cv})}{\,\,\D(\hbar\w)^2} \right] 
       \nonumber \\ &&+ \mathcal{O}(\sigma^4),
  \end{eqnarray}
where $\ve_{cv,T}^{\rm AH} = \ve_{c,T}^{\rm AH} - \ve_{v,T}^{\rm AH}$, and
$\ve_{m,T}^{\rm AH}$ is the temperature-dependent electron energy
given by Eq.~(\ref{eq.E.AH}).

In order to make Eq.~(\ref{eq.dir.1}) more compact, it is convenient to rewrite the second term
inside the square brackets by performing a Taylor expansion of the Dirac delta around $\ve_{cv,T}^{\rm AH}$,
and group the terms proportional to $\sigma_{\nu,T}^4$ and higher order inside 
the term $\mathcal{O}(\sigma^4)$ at the end. This step leads to:
     \begin{eqnarray}\label{eq.dir.2}
  &&\e_2(\w;T) = \frac{2 \pi }{ m_{\rm e} N_{\rm e} } \frac{\w_{\rm p}^2}{\,\w^2} \sum_{cv} | p_{cv}|^2
        \nonumber \\
    &&\,\,\, \times \left[ 1 + \frac{1}{2}\Gamma_{cv}^2 \frac{\D^2 }{\,\,\D(\hbar\w)^2} \right]                   
         \d\!\left(\ve_{cv,T}^{\rm AH} - \hbar\w \right) + \mathcal{O}(\sigma^4),
  \end{eqnarray}
having defined:
    \begin{equation}\label{eq.gamma}
   \Gamma_{cv}^2 = 
    {\sum}_\nu A_{cv\nu}^2 \sigma_{\nu,T}^2 = {\sum_\nu} \left(g_{cc\nu}-g_{vv\nu}\right)^2 (2n_{\nu,T}+1).
    \end{equation}
In the final step we note that the term $\D^2/\D(\hbar\w)^2$ in Eq.~(\ref{eq.dir.2})
acts so as to broaden the lineshape. This is seen by using the Fourier representation of the Dirac delta,
$\delta(\hbar\w) = (2 \pi\hbar)^{-1}\int dt \exp(-i\w t)$. We find:
  \begin{eqnarray}\label{eq.fourier.1}
    &&  \left[ 1 + \frac{1}{2}\Gamma_{cv}^2 \frac{\D^2 }{\,\,\D(\hbar\w)^2} \right]
         \d\!\left(\ve_{cv,T}^{\rm AH} - \hbar\w \right)  \nonumber \\
      && = 
       \frac{1}{2\pi\hbar} \int dt  \left[1-\frac{\Gamma_{cv}^2}{2 \hbar^2} t^2 \right]
       \exp\left(-i \frac{\ve_{cv,T}^{\rm AH} - \hbar\w}{\hbar}t\right)\!\!.\,\,
  \end{eqnarray}
The term within the square brackets corresponds to the first order Taylor expansion of a Gaussian 
(or alternatively a Lorentzian), therefore the last expression can be rewritten as:
   \begin{equation}\label{eq.fourier.2}
       \frac{1}{2\pi\hbar} \int dt  \exp\left[-\frac{\Gamma_{cv}^2}{2\hbar^2} t^2 \right]
       \exp\left(-i \frac{\ve_{cv,T}^{\rm AH} - \hbar\w}{\hbar}t\right) + \mathcal{O}(\sigma^4),
  \end{equation}
where we recognize the Fourier transform of a Gaussian.  An explicit evaluation of the integral yields
also a Gaussian, and the final result is Eq.~(\ref{eq.dir.3}).

\section{Williams-Lax expression for indirect absorption}\label{app.WL-ind}
In this Appendix we outline the derivation of Eq.~(\ref{eq.ind-final}) starting from 
Eq.~(\ref{eq.eps-dir-tmp2}). It is convenient to introduce the notation:
  \begin{equation}
  C_{cv\mu} = {\sum_n}^\prime \left[\frac{p_{cn}\, g_{nv\mu}}{\ve_v-\ve_n}
          +\frac{g_{cn\mu}\, p_{nv}}{\ve_c-\ve_n}\right] \!\frac{1}{l_\mu},
  \end{equation}
so that Eq.~(\ref{eq.eps-dir-tmp2}) can be written as:
  \begin{eqnarray}\label{eq.hello1}
  &&\!\!\!\e_2(\w;T)
   = \frac{2 \pi }{ m_{\rm e} N_{\rm e} } \frac{\w_{\rm p}^2}{\,\w^2} \sum_{cv}
    {\prod}_\nu \int\! dx_\nu \frac{\exp(-x_\nu^2/2\sigma_{\nu,T}^2)}{\sqrt{2\pi}\sigma_{\nu,T}} \nonumber \\
    && \!\!\!\times \sum_{\mu\mu'} C_{cv\mu}C_{cv\mu'} x_{\mu}x_{\mu'}
  \,\delta(\ve_{cv} + \Delta\ve_{cv}^x -\hbar\w) + \mathcal{O}(\sigma^4).\,\,\,\,\,
  \end{eqnarray}
We now write explicitly the expansion of the Dirac delta to second order in $x_\nu$,
using Eqs.~(\ref{eq.dex})-(\ref{eq.delta-exp}):
  \begin{eqnarray}\label{eq.b3}
  && \delta(\ve_{cv} + \Delta\ve_{cv}^x -\hbar\w) =  \delta(\ve_{cv} -\hbar\w) \nonumber \\ 
   && -\left.\frac{\D \d}{\D(\hbar\w)}\right|_{\hbar\w-\ve_{cv}} 
   \left[\sum_\nu A_{cv\nu} x_\nu + \sum_{\mu\nu} B_{cv\mu\nu} x_\mu x_\nu\right] \nonumber \\
  && + \frac{1}{2} \left.\frac{\D^2\d}{\D(\hbar\w)^2}\right|_{\hbar\w-\ve_{cv}}
  \left[\sum_{\mu\nu} A_{cv\mu} A_{cv\nu} x_\mu x_\nu \right] + \mathcal{O}(x^3). \,\,\,\,\,
  \end{eqnarray}
By replacing the last expression in Eq.~(\ref{eq.hello1}) and perfoming the integrations,
after a few lengthy but straightforward manipulations we find:
\hide
    \begin{eqnarray}
  &&\!\!\!\e_2(\w;T)
   = \frac{2 \pi }{ m_{\rm e} N_{\rm e} } \frac{\w_{\rm p}^2}{\,\w^2} \sum_{cv}
    {\prod}_\nu \int\! dx_\nu \frac{\exp(-x_\nu^2/2\sigma_{\nu,T}^2)}{\sqrt{2\pi}\sigma_{\nu,T}} \nonumber \\
    && \!\!\!\times \sum_{\mu\mu'} C_{cv\mu}C_{cv\mu'} x_{\mu}x_{\mu'}
  \left\{ \delta(\ve_{cv} -\hbar\w)
   -\left.\frac{\D \d}{\D(\hbar\w)}\right|_{\hbar\w-\ve_{cv}}
    \left[\sum_{\mu\nu} B_{cv\mu\nu} x_\mu x_\nu\right] 
    + \frac{1}{2} \left.\frac{\D^2\d}{\D(\hbar\w)^2}\right|_{\hbar\w-\ve_{cv}}
  \left[\sum_{\mu\nu} A_{cv\mu} A_{cv\nu} x_\mu x_\nu \right] + \mathcal{O}(x^3)
   \right\} + \mathcal{O}(\sigma^4).\,\,\,\,\,
  \end{eqnarray}
 \begin{eqnarray}
 &&\e_2(\w;T)
   = \frac{2 \pi }{ m_{\rm e} N_{\rm e} } \frac{\w_{\rm p}^2}{\,\w^2} \sum_{cv} 
  \left\{ \sum_\nu |C_{cv\nu}|^2 \sigma_{\nu,T}^2 \delta(\ve_{cv} -\hbar\w)
    -\left.\frac{\D \d}{\D(\hbar\w)}\right|_{\hbar\w-\ve_{cv}} 
    \left[ \sum_{\mu\nu} |C_{cv\mu}|^2 B_{cv\nu\nu} \sigma_{\mu,T}^2 \sigma_{\nu,T}^2
            + 2\sum_\nu |C_{cv\nu}|^2 B_{cv\nu\nu} \sigma_{\nu,T}^4 \right.\right.\nonumber \\
   && \left.\left.   + \sum_{\mu\ne \mu'}\sum_{\nu\ne \nu'} C_{cv\mu}C_{cv\mu'} B_{cv\nu\nu'} \left( 
      \sigma_{\nu,T}^2 \sigma_{\nu',T}^2 +\sigma_{\mu,T}^2 \sigma_{\mu',T}^2 \right)\right] \right. \nonumber \\
  && \left. + \frac{1}{2} \left.\frac{\D^2\d}{\D(\hbar\w)^2}\right|_{\hbar\w-\ve_{cv}}
   \left[ \sum_{\mu\nu} |C_{cv\mu}|^2 |A_{cv\nu}|^2 \sigma_{\mu,T}^2 \sigma_{\nu,T}^2
            + 2\sum_\nu |C_{cv\nu}|^2 |A_{cv\nu}|^2 \sigma_{\nu,T}^4 \right.\right.\nonumber \\
   && \left.\left.   + \sum_{\mu\ne \mu'}\sum_{\nu\ne \nu'} C_{cv\mu}C_{cv\mu'} A_{cv\nu}A_{cv\nu'} \left(
      \sigma_{\nu,T}^2 \sigma_{\nu',T}^2 +\sigma_{\mu,T}^2 \sigma_{\mu',T}^2 \right)\right]
  \right\} + \mathcal{O}(\sigma^4).
 \end{eqnarray}
We neglect interference-terms (i.e. all terms which are not appearing as $|\cdots|^2$), and we
rearrange:
  \begin{eqnarray}
 &&\e_2(\w;T)
   = \frac{2 \pi }{ m_{\rm e} N_{\rm e} } \frac{\w_{\rm p}^2}{\,\w^2} \sum_{cv}
   \sum_\nu |C_{cv\nu}|^2 \sigma_{\nu,T}^2 
      \left[ 1-\sum_{\mu} (1+2\d_{\mu\nu})B_{cv\mu\mu} \sigma_{\mu,T}^2 \frac{\D \d}{\D(\hbar\w)}
     + \frac{1}{2} \sum_{\mu} (1+2\d_{\mu\nu}) |A_{cv\mu}|^2 \sigma_{\mu,T}^2
         \frac{\D^2\d}{\D(\hbar\w)^2}
    \right]
     \delta(\ve_{cv} -\hbar\w) 
   + \mathcal{O}(\sigma^4).
 \end{eqnarray}
 \begin{eqnarray}
 &&\e_2(\w;T)
   = \frac{2 \pi }{ m_{\rm e} N_{\rm e} } \frac{\w_{\rm p}^2}{\,\w^2} \sum_{cv}
   \sum_\nu |C_{cv\nu}|^2 \sigma_{\nu,T}^2
      \left[ 1-\sum_{\mu} (1+2\d_{\mu\nu})B_{cv\mu\mu} \sigma_{\mu,T}^2 \frac{\D \d}{\D(\hbar\w)}
     + \frac{1}{2} \sum_{\mu} (1+2\d_{\mu\nu}) |A_{cv\mu}|^2 \sigma_{\mu,T}^2
         \frac{\D^2\d}{\D(\hbar\w)^2}
    \right]
     \delta(\ve_{cv} -\hbar\w)
   + \mathcal{O}(\sigma^6).
 \end{eqnarray}
 Here we can neglect the terms with the Kronecker delta for the following reason:
there are $\sim N$ such terms, but there are $\sim N^2$ similar terms coming from $\sum_\mu$.
Since both types of terms are positive definite (i.e. we are not comparing the diagonal and
off-diagonal elements of a matrix, but products of positive-definite terms), the $\sim N^2$
terms dominate over the $\sim N$ terms for large $N$. This is always the case in {\it extended} solids.
\unhide
   \begin{eqnarray}\label{eq.almost-final}
 \e_2(\w;T)
   &=& \frac{2 \pi }{ m_{\rm e} N_{\rm e} } \frac{\w_{\rm p}^2}{\,\w^2} \sum_{cv}
   \sum_\nu |C_{cv\nu}|^2 \sigma_{\nu,T}^2 \nonumber \\ 
   &\times& \left[ 1-\sum_{\mu} B_{cv\mu\mu} \sigma_{\mu,T}^2 \frac{\D \d}{\D(\hbar\w)} \right.
    \nonumber \\ &+& \left. \frac{1}{2} \sum_{\mu} |A_{cv\mu}|^2 \sigma_{\mu,T}^2
         \frac{\D^2\d}{\D(\hbar\w)^2}
    \right]
     \delta(\ve_{cv} -\hbar\w) \nonumber \\ &+& \mathcal{O}(\sigma^6).
 \end{eqnarray}
In this expression we neglected interference terms of the type
$\sum_{\mu\nu\ne \l\k} C_{cv\mu}C_{cv\l} B_{cv\nu\k} 
      \sigma_{\nu,T}^2 \sigma_{\k,T}^2$ next to positive-definite terms
of the same order, such as for example $\sum_{\mu\nu} |C_{cv\mu}|^2 B_{cv\nu\nu} \sigma_{\mu,T}^2 \sigma_{\nu,T}^2$.
This is justified since in the case of extended solids the former sum tends to zero as the
phases of each term cancel out in average. In a similar spirit, we neglected terms
like $2 B_{cv\nu\nu} \sigma_{\nu,T}^2$ next to $\sum_{\mu} B_{cv\mu\mu} \sigma_{\mu,T}^2$.
This is justified as each term in the sum is positive definite, and the number of these terms 
goes to infinity in extended solids.

Now Eq.~(\ref{eq.almost-final}) can be recast in the form given by Eq.~(\ref{eq.ind-final})
by using the Fourier representation of the Dirac delta, and following the same steps as 
in Appendix~\ref{app.WL}, Eqs.~(\ref{eq.fourier.1})-(\ref{eq.fourier.2}). In particular, the term 
$\D \d/\D(\hbar\w)$ yields a shift of the excitation energies which coincides with the
energy renormalization in the Allen-Heine theory, see Eq.~(\ref{eq.E.AH}). Similarly, the term $\D^2\d/\D(\hbar\w)^2$
yields a broadening of the absorption peak, which corresponds to the linewidth $\Gamma_{cv}$
of Eq.~(\ref{eq.gamma.main}).

\section{Calculation of the zero-point band gap renormalization using the joint density of states}~\label{app.equil_JDOS}

\begin{figure}[t!]
 \begin{center}
	 \hspace*{0.5cm}
 \subfloat{\includegraphics[width=0.305\textwidth]{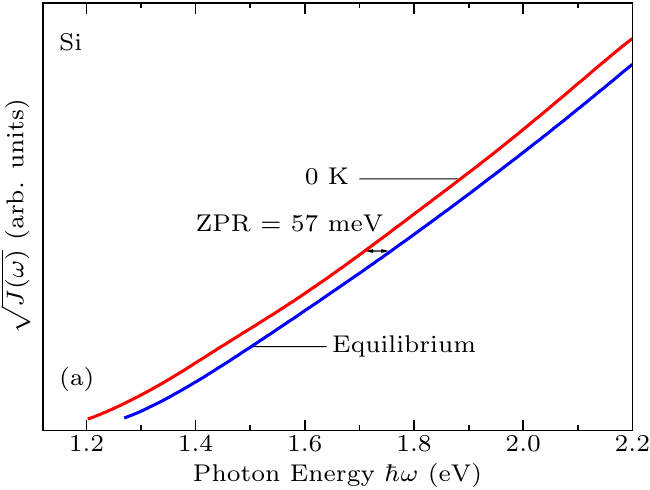}}
 \newline
 \vspace*{-0.1cm}
   \hspace*{0.1cm}
   \subfloat{\includegraphics[width=0.3\textwidth]{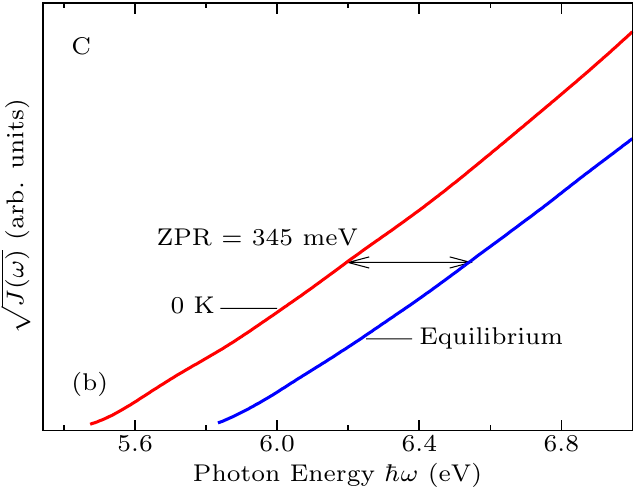}}
   \vspace*{-0.1cm}
   \hspace*{-0.8cm}
   \newline
   \subfloat{\includegraphics[width=0.305\textwidth]{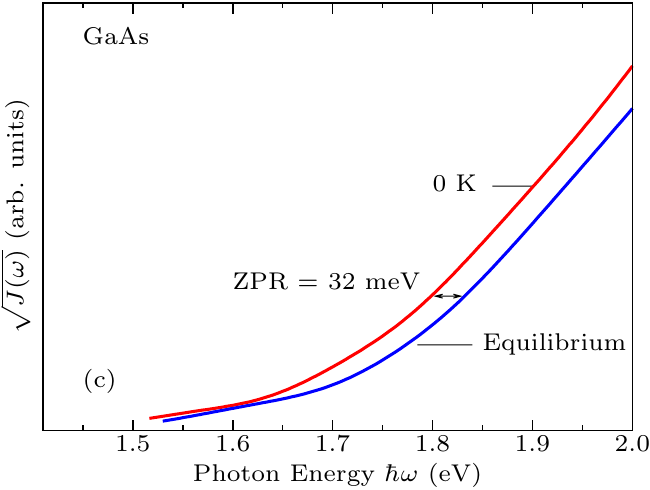}}
 \end{center}
\vspace*{-0.45cm}
\caption{\label{app.fig.fits_renorm}
  Square-root of the joint density of states of (a) silicon, (b) diamond, and (c) gallium arsenide.
  The calculations were performed with nuclei clamped in their equilibrium positions (blue lines),
  and using the one-shot WL method at $T=0$~K (red lines).
  The horizontal offset between blue and red curves corresponds to the zero-point renormalization
  of the band gap in each case.
  The calculations for the equilibrium structures were performed in the primitive unit cells,
  using 20480, 20480 and 51200 random {\bf k}-points for Si, C, and GaAs, respectively.
  The WL calculations were performed on $8\!\times\!8\!\times\!8$ supercells, using 40, 40,
  and 100 random {\bf k}-points for Si, C, and GaAs, respectively.
  A Gaussian broadening of 30~meV was used in all plots.
  }
\end{figure}

In this Appendix we discuss an accurate procedure for determining the zero-point band gap renormalization
within a supercell calculation.

We consider the joint density of states (JDOS), defined as the convolution between the
density of states of valence and conduction bands:\cite{Cardona_Book}
   \begin{equation}\label{eq.jdos}
J(\w) = \sum_{cv} \d\!\left(\ve_{cv} - \hbar\w \right).
\end{equation}
Within an energy range where band extrema are parabolic, it can be shown that $J(\w)={\rm const}\times
(\hbar\w-E_{\rm g})^2$. This relation holds both for direct-gap and indirect-gap semiconductors.
This is the basic relation underpinning the use of Tauc's plots in experimental spectra.\cite{Tauc,Parravicini}
Within the WL theory, the previous relation can be shown to remain essentially unchanged:
\begin{eqnarray}\label{eq.jdos_wl}
J(\w;T) = \sum_{cv}
\d\!\left(\ve_{cv,T}^{\rm AH} - \hbar\w \right) + \mathcal{O}(\sigma^4).
\end{eqnarray}
This result can be obtained by following the same prescription used to obtain Eq.~(\ref{eq.dir.2}). 
Therefore also in this case we have $J(\w,T)={\rm const}\times
(\hbar\w-E_{{\rm g},T})^2$. 

Using Eqs.~(\ref{eq.jdos}) and (\ref{eq.jdos_wl}) we can determine the zero-point
renormalization of the band gap as the horizontal 
offset between the curves $J(\w)^{1/2}$ and $J(\w,T=0)^{1/2}$.
This procedure is very accurate because we are comparing two calculations
executed under identical conditions; therefore the numerical errors arising
from the choice of the energy range, the Gaussian broadening and the Brillouin-zone sampling
tend to cancel out.

In  Figs.~\ref{app.fig.fits_renorm}~(a),~(b) and~(c) we show the square-root of the joint density of states
of silicon, diamond and gallium arsenide, calculated with the atoms at their relaxed positions
(red lines), and the corresponding one-shot WL calculations at 0~K (blue lines).
In all cases the two curves are parallel, and
the horizontal offset between the curves gives the zero-point renormalization.
Using this method, our values of the zero-point renormalization of the band
gaps of Si, C and GaAs are 57~meV, 345~meV and 32~meV, respectively.

We note that this new procedure critically requires the JDOS as opposed to the optical spectra,
since in the case of indirect-gap materials we have no optical absorption below the
direct gap in the equilibrium structure.

\section{Analysis of multi-phonon contributions to the band gap renormalization of diamond}~\label{app.WL_AH}

\begin{figure}[h!]
	\begin{center}
		\subfloat{\includegraphics[width=0.36\textwidth]{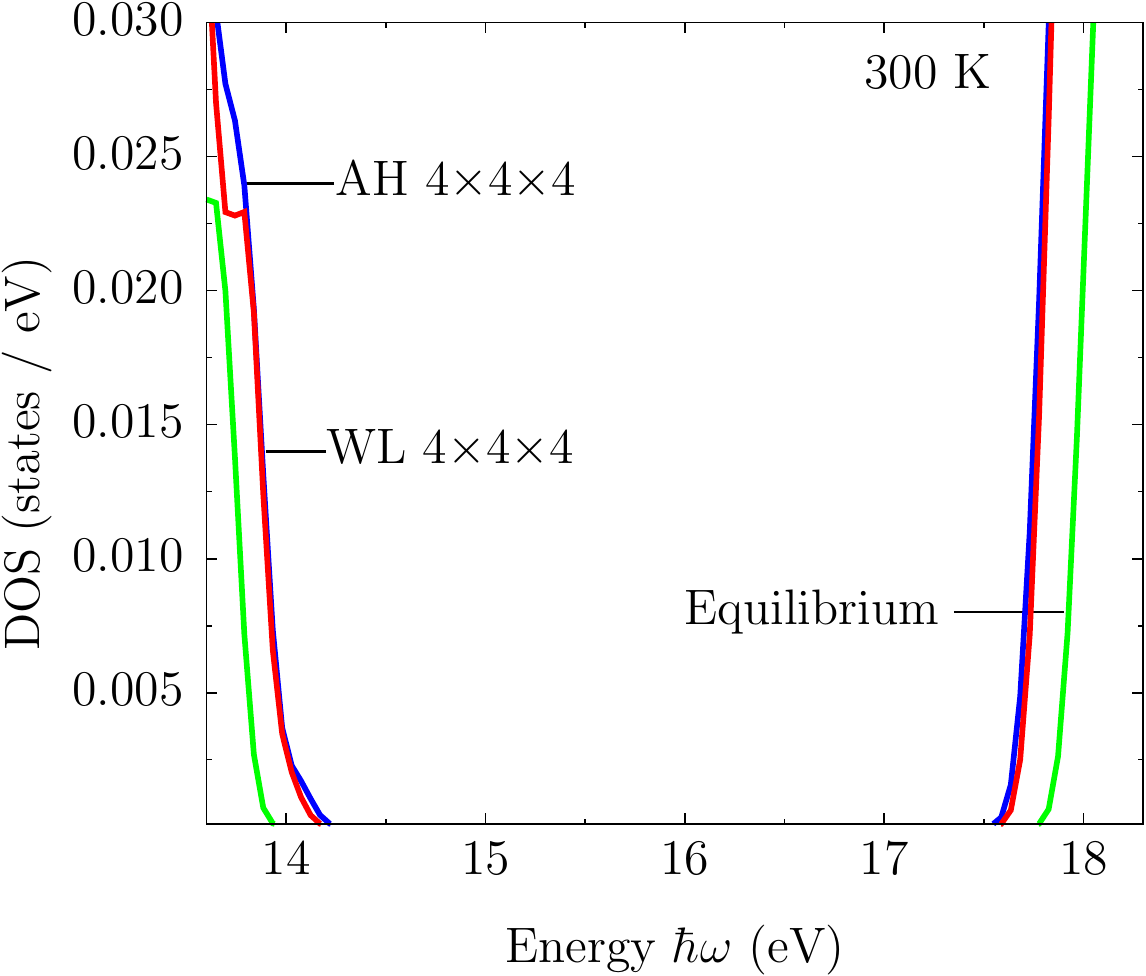}}
	\end{center}
	\vspace*{-0.5cm}
	\caption{\label{app.figD1}
		Density of states of diamond close to the valence and conduction band edges:
                calculation at the equilibrium geometry (green line), one-shot WL calculation
                (red line), and AH calculation (blue line). The latter two calculations were
                performed for $T=300$~K, using a $4\times4\times4$ supercell and
                65 {\bf k}-points in the Brillouin zone of the supercell.
		The calculation with the atoms at their equilibrium geometry was carried out 
                within the primitive unit cell, using 4160 {\bf k}-points.
                A Gaussian smearing of 30~meV was used for all curves.
}
\end{figure}

In this Appendix we first establish the link between our one-shot WL calculation
and multi-phonon processes in the AH theory, and then show that multi-phonon effects
are essentially negligible in diamond.

We perform an expansion of the Kohn-Sham energy $\ve_{n}^x$ in terms of
normal-mode coordinates to obtain: 
  \begin{eqnarray}\label{eq.exp_AH_4}
    \e_{n}^x &=& \e_n + \sum_\nu \frac{\D \e_n^x}{\D x_\nu} x_\nu   
      + \frac{1}{2}\sum_{\mu\nu} \frac{\D^2 \e_n^x}{\D x_\mu \D x_\nu} x_\mu x_\nu \nonumber \\
  &+& \frac{1}{3!}\sum_{\mu\nu\l} \frac{\D^3 \e_n^x}{\D x_\mu \D x_\nu \D x_\l} x_\mu x_\nu x_\l \nonumber \\
  &    +& \frac{1}{4!}\sum_{\mu\nu\l\k} \frac{\D^4 \e_n^x}{\D x_\mu \D x_\nu \D x_\l \D x_\k} x_\mu x_\nu x_\l x_\k  
  + \mathcal{O}(x^5).
\end{eqnarray}
The AH theory is obtained by taking the thermal averages of $x_\nu$,
$x_\mu x_\nu$, $x_\mu x_\nu x_\l$, and $x_\mu x_\nu x_\l x_\k$. The result is:
  \begin{eqnarray}\label{eq.exp_AH_T_4}
   \e_{n,T}^{\rm AH} &=& \e_n
  +\frac{1}{2}\sum_\nu \frac{\D^2 \e_n^x}{\D x_\nu^2} \sigma_{\nu,T}^2 
   + \frac{3}{4!}\sum_{\mu\ne\nu} \frac{\D^4 \e_n^x}{\D x_\mu^2 \D x_\nu^2}\sigma_{\mu,T}^2 \sigma_{\nu,T}^2 \nonumber \\
   &  +& \frac{3}{4!}\sum_{\nu} \frac{\D^4 \e_n^x}{\D x_\nu^4}\sigma_{\nu,T}^4 
     + \mathcal{O}(\sigma^6).
\end{eqnarray}
The second term on the r.h.s. represents the energy-level renormalization within the
AH theory, and accounts only for single-phonon processes. This was discussed 
in Sec.~\ref{sec.general-theory}. The third and fourth terms of the r.h.s. represent 
the two-phonon contribution to the energy-level renormalization in the AH theory. 

Now we show how our one-shot calculation correctly captures the two-phonon contribution.
By replacing our `optimal configuration' from Sec.~\ref{sec.method-results} inside
Eq.~(\ref{eq.exp_AH_4}), and taking the
limit of $N\rightarrow\infty$, we find:
\begin{eqnarray}\label{eq.exp_1C_4}
	  \e_{n,T}^{\rm 1C} &=& \e_n
        +\frac{1}{2}\sum_{\nu} \frac{\D^2 \e_n^x}{\D x_\nu^2} \sigma_{\nu,T}^2  
    + \frac{3}{4!}\sum_{\mu\ne\nu} \frac{\D^4 \e_n^x}{\D x_\mu^2 \D x_\nu^2}\sigma_{\mu,T}^2 \sigma_{\nu,T}^2 \nonumber \\
    &     +& \frac{1}{4!}\sum_{\nu} \frac{\D^4 \e_n^x}{\D x_\nu^4}\sigma_{\nu,T}^4 + \mathcal{O}(\sigma^6).
\end{eqnarray}
By comparing Eqs.~(\ref{eq.exp_AH_T_4}) and (\ref{eq.exp_1C_4}) we see that the one-shot
method and the fourth-order AH theory give the same result, except for a contribution
$\frac{2}{4!}\sum_{\nu} \frac{\D^4 \e_n^x}{\D x_\nu^4} \sigma_{\nu,T}^4$. In the limit of
$N\rightarrow\infty$ this contribution is negligible as compared to the other terms.
This reasoning is analogous to the discussion in Appendix~\ref{app.WL-ind}. 

Therefore we can conclude that the one-shot WL method captures not only the standard
one-phonon processes of the AH theory, but also multi-phonon contributions. The
two-phonon contributions are identical to what one would obtain from carrying the
AH theory to fourth-order in the atomic displacements.

In order to investigate the energy-level renormalization coming from two-phonon
and higher multi-phonon processes, we calculated the density of states (DOS) of diamond
within (i) the standard (one-phonon) AH theory, and (ii) the one-shot WL method. These quantities
are shown in Fig.~\ref{app.figD1}, for $T=300$~K, in blue and red, respectively.
For comparison, we also show the DOS calculated with the atoms in their equilibrium positions (green).
The AH correction to the DOS was obtained by evaluating the derivatives
$\D^2 \e_n^x/\D x_\nu^2$ by means of finite differences. This required
2N frozen-phonon calculations.

Figure~\ref{app.figD1} shows that the DOS obtained from the standard AH theory and 
from the one-shot WL method (which includes multi-phonon effects) essentially coincide. 
This demonstrates that multi-phonon contributions to the band gap renormalization 
are negligible in diamond.

\bibliography{references}{} 

\end{document}